\documentclass[aip,reprint,jcp]{revtex4-2}

\usepackage[T1]{fontenc}
\usepackage[utf8]{inputenc}

\usepackage[english]{babel}

\usepackage{amssymb,amsthm,mathtools}
\usepackage{amsfonts,dsfont}

\usepackage{mathptmx}

\usepackage{csquotes,xcolor}
\usepackage{graphicx,siunitx,setspace}
\usepackage{subcaption}

\usepackage[unicode,bookmarksnumbered]{hyperref}

\usepackage[capitalise]{cleveref}
\crefname{section}{Sec.}{Secs.}
\crefname{figure}{Fig.}{Figs.}

\usepackage{tikz}
\usetikzlibrary{arrows.meta}
\tikzset{>={Latex[width=2mm,length=2mm]}}

\newcommand{\eps}{\ensuremath \varepsilon}
\newcommand{\Id}{\ensuremath \mathrm{Id}}
\newcommand{\Z}{\ensuremath \mathbb{Z}}

\newcommand{\Medskip}{\medskip\noindent}

\theoremstyle{definition}
\newtheorem{example}{Example}

\begin{document}

\title[Hierarchical models for large chemical reaction networks]{Hierarchical models for large chemical reaction networks}

\author{Jérémie Unterberger}
\email{jeremie.unterberger@univ-lorraine.fr}
\affiliation{Université de Lorraine, CNRS, IECL, F-54000 Nancy, France}

\author{Ulysse Herbach}
\affiliation{Université de Lorraine, CNRS, Inria, IECL, F-54000 Nancy, France}

\author{Roxane Cellier}
\affiliation{Université de Lorraine, CNRS, Inria, IECL, F-54000 Nancy, France}

\date{\today}

\begin{abstract}
The quest for the origin of life, especially in the metabolism-first scenario brought forth by the celebrated Miller–Urey experiment, has triggered a research program dedicated to studying the emergence of complex dynamical behaviors in large chemical mixtures. Though autocatalysis, understood as the capacity of a system driven by chemical reactions to grow exponentially, has been recognized as a potential key factor driving instability or multistability, no quantitative theory has yet emerged, partly due to the lack of available kinetic data. We introduce a computational tool for large chemical reaction networks, based on a scale-splitting algorithm inspired by Wilson's renormalization group. We focus here on dilute regimes, i.e., the special case when species of interest have low concentration, so that non-unimolecular reactions may be neglected, and the time evolution is close to linear. The main feature of such networks is their ability to exhibit autocatalytic dynamics, depending on parameter thresholds. Our algorithm takes as input a network structure, and outputs (1) a simplified effective graph representation of the network containing exactly the main reaction pathways; the effective graph is hierarchical, meaning that its vertices are obtained by recursively coarse-graining particular subgraphs; (2) a finite number of analytical formulas in terms of kinetic rates for the dynamics of the system, called hierarchical formulas, which are approximate but easily interpretable, accurate when scale separation is effective, the coarse-graining offering a reliable picture of the dynamical behavior at different time-scales. The domain of validity of each formula can be interpreted as a kinetic phase of the network: each phase typically produces a different dynamical pattern of chemical composition. We show on a simple example that this approach allows fast and reliable statistical inference of kinetic rates from a time-series of concentration measurements. Hierarchical formulas have been implemented as a Python package, and are discussed on a simplified model of the formose reaction.
\end{abstract}

\maketitle


\section{Introduction}

Synthesis chemistry is traditionally concerned with finding a synthesis path with best possible yield for a given
compound. This point of view is however challenged in problems relating to complex metabolism, since the high diversity of molecules and potential pathways suggests instead a
detailed chemical analysis of the composition of samples produced in a one-pot experiment, possibly in a high-throughput framework, with a view to studying the dependence on chemical
conditions (pH, fluxes, catalysts...), see e.g., \onlinecite{Semenov2016,Muchowska2019,Robinson2022}. In presence of a large number of unknown reaction intermediates, mass spectrometers do not allow to disambiguate isomers, suggesting the use of techniques based on similarity indices and chemical expertise. These difficulties are compounded in experiments pertaining to the origin of life, since
there is no preferred synthesis goal, and time becomes an essential variable on the way to open-ended evolution.

In our view, what is lacking here is a systematic computational method to derive the composition of a large-diversity sample that
would capture
reliably the structure of the reaction network as a whole instead of trying to infer individual intermediates. This question can be taken both ways. Imagining one had a \enquote*{good} knowledge of the stoichiometry and kinetics of the reaction network, one would like to get a description of the main reaction pathways and time-dependent composition. Conversely, given the partial information on composition given by
mass spectroscopy data, one would like to infer the network (i.e., the main pathways) and kinetic rates. Implementing blindly the inference task using numerical ODE solvers would require to scan a large part of the rate space, which is impracticable for large
networks. Also, in presence of rates spanning several orders of magnitude, as is typically the case in chemistry, ODE solvers face a
well-known stiffness problem, and uncertainty on the rates can be amplified through the equation flow, yielding unreliable predictions \cite{Grassi2022}.

We fill in this gap here using an algorithm inherited from renormalization theory, which is
universally valid assuming a loose {\em scale separation hypothesis}, stating that kinetic rates
span many scales instead of forming large clusters. The general philosophy is that the \enquote*{bare}
(ab initio) network should be replaced by scale-dependent, effective networks describing more aptly and concisely the observed behavior at each time scale; the way these effective,
coarse-grained networks are constructed is through a particular inductive multi-scale analysis algorithm. The results presented here, however, are
limited to dilute regimes (see below), therefore they mainly describe the growth phase of autocatalytic networks. This is meant however to be the easier part of a program based on the same mathematical ideas, which lifts this restriction.

Various graph techniques going in the direction of a reduction of dimensionality are in use in the literature. Let us cite: (i) formal work on singular $\eps$-perturbation techniques based on the classification of either concentrations or reactions into fast/slow types, see \cite{Lee2010, Kan2016}, expanding the classical quasi-steady-state approximation; (ii) stochastic path integral techniques
\cite{E2007, Sinitsyn2009} giving accelerated stochastic simulation for stiff chemical kinetic systems; (iii) structural reduction based on topology alone \cite{Sinanoglu1975, Hirono2021}; (iv) data-driven model reduction techniques based on information theory \cite{Katsoulakis2020}; (v) and network renormalization techniques \cite{Gabrielli2025}, based either on Kadanoff's
 block-spin approach, where nodes are recursively replaced by node blocks, or on Laplacian
 renormalization \cite{Villegas2023, Villegas2025}, used either as a tool to describe
 the organization of the graph, in particular for community detection, or to model complex bio-molecular
 systems \cite{Henze2019}. Our work is ultimately based on perturbation theory, therefore closest to (i), but allows a whole spectrum of time-scales, leading to a recursive elimination of time-scales, from the fastest one to the slowest one. This is permitted in (ii), which is based on a much more complicated formalism than our paper (saddle-point equations of a cumulant-generating function), and requires Monte-Carlo sampling, and the previous knowledge of typical species concentrations and the associated time scales. Actually, one of the main strong points of our approach is that
 it automatically derives the effective time scales of the system, which, as turns out, are not
 directly accessible from the bare graph. Leaving out (iii) which does not take kinetics into account, and is mostly concerned with the identification of stationary states, (iv)-(v) are ultimately based on information theory, variational minimization, or on approximate diagonalization of the generator (Laplacian); they depend on numerical techniques
and require an exact knowledge of the kinetic rates. Furthermore, the reduced model representation in terms of extended quasi-modes or reconstruction in terms of fitting parameters make them not suitable for a
direct identification of reaction pathways or dynamics in compositional space.

Summarizing: our approach is unique in that it gives a {\em robust}, quantitative picture of the organization of kinetic chemical networks, which is directly interpretable by the chemist in terms of induction time scales, growth patterns, pathways and more generally, compositional dynamics. It is based on a {\em fully general algorithm that is implemented at a very low computational cost}. \enquote*{Robust} means here that only the stoichiometry and the {\em scales}
(orders of magnitude) of rates
are needed in input. Also, outputs are all expressed as functions of the {\em scale parameters};
formulas
hold on scale domains defined by sets of linear inequalities; scale domains, called {\em chemical phases}, are in finite number, and may be explored systematically
by the algorithm. Thus even the rate scales are not systematically needed in input; actually,
one of the main interests of hierarchical formulas is that they may be used to {\em infer kinetic rates}, starting from some very rough (mainly stoichiometric) information and
a time-series of concentration measurements obtained in a dilute regime through available analytical chemistry techniques.

\section{Theoretical framework and main examples}

This section is a cursory introduction to the concepts and techniques of the paper. It is illustrated by \cref{example_1,example_2} whose analysis
is pursued throughout the paper.

\subsection{A brief outline}

We discuss in this article only {\em dilute systems}. The framework is that of a solution containing initially a finite number of abundant species, called {\em external} species, plus a small (ideally, infinitesimal) quantity of {\em internal} species $\mathsf{S}_{\sigma}$ indexed by
a set $\Sigma$. The concentration of this \enquote*{seed} \cite{Peng2022} may amplify
exponentially in an initial time regime, a distinctive signature of autocatalytic behavior.
As long as it remains small enough, the concentration of abundant species may be considered as fixed, and the seed dynamics is well described in the kinetic limit by coupled linear equations for the chemical composition $X=(X_{\sigma})_{\sigma\in \Sigma}$, defined as the vector of concentrations,
\begin{equation}
\frac{dX_{\sigma}}{dt} = \sum_{\sigma'\in \Sigma} A_{\sigma,\sigma'} X_{\sigma'} , \qquad\qquad X_{\sigma} = [\mathsf{S}_{\sigma}] \label{eq:1}
\end{equation}
Assuming mass-action rates (or Michaelis–Menten rates, which have order one kinetics in this limit), off-diagonal coefficients $A_{\sigma,\sigma'}$ are the sums of kinetic rates of all $1-1$ or one-to-many reactions $\mathsf{S}_{\sigma'}\to \mathsf{S}_{\sigma} + \cdots$, hence positive. Such matrices, of the Perron-Frobenius form, have a generally unique 'top' eigenvalue/eigenvector satisfying the equation
$Av^* = \lambda^* v^*$, with $v^*>0$ (all components positive) and $\lambda^*=\lambda^*(A)$ real maximizing $\{\Re \lambda \ |\ \lambda $ eigenvalue of $A\}$. Letting $v^{\dagger,*}$
with $\langle v^{,*\dagger},v^*\rangle=1$ satisfy the left eigenvalue equation $A^{\dagger} v^{\dagger,*} = \lambda^* v^{\dagger,*}$, the large-time
behavior of \eqref{eq:1} is given by
\begin{eqnarray}
X(t) \sim \langle v^{\dagger,*},X(0)\rangle \ e^{\lambda^* t} v^* \label{eq:XvvX(0)}
\end{eqnarray}
Thus $\lambda^*$ is the {\em Lyapunov} (instability) {\em exponent} of the linearized dynamical system, or (in biological terms) its {\em growth rate}, while the {\em Lyapunov eigenvector}
$v^*$ reflects its long-term {\em asymptotic composition}. When
the only reactions of the network are detail balanced 1-1 reactions
\[\mathsf{S}_{\sigma}\underset{k_{\sigma'\to\sigma}}{\overset{k_{\sigma\to\sigma'}}{\rightleftarrows}} \mathsf{S}_{\sigma'} \;,\]
the matrix
$A^{\dagger}$ is the generator of a Markov chain \cite{Anderson1991}, $\lambda^*=0$, the vector $v^*$ is proportional to its equilibrium measure, and $v^{\dagger,*}={\bf 1}$, so that $\langle v^{\dagger,*},X(0)\rangle$ is
the total initial concentration $X_{\text{tot}}(0)=\sum_{\sigma\in\Sigma} X_{\sigma}(0)$. One is interested in determining
{\bf Lyapunov data}
$(\lambda^*,v^*,v^{\dagger,*})$ as a function of various parameters (pH, temperature, metallic catalyst concentration...) when $\lambda^*>0$. The network is then called {\em autocatalytic};
tight connection to the stoichiometric definition of autocatalysis (whereby it is meant that some combination of the
reactions has a strictly positive balance for all internal species) is shown in \cite{Unterberger2022}.

We shall focus here on two illustrative examples. The first one is an autocatalytic version of the Michaelis–Menten
model of enzymatic catalysis with species $\mathsf{S}_1$ (substrate), $\mathsf{S}_0$ (enzyme/substrate complex),
$\mathsf{S}_2$ (product). The second one is obtained by coupling through rates $k_{\pm}$ two autocatalytic
cycles $\mathsf{S}_1\to \mathsf{S}_2 \to 2\mathsf{S}_2\to 2\mathsf{S}_1$ and $\mathsf{S}_{\bar{1}}\to \mathsf{S}_{\bar{2}} \to 2\mathsf{S}_{\bar{2}}\to 2\mathsf{S}_{\bar{1}}$.

\begin{example}[One autocatalytic cycle]\label{example_1}
$$\mathsf{S}_1 \overset{k_{\mathrm{on}}}{\underset{k_{\mathrm{off}}}{\rightleftarrows}} \mathsf{S}_0, \qquad
\mathsf{S}_0 \overset{k_2}{\underset{\bar{k}_2}{\rightleftarrows}} \mathsf{S}_2, \qquad
\mathsf{S}_1 \overset{\nu_+}{\to} \mathsf{S}_0 + \mathsf{S}_0 $$
\end{example}

\begin{example}[Two coupled autocatalytic cycles]
\label{example_2}
\begin{eqnarray*}
\mathsf{S}_1 \overset{k_{\text{max}}}{\underset{k_{\text{min}}}{\rightleftarrows}} \mathsf{S}_2, \qquad
\mathsf{S}_{\bar{1}} \overset{\bar{k}_{\text{max}}}{\underset{\bar{k}_{\text{min}}}{\rightleftarrows}} \mathsf{S}_{\bar{2}}, \qquad
\mathsf{S}_1 \overset{k_+}{\underset{k_-}{\rightleftarrows}} \mathsf{S}_{\bar{1}} \\
 \mathsf{S}_2 \overset{k_+}{\to} \mathsf{S}_2 + \mathsf{S}_2, \qquad
\mathsf{S}_{\bar{2}} \overset{k_-}{\to} \mathsf{S}_{\bar{2}} + \mathsf{S}_{\bar{2}}
\end{eqnarray*}
\end{example}

\begin{figure*}
\centering\includegraphics[width=\textwidth]{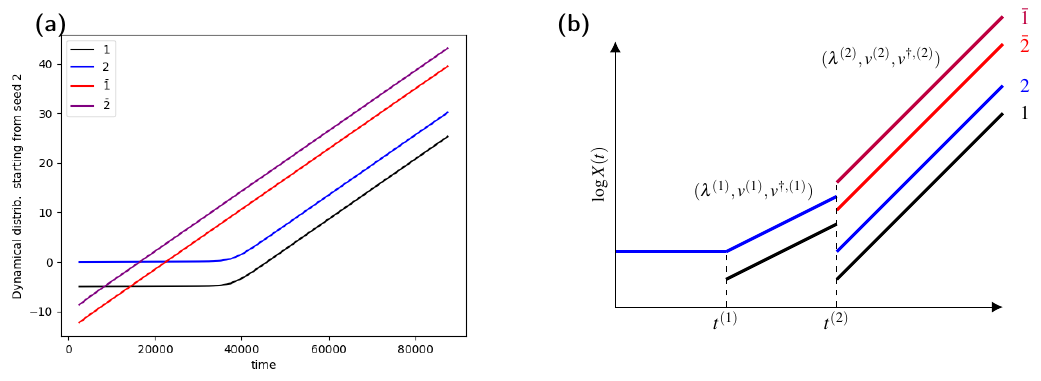}
\caption{Time evolution for two coupled autocatalytic cycles, with Lyapunov data $(\lambda^{(i)},v^{(i)},v^{\dagger,(i)})$. Left: numerical solution of dynamical equations (\ref{eq:1}) in \cref{example_1}. Right: schematized version based on hierarchical formulas.}
\label{fig_time_evolution}
\end{figure*}

Plotting measured log-concentrations of an autocatalytic network as a function of time, one should observe parallel, upward tilted lines with slope $\lambda^*$ and
relative offsets $\log(v^*_{\sigma}/v^*_{\sigma'})$. However, this asymptotic regime is not necessarily
accessible experimentally, since (depending on initial concentrations) the system may undergo a cross-over to a nonlinear regime before, typically a stationary state or cycle. Instead, regularly repeated measurements will typically (see \cref{fig_time_evolution}a) present themselves (for small enough $X(0)$, and after an initial transient regime) roughly in the form of broken lines, with slopes sharply increasing at well-defined transition times
$t^{(1)}\ll t^{(2)}\ll \cdots$. As time goes by,
more and more lines become parallel, following a merging pattern that will be explained below. \Cref{fig_time_evolution} shows a schematized example for the two coupled autocatalytic cycles $(1
\rightleftarrows 2) \rightleftarrows (\bar{1}\rightleftarrows \bar{2})$ of \cref{example_2}, in which the initial composition is some arbitrary mixture of species 1,2. Solving numerically the model (see again \cref{fig_time_evolution}a) shows that the broken lines actually interpenetrate instead
of following each other, and that the real solution interpolates smoothly between the broken lines at species-dependent {\em induction times} (see below (\ref{eq:XivdaggerX0lambdavi})). A log-log plot makes it easier to span several time scales simultaneously, but is ill-suited for
direct interpretation.

Here is, in words, what is shown in \cref{fig_time_evolution}b. Species 1 (in black) is not apparent before time $t^{(1)}$, because the dominant flux is from $\mathsf{S}_1$ to $\mathsf{S}_2$; similarly, species $\mathsf{S}_{\bar{1}},
\mathsf{S}_{\bar{2}}$ (in red/purple) before $t^{(2)}$ because dominant fluxes are directed from $\mathsf{S}_{\bar{1}},
\mathsf{S}_{\bar{2}}$ and not to $\mathsf{S}_{\bar{1}},
\mathsf{S}_{\bar{2}}$ . Between time $t^{(1)}$ and $t^{(2)}$, the slope $\lambda^{(1)}$ materializes an autocatalytic cycle coupling $1,2$, and $v^{(1)}=\left(\begin{array}{c} v^{(1)}_1 \\ v^{(1)}_2\end{array}\right)$ gives the
relative concentrations. The offset $\log(X_1/X_2) \simeq \log(v^{(1)}_1/v^{(1)}_2) = \log(v^{(2)}_1/v^{(2)}_2)$ remains constant for all $t>t^{(1)}$, therefore
$(\mathsf{S}_1,\mathsf{S}_2)$ may be considered a quasi-species \cite{Eigen1988}.
After time $t^{(2)}$, the slope $\lambda^{(2)}>\lambda^{(1)}$ materializes an autocatalytic cycle coupling $\bar{1},\bar{2}$. Simultaneously, fluxes from the quasi-species $(\bar{1},\bar{2})$ to $(1,2)$ produce a quasi-species $(1,2,\bar{1},\bar{2})$,
i.e. concentration ratios of all species are fixed, and equal to that of $v^{(2)}$; the \enquote*{fitter} (i.e. with larger growth rate) second autocatalytic cycle $(\bar{1},\bar{2})$ has \enquote*{absorbed} the first one. The system does not evolve at later times, so
that $(\lambda^{(\infty)},v^{(\infty)}) = (\lambda^{(2)},v^{(2)})\sim (\lambda^*,v^*)$ are the Lyapunov data of the network, whereas $(\lambda^{(1)},v^{(1)})$ are {\em \enquote*{space-time localized} Lyapunov data}, supported on the subspace
$\{1,2\}$.

Note that a simple strategy consisting in simply \enquote*{cutting-off} slow reactions may fail to identify the second autocatalytic cycle because the cycle-closing edge $\bar{2}\to \bar{1}$ is slower than the other
cycle edges; or, it may disregard the even slower fluxes from $(\bar{1},\bar{2})$ to $(1,2)$, which enhance the growth rate of $(1,2)$. Disentangling these kinetic effects is even more difficult when cycles are intertwined, or when several cycles are competing. Our algorithm, however, does the job in a systematic way.

The methods presented below make it possible to derive (1) the ordered sequence of merging patterns; (2) {\bf hierarchical formulas}, which are approximate formulas,
in the form of simple ratios of kinetic rates, for the Lyapunov data $(\lambda^{(i)},v^{(i)},v^{\dagger,(i)})$. Induction times, and transient effects, characterizing the time behavior between induction times, may also be precisely described with the same methods, but are only discussed on an example in the present work. They provide a chemically interpretable, mathematically simple though faithful, representation of the dynamics in terms of formulas, which allows for a very robust inference of kinetic parameters. The extended dynamical algorithm will be presented in a future work.

\subsection{Hierarchical graphs and multi-scale decompositions}
\label{sec_hierarchical_graphs}

Only kinetics can tell whether the sequence of events envisioned on \cref{fig_time_evolution} is correct;
shuffling kinetic rates can also produce a wholly different sequence, e.g. $(1,2)$ absorbing
$(\bar{1},\bar{2})$ instead of the contrary, or different pairings of species into cycles. Stoichiometry alone, or even thermodynamics, is not predictive; in fact, dynamical compositional stability of high-energy intermediates, apparently contradicting the laws of thermodynamics, but protected by kinetic barriers, has been suggested to allow the emergence of autonomous systems by an evolutionary process through entropy dissipation \cite{Pross2023} . On the other hand, knowing the {\bf kinetic scales}, i.e. the integer scales $n_{\sigma\to\sigma'}:=\lfloor
\log(A_{\sigma',\sigma}) \rfloor$ ($\lfloor\cdot\rfloor$ = integer part) makes it possible to predict with some degree of approximation the sequence of events, the scales of the {\bf characteristic elements},
$\lfloor \log(t^{(i)})\rfloor$, $\lfloor \log(\lambda^{(i)})\rfloor$, relative offsets $\lfloor \log(v^{(i)}_{\sigma}/v^{(i)}_{\sigma'})\rfloor$,
 $\lfloor \log(v^{\dagger,(i)}_{\sigma}/v^{\dagger,(i)}_{\sigma'}) \rfloor$, and finally, a time
 evolution for {\em composition scales} $\lfloor \log(X_{\sigma}(t))\rfloor$, defining a {\em dynamical distribution} in log-scale.

\begin{figure}
\includegraphics[width=0.45\textwidth]{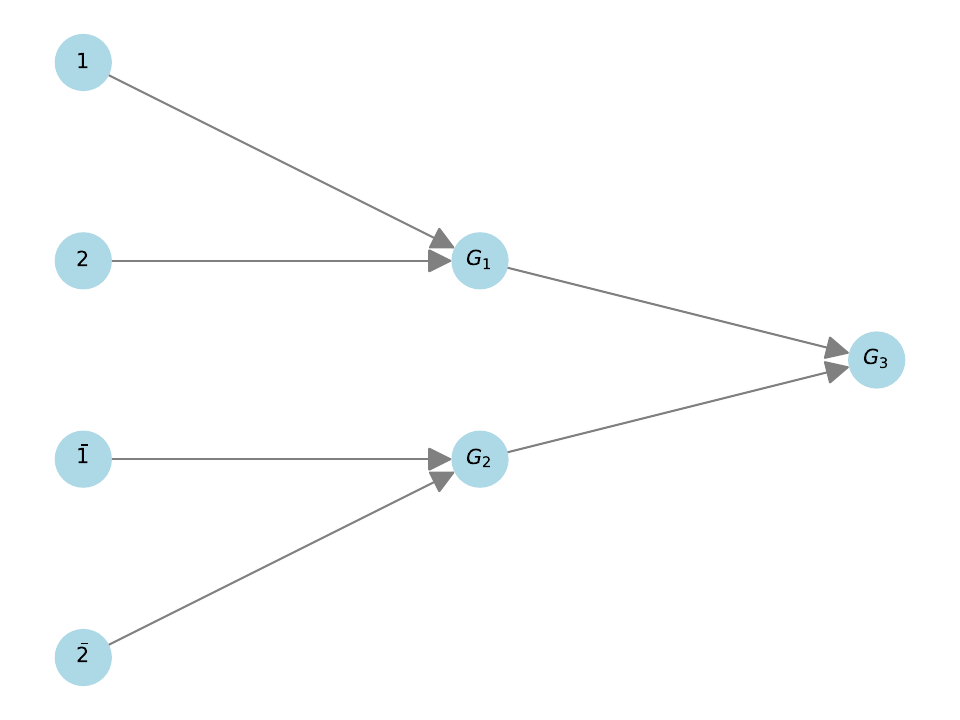}
\caption{Example of coalescence tree}
\label{fig_merging}
\end{figure}

The appropriate description is through {\em hierarchical structures}. {\em Merging patterns} are part
of them. A general pattern is a finite sequence of the form $\Sigma\equiv \Sigma^{(0)}\to \Sigma^{(1)}\cdots \to \Sigma^{(j_{max})}$, where elements of $\Sigma^{(j)}$ are either elements of $\Sigma^{(j-1)}$ or
subsets of $\Sigma^{(j-1)}$; e.g. $\Sigma=\{1,2,\bar{1},\bar{2}\}$, $\Sigma^{(1)} = \{\{1,2\},\bar{1},\bar{2}\}$,
$\Sigma^{(2)} = \{\{1,2\}, \{\bar{1},\bar{2}\}\}$, $\Sigma^{(3)}= \{\{\{1,2\},\{\bar{1},\bar{2}\}\}\}$. The recursive nesting, matryoshka-like, structure, may be drawn as a coalescence tree, see \cref{fig_merging}, with compound vertices given new names, $G_1=\{1,2\}, G_2=\{\bar{1},\bar{2}\}, G_3=
\{G_1,G_2\}$.

We now let elements of $\Sigma^{(j)}$, $j=0,\ldots,j_{max}$ be vertices of a graph $G^{(j)} = (\Sigma^{(j)},E(G^{(j)}))$, which is constructed from $G^{(0)}\equiv G$ and $\Sigma$ by successive {\em rewiring} steps as follows: if
$\sigma\to\sigma'$ is an edge of $G^{(j)}$ $(\sigma,\sigma'\in \Sigma^{(j)})$, then $G^{(j+1)}(\sigma)\to G^{(j+1)}(\sigma')$ is an
edge of $G^{(j+1)}$, where $G^{(j+1)}(\sigma)$ is the vertex of $G^{(j+1)}$ equal to or containing
$\sigma$. Self-edges $\sigma\to\sigma$ are removed. If e.g. $E(G^{(0)}) = \{1\rightleftarrows 2, \bar{1}\rightleftarrows \bar{2}, 1\rightleftarrows \bar{2}\}$, then $E(G^{(1)}) = \{\bar{1}\rightleftarrows \bar{2}, G_1\rightleftarrows \bar{2}\},\ E(G^{(2)}) = \{
G_1\rightleftarrows G_2\}$, $E(G^{(3)}) = \emptyset$. The compound vertex $\sigma=\{\sigma_1,\ldots,\sigma_n\}\in \Sigma^{(j)}$, $n\ge 2$ is formed only if $\sigma_1,\ldots,\sigma_n\in \Sigma^{(j-1)}$ form a {\em strongly connected component} (SCC) structure, i.e., there exists a path
connecting any pair $(\sigma_i,\sigma_{i'})$. However, only SCCs of a special type ({\em maximal
 dominant SCCs}, see below) are actually formed. Paths connecting the vertices inside
 $\sigma$ are the {\em dominant} (i.e. main) pathways of the network; therefore, they are the most
 relevant information for reactivity. Including them yields a {\em hierarchical graph structure}. However, only merging patterns, not individual dominant pathways, are useful in constructing the approximate dynamics.

\medskip {\em Input and output.} Transition state theory suggests the Ansatz $k_{\rho}\sim {\cal A} e^{-\Delta G^{\ddagger}_{\rho}/RT}$ $({\cal A} =$ Arrhenius constant, $\Delta G^{\ddagger}$ = activation energy) for the kinetic rate of reaction $\rho$. Since $\log_{10}(e^{-\Delta G^{\ddagger}/RT}) \simeq -\Delta G^{\ddagger}/1.36$ at room temperature if $\Delta G^{\ddagger}$ is measured in kcal/mol, chemical precision $(\pm 1$ kcal.mol) allows a determination of the kinetic scales $\log_{10} (k)$ up to one unit, so that integer data $n(k_{\rho})= \lfloor
\log_{10}(k_{\rho}) \rfloor$ characterize the input. Then the
 output, namely, space-time localized Lyapunov data and time evolution of concentrations, as deduced from the scales of
characteristic elements, would have the same order of precision. In practice, errors may
accumulate for large networks when scale separation is very weak or the size of dominant SCCs is
unbounded, see below (\ref{eq:edge-scales}) and discussion in \cref{sec_discussion}. Our simulations (see \cref{sec_results} and Suppl. Mat.)
show that the approximation scheme is good when properly used.

\subsection{Renormalization}

Hierarchical models are produced by a recursive algorithm which is directly inspired by Wilson's renormalization group
\cite{Wilson1983, Mastropietro2008, Unterberger2012} in quantum field theory, and its applications to statistical physics; see also \cite{Sinitsyn2009} for the use of related techniques in stochastic
networks. Namely, one starts from the highest kinetic scale
$n_1 = \max(n_{\sigma\to\sigma'})$, and then progressively incorporates lower scales one by one, merges {\em maximal dominant SCCs}, rewires edges following the above construction, and redefines the
kinetic scales of the new {\em effective graphs}. The outcome is known as first-order Michaelis–Menten dynamics for enzymatic catalysis
\[E+S \underset{k_{-1}}{\overset{k_{+1}}{\rightleftarrows}} ES \overset{k_{\text{cat}}}{\to} P\]
(synthesis of product $P$ from $S$ with intermediate bound state $ES$ involving enzyme $E$). Namely, if $k_{+1}[E],
k_{-1}\gg k_{\text{cat}}$ are fast, and $[S]$ is small enough (which defines the dilute regime), then the reaction velocity $d[P]/dt \simeq v_{max} \frac{[S]_0}{K_M} \simeq \frac{k_{\text{cat}} [E]_{tot}[S]}{k_{-1}/k_{+1}}$ is linear in
$[S]$. In our description, $E$ is an abundant external species, $\Sigma^{(0)}=\{S,ES,P\}, \Sigma^{(1)}= \{\{S,ES\},P\}$. We may assume $[E]=1$ (which amounts to a mere redefinition of $k_{+,1}$).
Simple computations (or our algorithm) show that the cycle $S \underset{k_{-1}}{\overset{k_{+1}}{\rightleftarrows}} ES$ may be merged
at times $t>t^{(1)}:= 1/\max(k_{+1},k_{-1})$, after which one obtains an effective graph
$\{S,ES\} \overset{k^{(1)}_{\text{cat}}}{\to} P$, with $k^{(1)}_{\text{cat}} \sim \begin{cases} k_{\text{cat}}, \qquad k_{+1} \gg k_{-1} \\ \frac{k_{+1}}{k_{-1}} k_{\text{cat}}, \qquad k_{+1} \ll k_{-1} \end{cases}$.
(The expression of $k^{(1)}_{\text{cat}}$ is usually obtained through the quasi-steady-state
approximation~\cite{Li2008}, i.e., by assuming that the concentration of the complex $ES$ does not change on the time scale over which the product formation is measured). The Lyapunov
exponent of the cycle is $0$ (catalysis is not autocatalysis), so that the associated eigenvector agrees with equilibrium, $\frac{[ES]}{[S]}\sim \frac{k_{+1}}{k_{-1}}$. The effective equation $dP/dt \sim
k^{(1)}_{\text{cat}} ([S]+[ES]) \sim \begin{cases} k^{(1)}_{\text{cat}} [ES], \qquad k_{+1} \gg k_{-1} \\
 k^{(1)}_{\text{cat}} [S], \qquad k_{+1} \ll k_{-1} \end{cases} $ is consistent with the Michaelis–Menten equation.

\medskip The model equation parallel to the enzymatic catalysis, but involving instead an autocatalytic cycle, is \cref{example_1},
$$ \mathsf{S}_0 \underset{k_{\mathrm{off}}}{\overset{k_{\mathrm{on}}}{\rightleftarrows}} \mathsf{S}_1, \qquad E+ \mathsf{S}_1 \overset{\nu_+}{\to} \mathsf{S}_0+ \mathsf{S}_0, \qquad \mathsf{S}_0 \overset{k_2}{\to} \mathsf{S}_2,
$$
where $E$ is an external species ensuring atom number conservation, and (to be specific)
$k_2, \nu_+ \ll k_{\mathrm{on}},k_{\mathrm{off}}$. We neglect here the reverse reaction $\mathsf{S}_2 \overset{\bar{k}_2}{\to}
\mathsf{S}_0$ for simplicity. The correspondence $(E,ES,S,P)\to (E,\mathsf{S}_0,\mathsf{S}_1,\mathsf{S}_2)$, $(k_{+1},k_{-1},k_{\text{cat}})\to (k_{\mathrm{off}},k_{\mathrm{on}},k_2)$ is misleading because of the extra doubling
reaction $\mathsf{S}_1\to \mathsf{S}_0+ \mathsf{S}_0$. Indeed, as in the catalytic case, the cycle $\mathsf{S}_0,\mathsf{S}_1$ may be merged at times $t>t^{(1)}$, after which the effective graph is
$\{\mathsf{S}_0,\mathsf{S}_1\} \overset{k^{(1)}_2}{\to} \mathsf{S}_2$, with $k^{(1)}_2 \sim \begin{cases} k_2, \qquad k_{\mathrm{off}} \gg k_{\mathrm{on}}\\ \frac{k_{\mathrm{off}}}{k_{\mathrm{on}}}k_2, \qquad k_{\mathrm{off}} \ll k_{\mathrm{on}} \end{cases}$, but the rest of the story is different. For definiteness, we choose $k_{\mathrm{off}}\gg k_{\mathrm{on}}$. Multiplying the stoichiometry matrix ${\mathbb{S}} = \left(\begin{array}{cccc}
-1 & +1 & +2 & -1 \\ 1 & -1 & -1 & 0 \\ 0 & 0 & 0 & 1 \end{array}\right)$ by the
flux vector $j= \left(\begin{array}{c} k_{\mathrm{on}} X_0 \\ k_{\mathrm{off}} X_1 \\ \nu_+ X_1 \\ k_2 X_0 \end{array}\right)$, one gets $dX/dt = AX$, with
$A = \left(\begin{array}{ccc} -k_{\mathrm{on}}-k_2 & k_{\mathrm{off}} + 2\nu_+ &
0 \\ k_{\mathrm{on}} & -k_{\mathrm{off}}-\nu_+ & 0 \\ k_2 & 0 & 0 \end{array}\right)$. Define {\em deficiency weights}
as $\eps_{\sigma} = \frac{\sum_{\sigma'} A_{\sigma',\sigma}}{\sum_{\sigma'\not=\sigma} A_{\sigma',\sigma}}$; they vanish when $A^{\dagger}$ is a Markov chain generator, in particular in the
enzymatic catalysis case ($\nu_+=0$). Here only $\eps_1 = \frac{\nu_+}{k_{\mathrm{off}}+2\nu_+}\sim \frac{\nu_+}{k_{\mathrm{off}}} $ is $\not=0$.
Note the contrary effects of the doubling reaction and the end reaction $\mathsf{S}_0\to \mathsf{S}_2$: at a
purely stoichiometric level, summing $\mathsf{S}_0\to \mathsf{S}_1, \mathsf{S}_1\to \mathsf{S}_0+\mathsf{S}_0$ increases $X_0$, while the end reaction
decreases $X_0$. If the effect of the end reaction dominates, then the autocatalytic cycle is
not viable, and the final composition is just pure $2$-state, i.e. $X_0(t),X_1(t)\to 0$; in the contrary case, one expects all components to increase exponentially with the same rate. The algorithm (or elementary computations)
proves that the autocatalytic threshold lies at $k_2\sim \eps_1 k_{\mathrm{on}} \sim \frac{k_{\mathrm{on}}}{k_{\mathrm{off}}}\nu_+$. For larger
$k_2$, the growth rate is $0$, the effect of the doubling reaction is negligible, and Michaelis–Menten equation is correct. However, for smaller $k_2$, the Lyapunov data are
$\lambda^* \sim \frac{k_{\mathrm{on}}}{k_{\mathrm{off}}}\nu_+$ and $v^*\sim \left(\begin{array}{c} 1 \\ \frac{k_{\mathrm{on}}}{k_{\mathrm{off}}} \\ \frac{k_2 k_{\mathrm{off}}}{k_{\mathrm{on}} \nu_+} \end{array}\right)$. The ratio
$v^*_0/v^*_1$ agrees with equilibrium, but not the ratios $v^*_i/v^*_2$, $i=0,1$ which depend on the
rate of the totally irreversible doubling reaction.

\section{Methods}
\label{sec_methods}

We discuss here the full hierarchical algorithm (\cref{sec_framework},
\cref{sec_hierarchical_formulas}), and the hierarchical model-based approach to statistical inference (\cref{sec_inference}).

\subsection{General framework}
\label{sec_framework}

We use here the formalism of {\em open chemical reaction networks} in the kinetic limit (see e.g. \onlinecite{Rao2016}, and \cref{sec_main_notation} for more details). We assume reaction rates to follow mass-action.
Here we have
\begin{itemize}
 \item a set of species $\{\mathsf{S}_{\sigma}\}$, with $\sigma$ varying
 in an index set $\Sigma^{ext}$ (indexing external species) or in $\Sigma$
 (indexing internal species, called species for short);
 \item a
stoichiometry matrix with columns representing the stoichiometry of reactions;
 \item for each reaction $\rho$, a kinetic rate, $k_{\rho}$.
\end{itemize}
Species concentrations $[\mathsf{S}_{\sigma}]$ are denoted $X_{\sigma}$. External species $\sigma$ are assumed to be chemostated, that is,
for each $\sigma\in \Sigma^{ext}$, a term is added to the r.-h.s. of the dynamical equations so that $dX_{\sigma}/dt = 0$. This is difficult to realize experimentally, but provided $X_{\sigma^{ext}}\gg
X_{\sigma}$ for all $\sigma\in \Sigma^{ext},\sigma\in \Sigma$ throughout the observation time window,
 $X_{\sigma^{ext}}$ can be considered as constant. Then, we define a reaction $\rho$ to be {\em uni-molecular if it has only one
internal reactant}.

\medskip {\em Dilute regime.} We consider the case when internal species have low concentration;
then reactions of
kinetic order $\ge 2$,
can be dismissed, since the associated fluxes are negligible. Equivalently, one may linearize the
system around 0, which makes sense even for more general rates, such as Michaelis–Menten rates.
The outcome is a linear system $dX/dt= AX$. Since we assumed mass-action rates, this is equivalent to keeping only
columns $\rho$ associated to unimolecular reactions. Degradation reactions are also included in the form
$\sigma\overset{\beta_{\sigma}}{\to}\emptyset$, where $\emptyset \in \Sigma$ is seen as a reservoir of unspecified, non-reacting species; mind the specific notation for the rates.

\medskip {\em Graph of split reactions.} The starting point
is the graph of split reactions $G=(\Sigma,E)$ (later on denoted: $G^{(0)}$, and called: bare graph, see below). For simplicity, we consider only reactions
with up to two products (which is chemically realistic); extension to reactions with
$\ge 3$ products would be straightforward. By construction,
$\sigma\to \sigma'$ $(\sigma\not=\sigma')$ is an edge in $E$ if either there exists
a $1 \to 1$ reaction $\sigma\to \sigma'$, or (ii) $\sigma
\to \sigma'$ is a {\em split reaction} coming from a $1\to 2$
reaction $\sigma\to \sigma' + \sigma''$. The rate $k_{\sigma\to\sigma'}$ is obtained by summing the kinetic rates
of all reactions (i), (ii) and equals the off-diagonal entry $A_{\sigma',\sigma}$. Each reaction (i), (ii) also
contributes $-k^{\rho}$ to $A_{\sigma,\sigma}$, therefore
diagonal coefficients of $A$ are $<0$.

\medskip {\em Autocatalysis.} We say that {\em $G$ is autocatalytic if $\lambda^*_G:=\lambda^*(A)>0$}. This
dynamical definition has been shown in \cite{Unterberger2022} to be equivalent to the usual stoichiometric one \cite{Blokhuis2020} for a sub-class of graphs including
strongly connected graphs with zero (or near-zero) degradation rates. Lyapunov eigenvectors are
zero eigenvectors of
\begin{equation} A(\alpha):=A -\alpha \Id
\end{equation}
with $\alpha = \lambda^*(A)$.

\medskip {\em Associated Markov chain.} We let $\tilde{A}=(\tilde{A}_{\sigma',\sigma})_{\sigma',\sigma\in \Sigma}$
be the matrix with off-diagonal coefficients $\tilde{A}_{\sigma',\sigma} =A_{\sigma',\sigma}$ $(\sigma\not=\sigma')$, and modified diagonal coefficients with absolute value

\begin{equation} -\tilde{A}_{\sigma,\sigma}=\sum_{\sigma'\not=\sigma} A_{\sigma',\sigma} = \sum_{\sigma'\not=\sigma}
k_{\sigma\to\sigma'} \label{eq:ksigma}
\end{equation}
(called: {\em total outgoing rate from $\sigma$}), also denoted $k_{\sigma}$.
 By construction \cite{Anderson1991}, $\tilde{A}$ is an adjoint Markov generator, so
that its Lyapunov exponent (i.e. eigenvalue with largest real part)
is $0$. The two matrices, $A$ and $\tilde{A}$, differ by the {\em deficiency rates} $\kappa_{\sigma} = A_{\sigma,\sigma} - \tilde{A}_{\sigma,\sigma}$, which are also equal to the sum of kinetic rates of all $1\to 2$ reactions
with reactant $\sigma$, and are therefore $\ge 0$.

\bigskip\noindent Next, we discuss our methodological contribution.

\subsection{Hierarchical formulas}
\label{sec_hierarchical_formulas}

\subsubsection{Lyapunov weights}

Though time is continuous, the {\em resolvent formula} (see: Suppl. Mat.) makes it possible to express all relevant quantities in terms of paths describing transitions between species, whose statistics are similar to those used in the Gillespie algorithm. Computations below hold for the corrected matrix $A(\alpha)$, with $\alpha\ge 0$.

\medskip {\bf Transition weights.} Let, for $\alpha\ge 0$,
\begin{equation} w(\alpha)_{\sigma\to\sigma'}= \frac{k_{\sigma\to\sigma'}}{|A_{\sigma,\sigma}| + \alpha} \ \ (\sigma\not=\sigma')
\end{equation}
In particular, $w_{\sigma\to\sigma'} := w(0)_{\sigma\to\sigma'}$. Similarly, we let
$\tilde{w}_{\sigma\to\sigma'} := k_{\sigma\to\sigma'} \,/\, |\tilde{A}_{\sigma,\sigma}| =
\frac{k_{\sigma\to\sigma'}}{k_{\sigma}}$, with $k_{\sigma}$ as in (\ref{eq:ksigma}). The latter ratio is equal to the jump probability $\sigma\to\sigma'$ of Gillespie's simulation algorithm for the associated Markov chain. Then, the {\em weight} $w(\alpha)_{\gamma}$ of a path $\gamma: \sigma_0\to\cdots\to \sigma_{\ell}$ of length $\ell\ge 0$ is the
product of the transition weights $w$ along its edges,
\begin{equation} w(\alpha)_{\gamma}= \prod_{i=1}^{\ell} w(\alpha)_{\sigma_{i-1}\to\sigma_i}.
\end{equation}

\medskip {\bf Stationary measure, stationary weights.} Let ${\bf 1}$ be the constant vector
with components ${\bf 1}_{\sigma}=1$. The zero-diagonal matrix $\tilde{\cal W}= (\tilde{w}_{\sigma\to
\sigma'})_{\sigma,\sigma'}$ is a Markov transition matrix since $\tilde{\cal W}{\bf 1} = {\bf 1}$ by construction. A stationary measure
of the discrete-time Markov chain with transition weights
$\tilde{w}$ is a positive distribution $\tilde{\pi}$ such that $\tilde{\pi} \tilde{\cal W} =
\tilde{\pi}$.
 Let $\tilde{\mu}_{\sigma} := k_{\sigma}^{-1} \tilde{\pi}_{\sigma}$;
plugging the definition of $\tilde{w}_{\sigma\to\sigma'} $ into the above eigenvector identity
yields $\sum_{\sigma\not=\sigma'} \tilde{\mu}_{\sigma} k_{\sigma\to\sigma'} = |\tilde{A}_{\sigma',\sigma'}|\, \tilde{\mu}_{\sigma'}$, from
which $\tilde{A}\tilde{\mu}=0$. Thus $\tilde{\pi}$ is a stationary
measure of the discrete-time Markov chain if and only if $\tilde{\mu}$ is a stationary
composition for $\tilde{A}$.
We work most of the time with discrete-time transition weights, and
our results will provide estimates for the associated discrete-time
weights, called {\bf Lyapunov weights},
\begin{equation} \pi^*_{\sigma} := (|A_{\sigma,\sigma}|+\lambda^*) v^*_{\sigma}. \label{eq:pi*}
\end{equation}
Then $Av^* = \lambda^* v^*$ if and only if $\sum_{\sigma\not=\sigma'}
\pi^*_{\sigma} w^*_{\sigma\to\sigma'} = \pi^*_{\sigma'}$, where
$w^*_{\sigma\to\sigma'} \equiv w(\lambda^*)_{\sigma\to\sigma'}$.

\subsubsection{Multi-scale analysis, dominant SCCs}
\label{sec_dominant}

\medskip {\bf Kinetic scales.} {\em Fix a} {\bf scale parameter} {\em $b>1$ once and for all.} If $(\sigma,\sigma')\in E$,
we let (with $\lfloor \, \cdot\, \rfloor$ denoting either the integer part, or rounding to the closest integer)
\begin{equation} n_{\sigma\to\sigma'} := \lfloor \log_{b} (k_{\sigma\to
\sigma'}) \rfloor \in \Z \label{eq:edge-scales}
\end{equation}
and split kinetic rates according to their scale.
Due to the associated truncation errors (see \cref{sec_discussion}), all our results are valid up to one or a few scale units. We
write $k\prec k'$, resp. $\preceq k'$ when $\lfloor\log_b (k/k')\rfloor\le -1$ (resp. $\le 0$), and use similarly curved
symbols $\succ, \succeq$; symmetrizing, $k,k'$ have {\em same order} if
\begin{equation} (k\sim k') \Leftrightarrow (k\preceq k'\preceq k)
\end{equation}
In particular, we often use the \enquote*{sum-max} substitution rule for positive quantities (kinetic rates or kinetic rate ratios)
\begin{equation} (k_1\succeq k_2\succeq\cdots\succeq k_m) \Rightarrow
(k_1+\cdots+k_m \sim k_1)
\end{equation}
 Thus,
$k\prec 1$ means in principle: $k$ small. The log-scale parameter, defined e.g. as $\ln(b)$, gives the minimum scale separation between two rates belonging to different scales.

\medskip {\bf Dominant edges.} An edge
$\sigma\to\sigma'$ is dominant if
\begin{equation} n_{\sigma\to\sigma'} = \max_{\sigma''\in \Sigma}
n_{\sigma\to\sigma''} \label{eq:dominant edges}
\end{equation}
Generalizing, if $\alpha>0$, $n_{\alpha} = \lfloor \log_b \alpha \rfloor$, then
$\sigma\to\sigma'$ is $\alpha$-dominant if $n_{\sigma\to\sigma'}\succeq n_{\alpha}$ and
$\sigma\to\sigma'$ is dominant (letting $\alpha\to 0$, the two notions coincide).
 As a general rule, $\alpha$-dominant edges appear as {\bf bold} lines on all our graphs.
An important particular case is when $\sigma$ is {\bf autocatalytic}, i.e. $\kappa_{\sigma}\succ k_{\sigma}$, for then
$n_{\sigma\to\sigma}$ is dominant and the equivalent doubling reaction $\sigma\overset{\kappa_{\sigma}}{\to} \sigma +\sigma$ ensures that $X_{\sigma}(t)$ increases
at least as fast as $\sim e^{\kappa_{\sigma}t}$ (possibly faster if there exists a path
from $\sigma'$ to $\sigma$, with $\kappa_{\sigma'}\succ \kappa_{\sigma}$).

\medskip {\bf Vertex scales.} A {\em vertex} $\sigma$ is an element
of $\Sigma$. The scale of $\sigma$ is
\begin{equation} n_{\sigma}:= \max_{\sigma'\in \Sigma}\, n_{\sigma\to\sigma'}
\label{eq: vertex scales}
\end{equation}
i.e. the maximum scale of all edges $\sigma\to \sigma'$
 of $G$ outgoing from $\sigma$.

\medskip Note that an edge $\sigma\to\sigma'$ is dominant if and only if
$n_{\sigma\to\sigma'} = n_{\sigma}$.
 By construction, $\log_{b} (k_{\sigma}) = \log_b (|\tilde{A}_{\sigma,\sigma}|)
\sim \log_b (|A_{\sigma,\sigma}|)
 \sim n_{\sigma}$ if $\sigma$ is {\em not} autocatalytic.

\medskip {\bf Cut-off graphs.} Split $\Sigma$ into a disjoint union
$\Sigma=\Sigma^{int}\uplus \Sigma^{ext}$ (internal species, versus external species).
The cut-off graph $G^{int} = (\Sigma^{int}, E^{int})$ has internal edge set
$E^{int} := \{(\sigma,\sigma')\in E\ |\ \sigma,\sigma'\in \Sigma^{int}\}
\subset E$, and degradation rates
 $\beta^{int}_{\sigma} := \beta_{\sigma} + \sum_{\sigma'\in \Sigma^{ext}} k_{\sigma\to\sigma'}$.
 Roughly speaking, external
 species have been 'frozen': influxes have been discarded, and outfluxes from $\Sigma^{int}$ to $\Sigma^{ext}$
 have been added to degradation rates.

\medskip
For the multi-scale analysis, we shall need in particular
{\em infra-red cut-off graphs.} Fix a scale $n$ (called:
{\em infra-red cut-off scale}, or simply, cut-off scale), let
\begin{eqnarray} && \Sigma^{\text{int}} \equiv \Sigma^{\ge n} := \{\sigma\in \Sigma\ |\ n_{\sigma}\ge n\}, \nonumber \\ && \qquad
\Sigma^{\text{ext}}\equiv \Sigma^{< n} := \{\sigma\in \Sigma\ |\ n_{\sigma} < n\} \label{eq:IRcutoff}
\end{eqnarray}
and denote $G_{\searrow n}$ the cut-off graph.
Even though the bare graph $G$ is connected, infra-red cut-off graphs $G^{int}$ are often disconnected.

{\em The cut-off graphs produced by our algorithm have by construction the following essential
properties: (i) every pair $\sigma,\sigma'$ of vertices in $\Sigma^{int}$ is connected by a
path of dominant edges of ${\cal G}^{int}$; (ii) no outgoing edge is dominant,}
which means that every outgoing edge $(\sigma,\sigma'_{ext}), \sigma\in \Sigma^{\text{int}}, \sigma'_{ext}\in
 \Sigma^{\text{ext}}$, is dominated by some internal edge $(\sigma,\sigma'), \sigma,\sigma'\in \Sigma^{\text{int}}$,
 i.e. $k_{\sigma\to\sigma'_{ext}}\prec k_{\sigma\to\sigma'}$. Such structures will be uncovered inductively by lowering a cut-off scale.

\medskip {\bf Deficiency weight.} Let $\alpha\ge 0$ be an overall degradation rate. The $\alpha$-deficiency
weight is the ratio
\begin{equation} \eps_{\sigma} : = \frac{\kappa_{\sigma}}{|A(\alpha)_{\sigma,\sigma}|} =
\frac{\kappa_{\sigma}}{|A_{\sigma,\sigma}| + \alpha}. \label{eq:deficiency-weight}
\end{equation}

\noindent For example, a couple consisting of a $1\to 1$ reaction $\sigma\overset{k_{\mathrm{off}}}{\to} \sigma'$ and a $1\to 2$ reaction
$\sigma\overset{\nu_+}{\to} \sigma' + \sigma'$ implies a deficiency rate, resp. weight
\begin{equation} \kappa_{\sigma} = \nu_+,
\qquad {\mathrm{resp.}} \qquad \eps_{\sigma} = \frac{\nu_+}{k_{\mathrm{off}}+\nu_+ + \alpha}
\end{equation}
If $\alpha\succ k_{\mathrm{off}},\nu_+$, then $\eps_{\sigma}\sim
\frac{\nu_+}{\alpha} \prec 1$; this regime is called {\em degradated}. In the contrary case,
 $\eps_{\sigma}\sim \min(\frac{\nu_+}{k_{\mathrm{off}}},1)$.

\medskip {\bf Scale ordering.} We let $(n_i)_{i=1,2,\ldots,n_{max}}$ be the
set of edge and vertex scales, and order them by decreasing order,
$n_1> n_2>\cdots$. Representing them {\em from top to bottom} yields a {\bf multi-scale graph} (see \cref{sec_results} for examples). By construction, there is
a reaction $\sigma\to\cdots$ of scale $n_{\sigma}$ {\em above} any
reaction with reactant $\sigma$. In particular, there is at least
one vertex of scale $n_1$.

\medskip
{\bf Dominant paths.} From \eqref{eq:dominant edges}, \eqref{eq: vertex scales}, the edge $\sigma\to\sigma'$ is
dominant if $n_{\sigma\to\sigma'}=n_{\sigma}$.
If $\sigma\not=\sigma'$, the condition is equivalent to
 $w_{\sigma\to\sigma'}\sim 1$. Iterating, we say
that $\sigma$ is connected to $\sigma'$ by a dominant path
if there exists a path $\gamma:\sigma=\sigma_1\to \cdots \to
\sigma_{\ell}=\sigma'$ whose edges $\sigma_i\to\sigma_{i+1}, \
i=1,\ldots,\ell-1$ are all dominant. This defines a notion of
{\em dominant path} (weak or strong) {\em connectivity}.

\medskip {\bf Dominant graph.} A dominant graph is a graph
connected by dominant paths, i.e. a graph $G=(V(G),E(G))$ such that all oriented
pairs of vertices $(\sigma,\sigma')$
are connected by dominant paths. The {\bf dominant subgraph} of a graph $G$ is obtained
by selecting only dominant edges of $G$; even if $G$ is connected, it may be disconnected.

 \medskip{\bf Dominant SCCs.} If $\sigma,\sigma' \in V(G)$,
we let $\sigma\sim\sigma'$ (including the case $\sigma=\sigma'$) if there exists a path
$\sigma=\sigma_1\to \cdots\to \sigma_{\ell}=\sigma'$ of dominant
edges of $G$ connecting $\sigma$ to $\sigma'$, and
similarly, a path of dominant
edges of $G$ connecting $\sigma'$ to $\sigma$. The
relation $\sim$ is an equivalence relation on vertices of
$G$. We call {\bf dominant SCCs} (dominant strongly
connected components) its equivalence classes. {\bf Maximal dominant SCCs} of $G$ are
dominant SCCs which are 'downstream', i.e. which are not the source of any dominant edge (in the terminology of
Markov chains \cite{Norris1997}, they are maximal classes of the dominant subgraph of $G$). {\bf Non-trivial dominant SCCs} are dominant SCCs
 containing at least two vertices; equivalently, containing a dominant cycle.

\medskip The main objects of consideration for renormalization are {\bf non-trivial maximal
dominant SCCs}, that is, maximal dominant SCCs which are non-trivial. Note that the presence
of a dominant cycle implies the existence of a non-trivial dominant SCC, but not necessarily
maximal. These structures will be uncovered inductively by lowering a cut-off scale, as we shall
presently see.

\subsubsection{Renormalization algorithm}
\label{sec_renormalization}

We refer to the Supplementary Materials both for heuristics and for a presentation of the interface. The mathematical proof will
be presented elsewhere \cite{Unterberger2025}. We content ourselves here with presenting the results.

The starting point is the split graph $G^{(0)}$ with reaction rates $k^{(0)}_{\sigma\to\sigma'} = k_{\sigma\to\sigma'},\ \beta^{(0)}_{\sigma} = \beta_{\sigma} = k_{\sigma\to\emptyset}$ obtained after linearizing at 0. Call $n^{(0)}_1> n^{(0)}_2>\cdots$ the reaction scales of $G^{(0)}$. We start from the cut-off scale $n^{(0)}_1$, i.e.
eliminate reactions with scale $< n^{(0)}_1$, so that all reactions in the infra-red cut-off graph
$G: = G^{(0)}_{\searrow n^{(0)}_1}$ (see \cref{sec_dominant})
have same scale $n^{(0)}_1$; in particular, they are all dominant. If $G$ contains no non-trivial maximal
dominant SCC, then we replace the cut-off scale by $n^{(0)}_2$.
The cut-off graph $G = G^{(0)}_{\searrow n^{(0)}_2}$ now contains reactions of both scales
$n^{(0)}_1,n^{(0)}_2$. We consider {\em only} those reactions which are dominant, yielding a
dominant subgraph, still denoted $G$. If $G$ contains no non-trivial maximal
dominant SCC, then we replace the cut-off scale by $n^{(0)}_3$, and so forth.

{\em One renormalization step $(i=1)$.} The process usually stops at some scale $n = n(i)$ (here $i=1$), called {\em step $i$ cut-off scale}, at which a non-trivial maximal
dominant SCC appears. By construction, $n(i)\equiv n^{(i-1)}_{j_i}$ is one of the scales of
the previous step graph. We first let $G^{(i-1)}_{cut} := G^{(i-1)}_{\searrow n^{(i-1)}_{j_i-1}}$ be the graph cut off at the scale just above $n$; this cut-off graph has no maximal dominant SCC. Then (adding scale $n$) we let $(G_p)_{p=1,2,\ldots}$ be the non-trivial maximal dominant SCCs of $G_{\searrow n}$. By construction, they involve only dominant edges, and their minimal scale is $n$; also, none of the vertices of $G_p$ is autocatalytic, because there is no dominant edge $\sigma\to\sigma'$, $\sigma'\not=\sigma$ when $\sigma$ is autocatalytic.
 Fix $p$. Adding to the internal edges
in $G_p$ the set ${\cal E}^{out}_p$ of outgoing edges $k_{\sigma\to\sigma'}$, $\sigma\in G_p, \sigma\not\in G_p$ yields a
new graph, ${\cal G}_p$, which is the same as the cut-off graph $G^{int}$, except for the fact that the target of outgoing edges is specified. By construction, {\em none of the outgoing edges is dominant}.
{\em We shall now collapse ${\cal G}_p$, $p=1,2,\ldots$ inside $G^{(0)}$}. Graphically, one (1) merges all vertices $\sigma$ in
$G_p$ into a compound vertex denoted $G_{i=1,p}$, so that $G^{(i)}(\sigma) = G_{i,p}$ (see discussion of
merging patterns in \cref{sec_hierarchical_graphs}); for all other vertices $\sigma$ (including
those in non-maximal dominant SCCs),
the map $G^{(i)}(\sigma)=\sigma$ is trivial; (2)
redirects every outgoing edge $\sigma\to\sigma'$ in ${\cal E}^{out}_p$ $(\sigma\in V(G_p),
\sigma'\not\in V(G_p))$ into an edge $G_{i,p}\to G^{(i)}(\sigma')$ from $G_{i,p}$; (3) and further,
redirects every ingoing edge $\sigma'\to\sigma$ $(\sigma\in V(G_p)$,
$\sigma'$ 'trivial', i.e. $G^{(i)}(\sigma')=\sigma'$) into an edge $\sigma'\to G_{i,p}$. We need now
specify the kinetic rates of the new coarse-grained graph $G^{(i)}$, called {\bf step $i$ effective graph}. Define first
\begin{equation} {\mathit{(characteristic\ rate)}} \qquad 1/\tau_{{\cal G}_p} = k_{p,min} \label{eq:char-rate}
\end{equation}
 where $k_{p,min}$ is any internal
rate with scale $n(i)$, equal to the lowest scale of all internal edges in $G_p$;
\begin{equation} {\mathit{(bare\ deficiency\ weight)}} \qquad \bar{\eps}_{{\cal G}_p} \sim \max\{\eps_{\sigma},\ \sigma\in V(G_p)\}; \label{eq:bare-def-weight}
\end{equation}
\begin{equation} {\mathit{(renormalization\ factor)}} \qquad
 Z(0)_{{\cal G}_p} \sim \max_{(\sigma,\sigma') \in {\cal E}_p^{out}}
\frac{k_{\sigma\to\sigma'}}{k_{\sigma}} \label{eq:Z(0)}
\end{equation}
which is $\prec 1$ since outgoing edges are non-dominant;
\begin{equation} {\mathit{ (external\ rate)}}
\qquad k^{ext}_{{\cal G}_p} \sim \frac{1}{\tau_{{\cal G}_p}} \ \times\
Z(0)_{{\cal G}_p} \label{eq:ext-rate}
\end{equation}
The {\bf resonance regime} is defined by $Z(0)_{{\cal G}_p}\sim \bar{\eps}_{{\cal G}_p}$;
in that regime, we cannot decide whether ${\cal G}_p$ is autocatalytic or not. Banning
this regime,
 the {\em Lyapunov\ exponent of} ${\cal G}_p$ is defined as
\begin{equation} \lambda_{{\cal G}_p} \sim
\begin{cases} 0, \qquad Z(0)_{{\cal G}_p} \succ \bar{\eps}_{{\cal G}_p} \\
 \bar{\eps}_{{\cal G}_p}/\tau_{{\cal G}_p} , \qquad Z(0)_{{\cal G}_p} \prec \bar{\eps}_{{\cal G}_p}
 \end{cases} \label{eq:lambdaGp}
\end{equation}

When $Z(0)_{{\cal G}_p} \succ \bar{\eps}_{{\cal G}_p}$ {\em (non-autocatalytic case)}, the Lyapunov exponent of $G^{int}_p$ is actually $\prec 0$, but this may change when ingoing edges are taken into account at step $(i+1)$.
However, when $Z(0)_{{\cal G}_p} \prec \bar{\eps}_{{\cal G}_p}$ {\em (autocatalytic case)}, the exponent is $\succ 0$,
and adding ingoing edges can only increase it.

The renormalized rates of the compound vertex $G_{i,p}$ are now
\begin{equation} {\mathit{(outgoing\ \ rates)}} \qquad
 k_{G_{i,p}\to\sigma'} \sim \frac{1}{\tau_{{\cal G}_p}} \, \times\, \max_{\sigma \in V(G_p)} \frac{k_{\sigma\to\sigma'}}{k_{\sigma}} \label{eq:ren-rates}
\end{equation}
\begin{equation} {\mathit{(deficiency\ rate) }}\qquad \kappa_{G_{i,p}} \sim \bar{\eps}_{{\cal G}_p}/\tau_{{\cal G}_p} \label{eq:ren-def-rate}
\end{equation}
\noindent giving rise to a self-edge $G_{i,p}\to G_{i,p}$, interpreted as a doubling reaction
$G_{i,p}\overset{\kappa_{G_{i,p}}}{\to} G_{i,p}+G_{i,p}$; note that, in the autocatalytic case, $\lambda_{{\cal G}_p}\sim
\kappa_{G_{i,p}} \succ k_{G_{i,p}}$ is dominant;
\begin{equation} {\mathit{(ingoing\ rates)}} \qquad k_{\sigma' \to
G_{i,p}} \sim \max_{\sigma \in V(G_p)} k_{\sigma'\to\sigma}. \label{eq:ren-ingoing-rate}
\end{equation}
Finally, we call {\em weight of $G_{i,p}$} the factor
\begin{equation} {\mathit{(weight)}}\qquad Z^{-1}_{G_p} \sim \Big(
\max(Z(0)_{{\cal G}_{i,p}}, \bar{\eps}_{{\cal G}_p}) \Big)^{-1} \label{eq:ren-weight}
\end{equation}
and let
\begin{equation} Z(\eps,\alpha)_{{\cal G}_p}:= Z(0)_{{\cal G}_p} - \bar{\eps}_{{\cal G}_p} + \alpha \tau_{{\cal G}_p}, \qquad \eps,\alpha\ge 0 \label{eq:Zepsalpha}
\end{equation}
Note that $Z(\eps,0)_{{\cal G}_p} \sim Z(0)_{{\cal G}_p}$ in the free regime. When $\alpha\succ Z_{G_p}/ \tau_{{\cal G}_p}$, $Z(\bar{\eps}_{{\cal G}_p},\alpha)_{{\cal G}_p}
\sim Z(0,\alpha)_{{\cal G}_p} \sim \alpha\tau_{{\cal G}_p}$ simply.

Looking at the new graph, we see that, by construction,
\begin{equation}
 k_{G_{i,p}} =
\sum_{\sigma'\not= G_{i,p}} k_{G_{i,p}\to \sigma'} \sim k^{ext}_{{\cal G}_p},
 \end{equation}
 from which we get
the renormalized deficiency weight,
\begin{equation}
\eps_{G_{i,p}}\sim \kappa_{G_{i,p}}/k_{G_{i,p}} \sim \frac{\bar{\eps}_{{\cal G}_p}}{Z(0)_{{\cal G}_p}}, \label{eq:ren-def-weight}
\end{equation}
and the new transition weights,
\begin{equation}
w(\alpha)_{G_{i,p}\to \sigma'} \sim \frac{k_{G_{i,p}\to\sigma'}}{k_{G_{i,p}}+\alpha}
\label{eq:new-transition-weights}
\end{equation}
In the non-autocatalytic case,
\begin{equation}
 w(\alpha)_{G_{i,p}\to \sigma'} \sim \begin{cases} w(0)_{G_{i,p}\to \sigma'} \sim
Z(0)_{{\cal G}_p}^{-1}\ \times\ \max_{\sigma\in V(G_p)} \frac{k_{\sigma\to\sigma'}}{k_{\sigma}}, \\ \qquad \qquad \qquad\qquad\qquad\qquad\qquad \alpha\preceq k_{G_{i,p}} \\ \frac{k_{G_{i,p}\to \sigma'}}{\alpha} \sim (\alpha\tau_{{\cal G}_p})^{-1}
\ \times\ \max_{\sigma\in V(G_p)} \frac{k_{\sigma\to\sigma'}}{k_{\sigma}}, \\
\qquad\qquad\qquad\qquad\qquad\qquad\qquad \alpha\succeq
 k_{G_{i,p}}
\end{cases}
\end{equation}
In the autocatalytic case, on the other hand, we only consider $\alpha\succeq
\lambda_{{\cal G}_p} \succ k_{G_{i,p}}$; then
\begin{equation}
 w(\alpha)_{G_{i,p}\to \sigma'} \sim \frac{k_{G_{i,p}\to \sigma'}}{\alpha}, \qquad \alpha\succeq
\lambda_{{\cal G}_p}. \label{eq:w(alpha)-free-autocata-regimes}
\end{equation}
The {\em free regime} is obtained in the non-autocatalytic case for $\alpha \prec k_{G_{i,p}}$.
The {\em autocatalytic regime} is obtained in the autocatalytic case for $\alpha \sim \lambda_{{\cal G}_p}$. In both cases,
\begin{eqnarray} && w(\alpha)_{G_{i,p}\to \sigma'} \sim w(\lambda_{{\cal G}_p})_{G_{i,p}\to \sigma'} \sim
Z_{G_p}^{-1} \ \times\ \max_{\sigma\in V(G_p)} \frac{k_{\sigma\to\sigma'}}{k_{\sigma}}
\nonumber\\ && \qquad {\mathit{(free\ and\ autocatalytic\ regimes)}} \label{eq:walphaGsigma'Z}
\end{eqnarray}
This defines a threshold value, $\alpha_{\mathrm{thr}} \sim k_{G_{i,p}}$ (non-autocatalytic case),
$\alpha_{\mathrm{thr}}\sim \lambda_{{\cal G}_p}$ (autocatalytic case); when $\alpha\succ \alpha_{\mathrm{thr}}$, defining the {\em degraded regime},
the denominator in the expression (\ref{eq:new-transition-weights}) behaves like $\alpha$. {\em Note
also that, in all three regimes, the prefactor in $w(\alpha)_{G_{i,p}\to\sigma'}$ is
$Z(0,\alpha)_{{\cal G}_p}^{-1}$.} This is to be remembered when considering (\ref{eq:pisigma1}),
(\ref{eq:pisigma2}) below.

Since $k_{G_{i,p}} \prec k_{p,min}$, the next-step cut-off scale will be $<n(i)$, allowing
downward induction on $n$. Note also that, when $G_{i,p}$ is autocatalytic, $\kappa_{G_{i,p}}
\succ k_{G_{i,p}}$ is the highest reaction scale with reactant $G_{i,p}$, see below (\ref{eq:dominant edges}); therefore, if $\alpha\succeq \kappa_{G_{i,p}}$,
$w(\alpha)_{G_{i,p}\to \sigma'} \sim
\frac{k_{G_{i,p}\to\sigma'}}{\alpha} \prec 1$.
Thus we have this essential fact: {\em edges outgoing from an autocatalytic vertex are small}.

\medskip\noindent {\em Final step.} The next renormalization steps $(i=2,3,\ldots)$ are exactly similar.
 Renormalization stops when there are no more non-trivial maximal
dominant SCCs. Each step involves a non-trivial nesting step $\Sigma^{(i)}\to \Sigma^{(i+1)}$, hence the total number of steps, $i_{\mathrm{max}}$ is less than the number of species. The step $i$ effective
graph $G^{(i)}=(V(G^{(i)}),E(G^{(i))})$ is obtained from the bare graph $G^{(0)}$ by
successive merging/rewiring steps.

\medskip\noindent {\em Step $i$ Lyapunov data.} In order to follow the evolution of the system as slower and slower transitions are incorporated, we consider the sequence of cut-off graphs
$G_{\mathrm{cut}}(i)$, $i=0,\ldots,i_{\mathrm{max}}$, and compute their Lyapunov data (for $i<i_{\mathrm{max}}$, these
may be understood as \enquote{transient} Lyapunov data, though we do not discuss dynamics here). By definition, $E(G_{\mathrm{cut}}(i))$ is
the set of bare edges (edges of $G^{(0)}$ with their rates) involved in the formation of the
effective graph $G^{(i)}_{\mathrm{cut}}$ cut-off just before scale $n(i+1)$, and $V(G_{\mathrm{cut}}(i))\subset\Sigma$ the subset of sources and targets of edges in $E(G_{\mathrm{cut}}(i))$. By construction, $G_{\mathrm{cut}}(i_{\mathrm{max}})=G^{(0)}$, and $\lambda^*(G_{\mathrm{cut}}(i))\le
\lambda^*(G_{\mathrm{cut}}{(i+1)})$, since cutting edges reduces the growth rate.

\medskip\noindent We wish to approximate the Lyapunov weights of $G_{\mathrm{cut}}(i)$; when the latter is not strongly connected, the Lyapunov vector is not uniquely determined, which leads us to the following \enquote{initial condition dependent}
construction. Fix $\sigma_0\in V(G_{\mathrm{cut}}(i))$; we let $(G_{\mathrm{cut}}(i))_{\sigma_0} \subset
G_{\mathrm{cut}}(i)$ be the subgraph with vertex subset $V((G_{\mathrm{cut}}(i))_{\sigma_0})= \{\sigma\in V(G_{\mathrm{cut}}(i))\ |\ \sigma$ accessible from $\sigma_0\}$, where \enquote{$\sigma$ accessible from $\sigma_0$} means: $\sigma$ is connected to $\sigma_0$ by some path; if we assume that
the initial condition is $X_{\sigma}(0) = \delta_{\sigma,\sigma_0}$, only this subgraph
can be reached. The {\bf $\sigma_0$-SCCs} of $G_{\mathrm{cut}}(i)$ are the maximal dominant
SCCs $G_q, q=1,2,\ldots$ of $G_{\mathrm{cut}}(i)$ included in $(G_{\mathrm{cut}}(i))_{\sigma_0}$.

\medskip {\em Lyapunov exponent of $(G_{\mathrm{cut}}(i))_{\sigma_0}$.} Each $G_q$ has a threshold rate $\alpha_q$, which
 is (by definition) $0$ if $G_q$ is not autocatalytic, otherwise gives the order of magnitude
 of the Lyapunov exponent $\lambda^*(G_q)$, with logarithm equal to the deficiency scale of $G_q$. The $G_q$ are connected between themselves in
 various ways, but a non-autocatalytic $G_q$ has no outgoing edge in $(G_{\mathrm{cut}}(i))_{\sigma_0}$ (otherwise $G_q$ would not be maximal). The {\bf threshold scale}
$\lfloor \log_b \alpha \rfloor$ of
$(G_{\mathrm{cut}}(i))_{\sigma_0}$ (logarithm of the {\bf threshold rate}) is the maximal deficiency scale (if any), $-\infty$ else ($\alpha=0$). If $\alpha>0$,
\begin{equation} \lambda^*((G_{\mathrm{cut}}(i))_{\sigma_0}) \sim \alpha \end{equation}
The growth rate of the graph $G_{\mathrm{cut}}(i))$ started from $\sigma_0$ will
be $\sim \alpha$. Otherwise (banning resonance cases), $(G_{\mathrm{cut}}(i))_{\sigma_0}$ is not
autocatalytic.

\medskip {\em Cores.} Cores are maximal dominant $\sigma_0$-SCCs maximizing
the set $\{\alpha_q,q=1,2,\ldots\}$. If $\alpha=0$, all $G_q,q=1,2,\ldots$ are cores, and
$(G_{\mathrm{cut}}(i))_{\sigma_0}$ is not autocatalytic. In the contrary case ($(G_{\mathrm{cut}}(i))_{\sigma_0}$
autocatalytic), we ban the {\bf resonance regime} defined by the case when there exist
$\alpha_{q},\alpha_{q'}$ with $q\not=q'$ such that $\alpha_q\sim\alpha_{q'}\sim \alpha$ are
both maximal. Thus (by reindexing), we may assume that $\alpha\sim \alpha_1\succ \alpha_q, q\not =1$, and
$G_1$ is the only core. In
the non-autocatalytic case, $\alpha_1=\cdots=\alpha_q=\alpha=0$, so that all $G_q$ are cores.

\medskip \emph{Hierarchical formulas for Lyapunov vector/weights of $(G_{\mathrm{cut}}(i))_{\sigma_0}$.}
We approximate the Lyapunov eigenvector of $(G_{\mathrm{cut}}(i))_{\sigma_0}$
by $v_{\sigma}\sim (k_{\sigma} + \alpha_{\sigma_0})^{-1} \pi_{\sigma}$, where
\begin{equation} \pi_{\sigma} \sim \prod_{\sigma'\supsetneq \sigma}
Z^{-1}_{\sigma'} ,
\qquad \sigma\subset G_q\ {\mathrm{core}} \label{eq:pisigma1}
\end{equation}
\begin{eqnarray}
&& \pi_{\sigma} \sim\Big(\max_{\gamma_: G_1\to G^{(i)}(\sigma)} w(\alpha_1)_{\gamma} \Big) \times \nonumber\\
 && \qquad
 \times \prod_{\sigma\subsetneq\sigma'\subset \bar{\sigma}} (Z(0,\alpha_1)_{\sigma'} )^{-1} \qquad
{\mathrm{else}} \label{eq:pisigma2}
\end{eqnarray}
if $\sigma\in V((G_{\mathrm{cut}}(i))_{\sigma_0})$, where:
\begin{itemize}
\item in (\ref{eq:pisigma1}),
\enquote{$\sigma\subset G_q$ core} means: $\exists q,\ \sigma\in V(G_q)$. Then the product $\prod_{\sigma'\supsetneq
\sigma}(\cdots)$ is over the chain of merged vertices containing $\sigma$ (if any);

\item in (\ref{eq:pisigma2}), it is assumed that $\sigma$ is not in a core, but
there exists a core $G_q$ and a path $\gamma: G_q\to G^{(i)}(\sigma)$ (otherwise
$\pi_{\sigma}=0$); by construction,
$G_q$ is autocatalytic, so that $q=1$. Then $\bar{\sigma}$ is the maximum compound
vertex containing $\sigma$, and $\prod_{\sigma\subsetneq\sigma'\subset \bar{\sigma}} (\cdots)$ is the product
over the chain of merged vertices $\sigma'$ contained in $\bar{\sigma}$.
\end{itemize}

\Medskip {\em Remark.} When $\sigma$ is not nested in a core, see (\ref{eq:pisigma2}), paths $\gamma$ cannot, by construction, form an $\alpha$-dominant cycle. Though there is no upper bound over the length of paths $\gamma$, which
may be arbitrarily large if there are cycles, the maximum order of magnitude is attained over the
finite subset of simple excursions (paths with no self-intersection). Moreover, one can write a generalization of Dijkstra's algorithm for finding the shortest paths between nodes in a weighted graph, which returns the set of maximal weight $\gamma$'s
in the form of a {\bf directed acyclic graph (DAG)} rooted in $G_1$.

\Medskip {\em Hierarchical formula for adjoint Lyapunov eigenvector.} The estimate for $v^{\dagger}$ is simpler. Reversing edges, let $\bar{\Sigma}_{\sigma_0}\subset \Sigma$
be the subset of bare vertices from which a $\sigma_0$-core is accessible. Then
\begin{equation}
v^{\dagger}_{\sigma} \sim \max_{G_q\, {\mathrm{core}},\, \gamma^{\dagger}:G^{(i)}(\sigma)\to G_q}
w(\alpha_1)_{\gamma^{\dagger}} \label{eq:vdaggersigma}
\end{equation}
In (\ref{eq:vdaggersigma}), it is assumed that there exists a core $G_q$ and a reverse path $\gamma^{\dagger} : G^{(i)}(\sigma) \to G_q$ (otherwise $v^{\dagger}_{\sigma}=0$).
In particular, $v^{\dagger}_{\sigma}\sim 1$ if $\sigma$ is a core. The maximum order
of magnitude, as in the case of the hierarchical formula for $\pi$, is attained over the
finite subset of simple excursions, and may be made explicit in terms of a {\bf reverse DAG}
with edges oriented towards the root $G_1$ (instead of away from $G_1$).

\Medskip The key formulas (\ref{eq:pisigma1})--(\ref{eq:vdaggersigma}) constitute what
we call the {\bf hierarchical formulas}. The dominant SCCs, DAG and reverse DAG constitute the {\bf hierarchical graph}. An example is produced in Result section (\cref{sec_example_3}), with edges of the DAG,
resp. reverse DAG, drawn in red, resp. blue. The analogy is with blood circulation (arteries/veins),
with edges from, resp. to, the cores, drawn red, resp. blue.

\subsubsection{Phases and thresholds}
\label{sec_phases_thresholds}

{\em Phases.} The multi-scale analysis requires previous knowledge of the {\em scale ordering}
of the edges (including self-edges),
but {\em not} of the scales themselves. Renormalized rates as defined in (\ref{eq:ren-rates})
rewire the network and shuffle the scales in a way that is difficult to anticipate. The different
possible reorderings define as many {\em chemical phases}; a phase $\phi$ is defined by a system of
explicit inequalities in the space of $\log(k)$ parameters, with geometric image ${\cal D}(\phi)$; see \cref{sec_discussion}. Iterating renormalization steps possibly produces a very large number of phases. We write $k\sim\phi$ if $k$ satisfies the inequalities defining $\phi$. If $\phi_1,\phi_2$
are two neighboring phases, i.e. if their geometric images ${\cal D}(\phi_1)$, ${\cal D}(\phi_2)$ have a common
boundary, we write similarly $\phi_1\sim\phi_2$. Connecting neighboring phases by an edge yields an adjacency graph. The hierarchical algorithm allows a fast {\em local} exploration of this graph, but getting a global picture for large networks may be very time-consuming
in general.

\medskip \noindent {\em Viability thresholds.} Networks with positive degradation rates $\beta_{\sigma}>0$
may be analyzed exactly as ordinary networks with a special species denoted $\emptyset$
and reaction rates $\sigma\overset{\beta_{\sigma}}{\to} \emptyset$. They modify the network
only if they become dominant at some step; see analysis of Toy Formose IIb. Assume that
$\sigma$ belongs to a step $i$ non-trivial maximal dominant SCC, then a dominant step $i$ edge $\sigma\overset{\beta^{(i)}_{\sigma}}{\to} \emptyset$ makes it non maximal, in particular the
vertices in the SCC are not merged any more, and do not define an autocatalytic component.
Conversely, an autocatalytic SCC whose support does not contain $\sigma$ remains autocatalytic
whatever the value of $\beta_{\sigma}$.

\subsubsection{Simplified time evolution}
\label{sec_time_evol}

We again assume that the initial condition is $X_{\sigma}(0) = \delta_{\sigma,\sigma_0}$ (only $\sigma_0$ is
present in the mixture). At step $i$ (kinetic scale $n(i)$), the asymptotic formula (\ref{eq:XvvX(0)}) suggests the following approximation if $\sigma_0\in V(G_{\mathrm{cut}}(i))$,
\begin{equation} X^{(i)}(t) \sim \langle v^{\dagger,(i)}, X(0)\rangle
 e^{\lambda^{(i)} t} v^{(i)} \label{eq:XivdaggerX0lambdavi}
\end{equation}
with $\lambda^{(i)}\sim \alpha_{\sigma_0}$, see (\ref{eq:pisigma1}, \ref{eq:pisigma2},
\ref{eq:vdaggersigma}).
A proper normalization is ensured by choosing $\max_{\sigma} (v^{\dagger,(i)}_{\sigma}) = 1$ and
fixing $v^{(i)}$ so that $\langle v^{\dagger,(i)},v^{(i)}\rangle \sim 1$.
Mind that Lyapunov data $(\lambda^{(i)}, v^{(i)}, v^{\dagger,(i)})$ are actually dependent
on $\sigma_0$ in general (though $\Sigma_{\sigma_0}=
\bar{\Sigma}_{\sigma_0}=\Sigma$ in most of our examples, because $G^{(i)}$ is strongly connected). If $\sigma_0\not\in V(G_{\mathrm{cut}}(i))$, we simply let $X^{(i)}_{\sigma}(t)\sim \delta_{\sigma,\sigma_0}$ (trivial time evolution). The {\bf match condition}
\begin{equation} X^{(i-1)}_{\sigma}(t) \sim X^{(i)}_{\sigma}(t) \label{eq:match-condition}
\end{equation}
defines a {\bf species-dependent transition time} $t^{(i)}_{\sigma_0\to\sigma}$ between step $(i-1)$ and step $i$. By construction, $t^{(i)}_{\sigma_0\to\sigma}$ is upper-bounded by {\bf induction times} $t^{(i)} \sim b^{-n(i)}$, but very often (see discussion in \cref{sec_framework}, and Examples below),
it is much smaller. After the final step, $v^{(i)}\equiv
v^{(\infty)}, v^{\dagger,(i)}\equiv v^{\dagger,(\infty)}$ become constant, and provide
$t\to\infty$ asymptotics; $\lambda^{(\infty)}$ is an approximation for the growth rate $\lambda^*$ of the
bare network. In the particular case when $G$ has no degradation rates, no irreversible $1\to 2$
reactions and is detail balanced, it is easy to see that $v^{(\infty)}$ approximates the equilibrium measure, and $v^{\dagger,(\infty)}\sim {\bf 1}$, where ${\bf 1}_{\sigma}=1,
\sigma\in \Sigma$.

We thus get simplified dynamics $X^H(t|k,\phi)$ for $k$ within the geometric domain of a given phase $\phi$.
Transient dynamics between transition times are not adequately covered in general by this rapid discussion, and
requires a more sophisticated
study which is left for future work. However, match conditions turn out to be sufficient to get the dynamics of the examples treated below.

\subsection{Kinetic inference based on hierarchical models}
\label{sec_inference}

We recast here our hierarchical formulas in a new light.
Namely, instead of considering the output $X^H(t|k,\phi)$ of the hierarchical algorithm as an approximation of the true concentration vector $X(t)$, we use it as the main building block of a family of models indexed by $k$ for the time-dependent concentrations. Then, given the analytical formulas, we consider statistical inference of chemical network as a model.

As a proof of concept, we consider in this article a simple MCMC approach to optimize the set of chemical rates $k$ based on time series of chemical concentration measurements.

For each species $\sigma$, we consider a time series of measurements $x_\sigma(t)$ obtained from a simple log-normal model with
variance $\eps^2$ centered on $X(t)$, $\log x_{\sigma}(t) \sim {\mathcal{N}}(\log X_{\sigma}(t),\eps^2)$, with density
\begin{equation}
p(x|k,\phi) = \prod_{t\in T} \prod_{\sigma\in\Sigma} \frac{e^{-\log^2(x_{\sigma}(t) / X^H(t|k,\phi)) }} {x_{\sigma}(t)\sqrt{2\pi \eps^2}}
\end{equation}
where $T$ is a finite set of measurement times.

\medskip As a general principle, different phases have a qualitatively different dynamical behavior, which allows one to
recover the phase from a time series of observed concentrations. In a Bayesian approach, starting from a
law on $(k,\phi)$ (prior based on chemical expertise), observations should then yield a law $p(k,\phi|x) \equiv p(k,\phi|(x(t))_{t\in T})$ (posterior) which puts
a negligible marginal weight $p(\phi|x)$ on all phases except one, $(\phi\not=\phi^*)\Rightarrow (p(\phi|x)/p(\phi^*|x)\ll 1)$. Furthermore, $p(k,\phi^*|x)$ should concentrate in a small volume around some optimal $k=k^* \sim \phi^*$ in the $\log(k)$ variables. Since $X^{H}(\cdot;k^*,\phi^*)\not = X(\cdot;k^*,\phi^*)$, the inferred values
$(k^*,\phi^*)$ should not be interpreted as the best estimates for the kinetic rates given the observations, rather, as
the parameters of the family of hierarchical models reproducing most accurately the observations. However, because
the hierarchical approximation is generally good, $k^*$ should be a good proxy for the actual kinetic rates. Since these are unknown in practice, the distinction is not very relevant.

\medskip Given that an a priori exploration of all phases is impractical for large networks, a simple and natural strategy to get $(k^*,\phi^*)$ is the following. (1) Start from some not implausible $k=k_0$. Apply the hierarchical
algorithm to get $\phi_0=\phi(k_0)$ and $X^{H}(t|k,\phi_0), k\sim\phi_0$. Then compute $k^*(\phi_0):=
$argmax$_{k\sim\phi_0} \log p(x|k,\phi_0)$ by a gradient descent algorithm, to obtain the maximum log-likelihood
$\log$ MLE$(\phi_0) = \log p(x|k,\phi_0)$ in the phase $\phi_0$. (2) In the next step, we explore the neighboring phases
$\phi\sim\phi_0$ by the hierarchical algorithm, and compute similarly $k^*(\phi)$ and MLE$(\phi_0)$ for all of them.
Then we move randomly from $\phi_0$ to a neighboring phases by a Metropolis algorithm based on probability ratios
$({\mathrm{MLE}}(\phi)/{\mathrm{MLE}}(\phi_0))$. Steps (1) and (2) are repeated in parallel on $N$ trajectories.
The phase random walk stops when the empirical distribution has reached equilibrium.

\medskip

\section{Results}
\label{sec_results}

We illustrate the algorithm described in \cref{sec_methods} with
 three examples. More material (in particular, a comparison to simulations) can be found in
 Supplementary Materials. Reactions are listed from first to last according to scale ordering. Lyapunov data are systematically derived; transition times, which follow immediately, are
 discussed only in a few cases, for a pure initial state $X_{\sigma}(0)=\delta_{\sigma,\sigma_0}$. In general, dominant edges are boldface, and so is a vertex $\sigma$ at scale $n_{\sigma}$. Asymptotic Lyapunov data (as in \cref{sec_framework}) are denoted $(\lambda^{(\infty)}_{\phi},
 v^{(\infty)}_{\phi}, v^{\dagger,(\infty)}_{\phi})$, where $\phi$ is a phase index. Lyapunov
 eigenvectors $v$ are normalized as explained below (\ref{eq:XivdaggerX0lambdavi}).
Dynamical plots combine numerical solution (full lines) with our hierarchical formulas
(dashed lines).


\subsection{Example 1: one cycle}
\label{sec_example_1}

\subsubsection{Hierarchical formulas}
\label{sec_example_1_hf}

We consider here the model of \cref{example_1}, and choose the multi-scale ordering $k_{\mathrm{off}}\gg k_{\mathrm{on}}\gg k_2 \gg \bar{k}_2$. The associated multi-scale graph is

\begin{center}
\begin{tikzpicture}

\draw[dotted](0,-1)--(0,-4.5); \draw[dotted](1,-3.3)--(1,-4.5);
\draw[dotted](2,-1.)--(2,-2);

\begin{scope}[shift={(0,-1)}]
\draw[dashed](-0.8,-1.4) rectangle (2.85,0.85); \draw(0.8,1.2) node {${\cal G}_1$};

\draw[<-, ultra thick](0,0)--(2,0); \draw(-0.3,0) node {$\mathsf{S}_0$};
\draw(1,0.3) node {$k_{\mathrm{off}}$};
\draw(2.35,0) node {$\mathsf{S}_1$};
\draw(5,0) node {$n^{(0)}_1$};
\end{scope}

\draw[dashed](-1,-1.5)--(4,-1.5);
\draw[->,ultra thick](0.,-2)--(2,-2);
\draw(1,-1.7) node {$k_{\mathrm{on}}$};
\draw(5,-2) node {$n^{(0)}_2$};

\draw[dashed](-1,-2.5)--(4,-2.5);

\draw[dashed](-1,-3.5)--(4,-3.5);
\draw[->](0,-3.3)--(1,-3.3); \draw[->, ultra thick](1,-4.5)--(0,-4.5);
\draw(0.5,-3) node {$k_2$};
\draw(0.5,-4.2) node {$\bar{k}_2$};
 \draw(1.35,-3.3) node {$\mathsf{S}_2$};

\draw(5,-3) node {$n^{(0)}_3$};

 \draw(5,-4.5) node {$n^{(0)}_4$};

\end{tikzpicture}
\end{center}

plus the doubling reaction, $\mathsf{S}_1\overset{\nu_+}{\to} \mathsf{S}_0+ \mathsf{S}_0$.
 Bare scales are $n^{(0)}_1 = \lfloor \log(k_{\mathrm{off}}) \rfloor$,
$n^{(0)}_2 = \lfloor \log(k_{\mathrm{on}}) \rfloor$, $n^{(0)}_3 = \lfloor \log(k_2) \rfloor$, $n^{(0)}_4 = \lfloor \log(\bar{k}_2) \rfloor$; the scale $\lfloor \log(\nu_+)\rfloor$ is as yet unspecified but
chosen $<n_1^{(0)}$. Dominant edges are $0 \underset{k_{\mathrm{off}}}{\overset{k_{\mathrm{on}}}{\rightleftarrows}} 1$ and $2\to 0$. At step (0), only the $n_1$-scale reaction $1\to 0$ is available.
 Then $n(1) = n_2^{(0)}$ (i.e. above the dashed line separating $(k_{\mathrm{off}})$ from $(k_{\mathrm{on}})$), the unique non-trivial maximal dominant SCC is
 $G_1 = (0\rightleftarrows 1)$, ${\cal G}_1$ is $G_1$ with the unique outgoing edge $0\overset{k_2}{\to} 2$. Collapsing $G_1$ yields a new set of vertices $\Sigma_1= \{\{0,1\},2\} \equiv \{G_1,2\}$ and the following scale-ordered reactions,
\begin{eqnarray*}
&& (k_2) \qquad G_1\to 2 \\
&& ........................... \\
&& (\bar{k}_2) \qquad {\bf 2} \to G_1
\end{eqnarray*}
and an equivalent doubling reaction, $G_1 \overset{\kappa_{G_1}}{\to} G_1 + G_1$, which may be
above $k_2$ (then $G_1$ is autocatalytic) or not. The new edge scales of $n_1^{(1)} = n^{(0)}_3$
and $n_2^{(1)}=n^{(0)}_4$.

We now discuss eqs. (\ref{eq:char-rate}--\ref{eq:ren-def-weight}). First, $1/\tau_{G_1} = k_{\mathrm{on}}$,
$\bar{\eps}_{{\cal G}_1} = \eps_1 \sim \frac{\nu_+}{k_{\mathrm{off}}}$, $Z(0)_{{\cal G}_1} \sim
\frac{k_2}{k_{\mathrm{on}}}$,
$k^{ext}_{{\cal G}_1} \sim k_2$. Rates of $G_1$ are $k_{G_1}\sim k_{G_1\to 2}
\sim k_2$, $\kappa_{G_1} \sim \frac{k_{\mathrm{on}}}{k_{\mathrm{off}}} \nu_+$, $k_{2\to G_1} \sim \bar{k}_2$, and $\eps_{G_{1,1}}\sim \frac{\nu_+/k_{\mathrm{off}}}{k_2/k_{\mathrm{on}}}$. The graph ${\cal G}_1$ is autocatalytic if $\frac{k_2}{k_{\mathrm{on}}} \prec \frac{\nu_+}{k_{\mathrm{off}}}$,
 i.e. $k_2 \prec \frac{\nu_+}{k_{\mathrm{off}}} k_{\mathrm{on}}$. In this regime {\em (phase $a$)} -- where the
 dotted line is irrelevant --, we get $ \lambda_{{\cal G}_1}\sim \kappa_{G_1}$, the effective step 1
 graph $ G^{(1)} : G_1 \underset{\bar{k}_2}{\overset{k_2}{\rightleftarrows}} 2$ has a
 single dominant non-trivial edge $2\to G_1$ (because the self-edge $G_1\overset{\kappa_{G_1}}{\to} G_1$ is dominant), so the algorithm stops. Assume instead $k_2 \succ \frac{\nu_+}{k_{\mathrm{off}}} k_{\mathrm{on}}$ {\em (phase $b$)}. Then the two edges of $G^{(1)}$ form at scale $n^{(2)}=n^{(0)}_4$
 (i.e. below the dotted line) the maximal dominant SCC
 $G_2 = (G_1 \rightleftarrows 2)$, which is autocatalytic (there are no outgoing edges), with exponent
 $\lambda_{{\cal G}_2}\sim \eps_{G_{1,1}}/\tau_{G_{2,1}} \sim \nu_+ \frac{k_{\mathrm{on}}}{k_{\mathrm{off}}}
 \frac{\bar{k}_2}{k_2}$.

\medskip\noindent Let us now discuss $v^*, v^{\dagger,*}$, DAGs, and dynamics from
(\ref{eq:pisigma2}), (\ref{eq:vdaggersigma}), using the match conditions (\ref{eq:XivdaggerX0lambdavi}, \ref{eq:match-condition}).

\bigskip (0) {\em At step 0,}
$G_{\mathrm{cut}}(1)= \{1\to 0\}$, $V(G_{\mathrm{cut}}(0))=\{0,1\}$. Choosing $\sigma_0=1$ so that $G_{\sigma_0}(0)=\{1\to 0\}$,
the root is the only core, $0$, the reverse DAG is $1\to 0$, and $v^{(0)}_i=\delta_{i,0}$ $(i=0,1)$, $v^{\dagger,(0)} =
{\bf 1}$. To leading order in the initial regime, $X^{(0)}_i(t) \sim X_{\text{tot}}(0) \delta_{i,0}$, where
$X_{\text{tot}}(0) = \sum_{i=0,1} X_i(0)$ is the total initial concentration. On the other hand, choosing $\sigma_0=2 \not \in
 V(G_{\mathrm{cut}}(0))$,
 $X^{(0)}_2(t)\sim 1$.

\bigskip (1) {\em (step 1)}

\medskip {\em In phase $a$}, the unique core $G_1$ has
weight $Z^{-1}_{G_1} \sim
(\bar{\eps}_{{\cal G}_1})^{-1}$, threshold rate $\alpha_{G_1}\sim \kappa_{G_1}\sim
\lambda_{{\cal G}_1}$, and $w(\lambda_{{\cal G}_1})_{G_1\to 2} \sim \frac{k_{G_1\to 2}}{\lambda_{ {\cal G}_1}}\sim \frac{k_2}{\lambda_{{\cal G}_1}}$, $w(\lambda_{{\cal G}_1})_{2\to G_{1,1}} \sim \frac{\bar{k}_2}{\lambda_{{\cal G}_1}}$. We get: $\pi_0,\pi_1\sim Z_{G_1}^{-1}$, and
$\pi_2\sim w(\lambda_{{\cal G}_1})_{G_1\to 2}$, whence (dividing by $k_{\sigma} + \lambda_{{\cal G}_1}\sim \max(k_{\sigma},\lambda_{{\cal G}_1})$, and multiplying by $\lambda_{{\cal G}_1}$ in the end to ensure proper normalization)
$v^{(\infty)}_a\equiv v^{(1)}_a \sim \lambda_{{\cal G}_1} \left(\begin{array}{c} k_{\mathrm{on}} \\ k_{\mathrm{off}} \\ \lambda_{{\cal G}_1} \end{array}\right)^{-1}
\times \left(\begin{array}{c} Z_{G_{1,1}}^{-1} \\ Z_{G_{1,1}}^{-1} \\ w(\lambda_{{\cal G}_1})_{G_{1,1}\to 2} \end{array}\right) \sim \left(\begin{array}{c} 1 \\ \frac{k_{\mathrm{on}}}{k_{\mathrm{off}}} \\ \frac{k_2 k_{\mathrm{off}}}{k_{\mathrm{on}}\nu_+} \end{array}
\right)$, whereas $v^{\dagger,(1)}_a\sim \left(\begin{array}{c} 1 \\ 1 \\ w(\lambda_{{\cal G}_1})_{2\to G_{1,1}} \end{array}\right) \sim \left(\begin{array}{c} 1 \\ 1 \\ \frac{\bar{k}_2 k_{\mathrm{off}}}{\nu_+ k_{\mathrm{on}}}
\end{array}\right).$ Thus
\begin{equation}
X^{(1)}(t) \sim v^{\dagger,(1)}_{a,\sigma_0} e^{\lambda_{{\cal G}_1}t} v^{(1)}_a
\label{eq:Ex1-a-X(1)}.
\end{equation}

\medskip {\em In phase $b$}, (1) $G_{\mathrm{cut}}(1)= \{1\to 0, 0\to 1, 0\to 2\}$, $V(G_{\mathrm{cut}}(1))=\{0,1,2\}$; we now have $Z^{-1}_{G_1} \sim Z(0)^{-1}_{{\cal G}_1}$. The unique core
is now $2$, so that
$v^{(1)}_b\sim \left(\begin{array}{c} 0 \\ 0 \\ 1 \end{array}\right)$,
$v_b^{\dagger,(1)} \sim {\bf 1}$, and
$X^{(1)}(t) \sim \left(\begin{array}{c} 0 \\ 0 \\ 1 \end{array}\right)$, compare with
(\ref{eq:Ex1-a-X(1)}). (2) Next,
$G_{\mathrm{cut}}(2)= G^{(0)}$,
$Z^{-1}_{G_2}\sim \bar{\eps}_{{\cal G}_2}^{-1} \sim \eps_{G_1}^{-1}$. Then
$v^{(\infty)}_b\equiv v_b^{(2)} \sim \lambda_{{\cal G}_2} \left(\begin{array}{c} k_{\mathrm{on}} \\ k_{\mathrm{off}} \\ \bar{k}_2 \end{array}\right)^{-1}
\times \left(\begin{array}{c} Z_{G_1}^{-1} Z_{G_2}^{-1} \\ Z_{G_1}^{-1}
Z_{G_2}^{-1} \\ Z_{G_2}^{-1} \end{array}\right) \sim \left(\begin{array}{c} \frac{\bar{k}_2}{k_2} \\ \frac{k_{\mathrm{on}} \bar{k}_2}{k_{\mathrm{off}} k_2} \\ 1
\end{array}\right)$ and $v^{\dagger,(2)}_b \sim {\bf 1}$. Thus
$X^{(2)}(t) \sim e^{\lambda_{{\cal G}_2}t} v^{(2)}_b$,
 compare with (\ref{eq:Ex1-a-X(1)}).

\subsubsection{Inference}
\label{sec_example_1_inference}

We choose here $k = k^{\text{ref}}$ inside phase $a$, and
show the result of the inference algorithm sketched above. Numerical parameter values are $b=4$ and
\begin{equation}
k^{\text{ref}}_{\text{off}} = b^{-0.5},\ k^{\text{ref}}_{\text{on}} = b^{-3.5}, \ k^{\text{ref}}_2 = b^{-6.5},\ k^{\text{ref}}_2 = b^{-8.5}, \
\nu^{\text{ref}}_+ = b^{-2.5}
\end{equation}

We write $X^{H}(t;k,\phi,i)$, $\phi = a,b$, $i = 0,1,2$ for the dynamically extended
hierarchical formulas for the phase $\phi$ dynamics started from the seed $x_j(t=0)=\delta_{i,j}$ (only species $i$
is present initially). Then (see Suppl. Math. \S II B. for a derivation of the formulas):
\begin{equation}
 X^{H}(t\, |\, k, a, 0) \sim e^{\lambda_{{\cal G}_1}t} \left(\begin{array}{c} 1 \\
\frac{k_{\mathrm{on}}}{k_{\mathrm{off}}} \tanh(k_{\mathrm{off}}t) \\
\frac{k_2 k_{\mathrm{off}}}{\nu_+ k_{\mathrm{on}}} \tanh(\frac{\nu_+ k_{\mathrm{on}}}{k_{\mathrm{off}}}t)
\end{array}\right),
\end{equation}

\begin{equation}
X^{H}(t\, |\, k,b,0) \sim e^{\lambda_{{\cal G}_2}t} \left(\begin{array}{c} e^{-k_2 t} + \frac{\bar{k}_2}{k_2} \\ \frac{k_{\mathrm{on}}}{k_{\mathrm{off}}} \tanh(k_{\mathrm{off}}t)\, \times\, (e^{-k_2 t} + \frac{\bar{k}_2}{k_2} )
 \\
\tanh(k_2 t)
\end{array}\right)
\end{equation}

when the seed is species $0$, and

\begin{equation}
 X^{H}(t\, |\, k,a,2) \sim \left(\begin{array}{c} \frac{\bar{k}_2 k_{\mathrm{off}}}{k_{\mathrm{on}}\nu_+}
 \tanh (\frac{\nu_+ k_{\mathrm{on}}}{k_{\mathrm{off}}}t)\, e^{\lambda_{{\cal G}_1}t} \\
 \qquad\qquad \frac{\bar{k}_2}{\nu_+}
 \tanh (\frac{\nu_+ k_{\mathrm{on}}}{k_{\mathrm{off}}}t)\, e^{\lambda_{{\cal G}_1}t} \\
 1+ k_2 \bar{k}_2 (\frac{k_{\mathrm{off}}}{k_{\mathrm{on}}\nu_+})^2 \, e^{\lambda_{{\cal G}_1}t}
\end{array}\right),
\end{equation}

\begin{equation}
X^{H}(t\, |\, k,b,2) \sim e^{\lambda_{{\cal G}_2}t} \left(\begin{array}{c} \frac{\bar{k}_2}{k_2} \tanh (k_2 t) \\
\frac{k_{\mathrm{on}}}{k_{\mathrm{off}}} \frac{\bar{k}_2}{k_2} \tanh (k_2 t) \\ 1
\end{array}\right)
\end{equation}
when the seed is species $2$. Any adequate time-window should include the transition time $\tau = b^{5.5} \simeq 2\times 10^3$
defined by $\lambda_{{\cal G}_1}\tau \equiv 1$, around which we start seeing the asymptotic slope of log-concentrations curves (formulas $X^H(t|k,\phi,i)$ are actually phase-independent
to lowest order when $t\ll \tau$). On the other hand, the equilibration time between species 0 and 1 is $\sim 1$, which is much smaller. We assume that we do not have such short-time concentration measurements, and decide instead that $T$ is made up of
regularly spaced values ranging in $(b^{4.5},b^{6.6}) \simeq (5\times 10^3, 9\times 10^4)$. A number of time measurements comparable to the number of reactions is enough for the inference; we chose for the simulation
dim$(T)=12$.

\medskip {\em Exact inference.} We first maximize $p(x_{SEED}|k,\phi=a)$ with $x_{{\mathrm{SEED}}} = X^H(\cdot|k_{\mathrm{ref}},a,{\mathrm{SEED}})$, SEED$=0,1,2$, starting from some arbitrary initial rates $k^{init} \equiv k^{init}[a]$,
$k^{init}_{off} = b^{-0.5}, k^{init}_{on} = b^{-2.5}, k^{init}_2 = b^{-4.5}, \bar{k}^{init}_2 = b^{-5.5},
\nu^{init}_+ = b^{-1.5}$ located inside the geometric domain of phase $a$. Since $x$ is the time-trajectory of a
hierarchical model, it is in theory possible to get an exact inference, with $k^{(n)}$ ($n$-th iteration of the
solve) converging to $k_{\mathrm{ref}}$, and $x^{(n)}(\cdot) = X^H(\cdot|k^{(n)},a,i)$ converging to $x_i$.

\begin{figure*}
\centering\includegraphics[width=\textwidth]{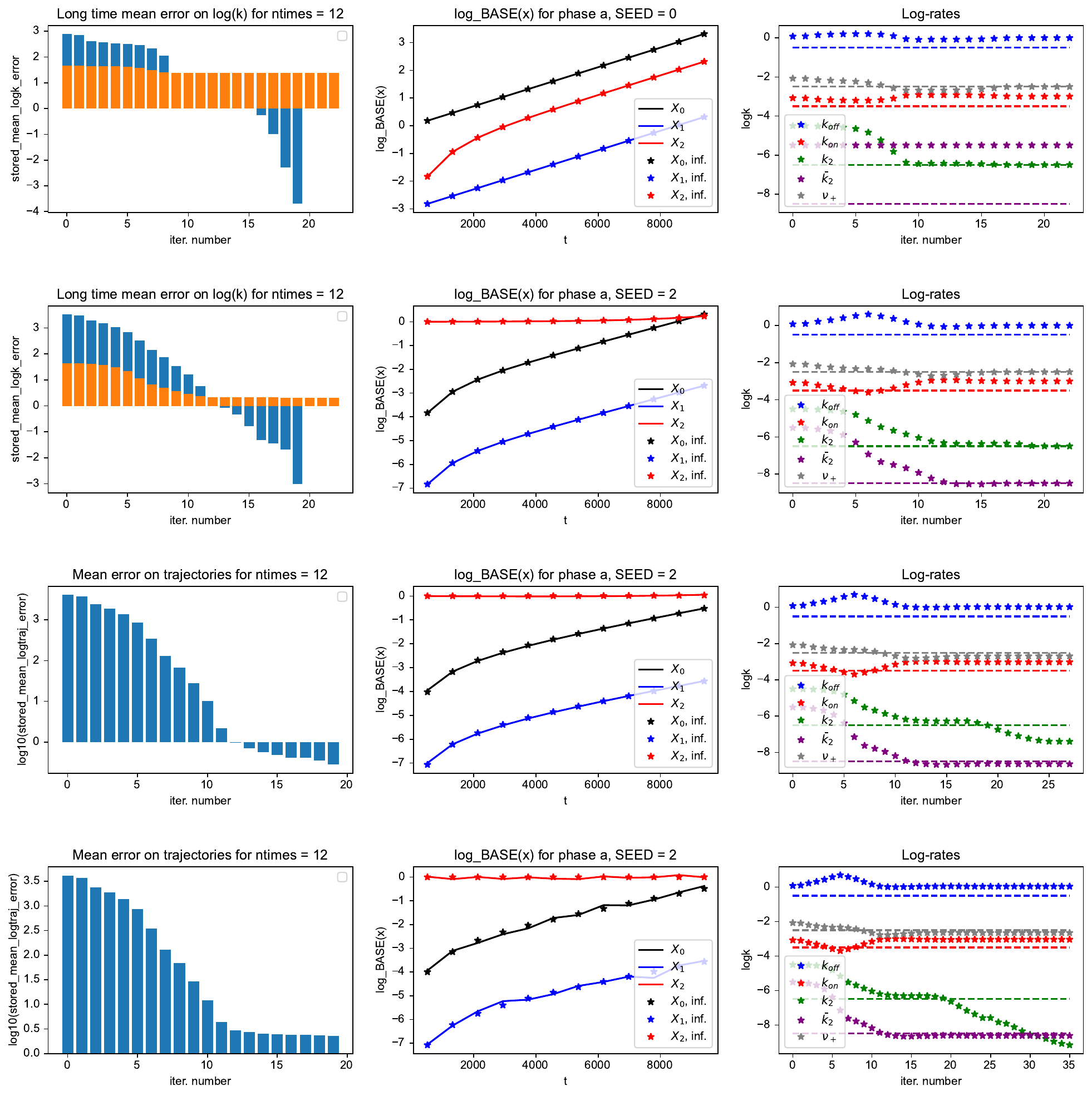}
\caption{Inference results}
\label{fig_inference_results}
\end{figure*}

The results are shown in \cref{fig_inference_results} when $i=0$ (top), $2$ (bottom). We use the L-BFGS-B method with gtol = $10^{-1}$ stopping condition ($L^{\infty}$ norm on the gradient of the objection function). The algorithm terminates after $<20$ iterations and is very quick (a few seconds). The first plots give $\log_{10}(MSE_i)$, where $MSE_i$ is the mean square error on trajectories and may be used as objective function in place of the log-likelihood since the two differ only by a trivial affine transformation,

\begin{equation}
MSE_i = \frac{1}{{\mathrm{dim}}(T)} \sum_{t\in T} \sum_{\sigma\in\Sigma} \log^2(x_{\sigma}(t)/X^H(t|k,a,i))
\end{equation}

The values of the last iterations are very negative, suggesting that the model reproduces quite closely the
time trajectories, which is confirmed by the second log-plots, with $\log_b(x)$ on the $y$-axis (lines for
$x = x^{\mathrm{ref}}$, bullets for the result of the inference, with $k=k^*$ equal to the values of the rates at the last iteration).

Then the third plots show the log-rates as a function of the iteration number (stars), and the reference values (dotted lines). A few things seem to go wrong -- but this is only due to the lack of identifiability of the model. There are actually two problems:
\begin{enumerate}
\item Because of the lack of short-time measurements, substitutions $(k_{\mathrm{off}},k_{\mathrm{on}}) \to (Ck_{\mathrm{off}},Ck_{\mathrm{on}})$ with $C$ constant do not change the model.
There is nothing we can do against it if short-time measurements are not possible, but note that such partial rate rescalings can be predicted {\em beforehand}, and do not harm the quality of the inference.
For the same reason, we did not provide dynamically extended hierarchical formulas for $X^H(t|k,\phi,1)$ started from SEED $=1$, since they are in practice indistinguishable from those for $X^H(t|k,\phi,0)$.
\item The value of $\bar{k}_2$ is accurately inferred when the seed is $2$, but totally wrong when the seed is $0$.
This, again, can be predicted from the hierarchical formulas themselves: $X^H(\cdot|k,a,0)$ is independent of
$\bar{k}_2$, $X^H(\cdot|k,a,2)$ is not. The fourth plots, which represent the MSE on the log-rates,
$\frac{1}{5} \sum_i \log^2(k^*_i/k^{\text{ref}}_i)$, confirm this analysis.
\end{enumerate}

The third column gives the result of the inference when $x = x_{\mathrm{num}}$ is the true value of the
concentrations obtained by numerically solving the dynamics. We chose SEED $=2$ since the inference is better starting from that initial condition in the given time-regime. Curves are unsurprisingly very similar. We kept the
dotted lines on the log-rate plot, but rates are now \enquote*{effective} rates and should not converge exactly to $k_{\mathrm{ref}}$ any more; this is particularly noticeable for $k_2$ (in green). The fourth column gives the variant when
$x=x_{noise}$ is log-normal, centered upon $x_{num}$, $\log x_{noise}(t) \sim {\cal N}(x_{num}(t),\eps^2)$
with standard deviation $\eps=0.1$. The fluctuations are clearly visible on the $\log_b x$ curves shown on the second plot, but have little effect on the inferred rate coefficients.

This idea that \enquote*{effective} rates are not equal to the original rates is very general, and justifies in
particular the multiplicative \enquote*{corrections} of order 1 applied to Lyapunov exponents
$\lambda_{\cal G}$ in \S II B., C. of Suppl. Mat. to improve the fit with numerical trajectories,
since they may be generated from the hierarchical formulas by simply redefining rates.

We computed the probability ratios $MLE(\phi=b)/MLE(\phi=a)$ by starting instead from
$k^{init} \equiv k^{init}[b]$ in the geometric domain of phase b, $k^{init}_off = b^{-0.5}, k^{init}_{on} = b^{-2.5},
k^{init}_2 = b^{-3.5}, \bar{k}_2 = b^{-5.5}, \nu^{init}_+ = b^{-2.5}$ by using the results of the two
alternative maximizations. The one started from $k^{init}[b]$ finds optimal log-rates $\log_b k^*_{\mathrm{off}}[b] = 0.0,
\log_b k^*_{\mathrm{on}}[b] = -3.0, \log_b k^*_2[b] = -3.3, \log_b \bar{k}^*_2[b] = -5.7, \log_b \nu^*_+[b] = -0.8$
located inside phase $b$, with $\ln (MLE(\phi=b)/MLE(\phi=a)) \simeq -1.5\times 10^3$. The odds are so overwhelming that the optimal solution found in phase $b$ may be dismissed as irrelevant.

\subsection{Example 2: two coupled cycles}
\label{sec_example_2}

We consider here \cref{example_2}.

\begin{center}
\begin{tikzpicture}[scale = 0.8]

\draw[dashed](-0.55,-2.4) rectangle(2.55,0.5);
\draw[dashed](4-0.55,-3.4) rectangle(4+2.55,-0.2);
\draw(-1,-1) node {${\cal G}_1$};
\draw(3,-3) node {${\cal G}_2$};

 \draw(2,1) node {$1$};
\draw[->, ultra thick](2,0)--(0,0); \draw(1,0.3) node {$k_{\text{max}}$};
\draw(0,1) node {$2$};

\draw(4,1) node {$\bar{1}$};
\draw[->, ultra thick](4,-1)--(6,-1); \draw(5,-1+0.3) node {$\bar{k}_{\text{max}}$};
\draw(6,1) node {$\bar{2}$};

\draw(8,0) node {$n_1^{(0)}$};
\draw[dashed](-1,-0.5)--(8,-0.5);

\draw[dashed](-1,-2.5)--(8,-2.5); \draw(8,-2) node {$n_3^{(0)}$};
\draw(8,-3) node {$n_4^{(0)}$};

 \draw[->, ultra thick](0,-2)--(2,-2);
\draw(1,-1-0.7) node {$k_{\text{min}}$};

 \draw[->, ultra thick](6,-3)--(4,-3);
\draw(5,-2-0.7) node {$\bar{k}_{\text{min}}$};
\draw(8,-1) node {$n_2^{(0)}$};
\draw[dashed](-1,-1.5)--(8,-1.5);

\draw[->](2,-4)--(4,-4); \draw(3,-2-1.7) node {$k_+$};
\draw(8,-4) node {$n_5^{(0)}$};
\draw[dashed](-1,-3.5)--(8,-3.5);

\draw[<-](2,-5)--(4,-5); \draw(3,-2-2.7) node {$k_-$};
\draw(8,-5) node {$n_6^{(0)}$};
\draw[dashed](-1,-4.5)--(8,-4.5);

\draw[dotted](0,0)--(0,-2);
\draw[dotted](2,0)--(2,-5);
\draw[dotted](4,-1)--(4,-5);
\draw[dotted](6,-1)--(6,-3);

\end{tikzpicture}
\end{center}

\noindent with $k_{\text{max}} = b^0, \bar{k}_{\text{max}} = b^{-2}, \bar{\nu}_{+,1}= b^{-3},
k_{\text{min}} = b^{-5}$, $\bar{k}_{\text{min}} = b^{-6}, \nu_{+,1} = b^{-7}, k_+ = b^{-12}$
and $k_-= b^{-16}$. We chose $b=3$, and obtain the plot on \cref{fig_time_evolution} (right) for the
log-distribution $\log_b X(t)$ starting from $X_i(0)=\delta_{i,2}, i=1,2,\bar{1},\bar{2}$.
The plot becomes very caricatural if $b$ is large, say $b=10$, the first slope $\lambda_1$
being hard to discern.

\bigskip Following our analysis, we have $\eps_1 \sim \frac{\nu_{+,1}}{k_{\text{max}}}= b^{-7}, \eps_2 \sim
\frac{k_+}{k_{\text{min}}} = b^{-7}, \eps_{\bar{1}} \sim \frac{\bar{\nu}_{+,1}}{\bar{k}_{\text{max}}} =
b^{-1}, \eps_{\bar{2}} \sim \frac{k_-}{\bar{k}_{\text{min}}} = b^{-10}$. The original scales (from top to bottom) are
$n^{(0)}_{1,\cdots,8} =n_{max}, \bar{n}_{max}, \bar{n}_{+,1}, n_{min}, \bar{n}_{min}, n_{+,1},
n_+,n_-$.

\medskip (0) At cut-off scale $n^{(0)} = \bar{n}_{+,1}$, edges $1\to 2, \bar{1}\to \bar{2}$ define a two-component reverse DAG with
components rooted at $2,\bar{2}$.

\medskip (1) At cut-off scale $n^{(1)} = n_{min}$ (just below the upper dashed line), the cycle
$1 \underset{k_{\text{min}}}{\overset{k_{\text{max}}}{\rightleftarrows}} 2$ defines a dominant SCC
$G_1 = \{1 \rightleftarrows 2\}$ with renormalized weight $Z(0)_{{\cal G}_1} \sim \frac{k_+}{k_{\text{max}}} \sim b^{-12}$ and
deficiency weight $\bar{\eps}_{G_1} \sim \max(\eps_1,\eps_2)\sim b^{-7} \succ Z(0)_{{\cal G}_1}$. Thus $G_1$ is autocatalytic, and $\lambda_{{\cal G}_1} \sim \bar{\eps}_{G_1} k_{\text{min}} \sim k_+ \sim
b^{-12}$, $Z_{G_1}\equiv Z^{(1)}_{G_1}\sim \bar{\eps}_{G_1}$. The new effective graph has vertices $\{G_1,\bar{1},\bar{2}\}$ and
edges $k_{\bar{1}\to G_1}
\sim k_{\bar{1}\to 1} \sim b^{-16}$, $k_{G_1\to \bar{1}} \sim \frac{k_+}{k_{\text{max}}} k_{\text{min}} \sim b^{-17}$ below the edges connecting $\bar{1},\bar{2}$. The two cores are $G_1$ and $\bar{2}$.
Choosing $\sigma_0=1,2$, $G_{cut,\sigma_0}(1)=\{1\rightleftarrows 2\}$ has Lyapunov eigenvector components
$v^{(1)}_1 \sim k_{\text{max}}^{-1} Z_{G_1}^{-1} \sim k_{\text{min}}/(k_+ k_{\text{max}}) \sim b^{7},
v^{(1)}_2\sim k_{\text{min}}^{-1} Z_{G_1}^{-1}\sim k_+^{-1}\sim b^{12}$; dividing by $b^{12}$ for normalization, $v^{(1)} \sim \left(\begin{array}{c} b^{-5} \\ 1 \\
0 \\ 0 \end{array}\right)$, while $v^{\dagger,(1)}\sim \left(\begin{array}{c} 1 \\ 1 \\ 0 \\
0 \end{array}\right)$.

\medskip (2) At cut-off scale $n^{(2)} =
\bar{n}_{min}$ (just below the lower dashed line), the cycle
$\bar{1} \underset{\bar{k}_{\text{min}}}{\overset{\bar{k}_{\text{max}}}{\rightleftarrows}} \bar{2}$ defines a dominant SCC
$G_2 = \{\bar{1},\bar{2}\}$ with renormalized weight $Z(0)_{{\cal G}_2} \sim \frac{k_{\bar{1}\to G_1}}{\bar{k}_{\text{max}}} \sim b^{-14}$ and
deficiency weight $\bar{\eps}_{G_2} \sim \max(\eps_{\bar{1}},\eps_{\bar{2}})\sim b^{-1} \succ Z(0)_{{\cal G}_2}$. Thus $G_2$ is autocatalytic, and $\lambda_{{\cal G}_2} \sim \bar{\eps}_{G_2} \bar{k}_{\text{min}} \sim \frac{\bar{\nu}_{+,1}}{\bar{k}_{\text{max}}} \bar{k}_{\text{min}} \sim
b^{-7}$, $Z_{G_2}\sim \bar{\eps}_{G_2}$. The new effective graph has vertices $\{G_1,G_2\}$
and edges $k_{G_1\to G_2} \sim k_{G_1\to \bar{1}}$, $k_{G_2\to G_1} \sim \frac{k_{\bar{1}\to G_1}}{k_{\bar{1}}} \bar{k}_{\text{min}} \sim \frac{k_-}{\bar{k}_{\text{max}}} \bar{k}_{\text{min}} \sim b^{-20}$.
Since $\lambda_{{\cal G}_2}\succ \lambda_{{\cal G}_1}$, the new root is $G_2$, and now
$Z_{G_1}\equiv Z^{(2)}_{G_1} \sim Z(\lambda_{{\cal G}_2})_{{\cal G}_1} \sim \frac{\lambda_{{\cal G}_2}}{k_{\text{min}}} \sim b^{-2}$, $w(\lambda_{{\cal G}_2})_{G_2\to G_1} \sim \frac{k_{G_2\to G_1}}{\lambda_{{\cal G}_2}}\sim
\frac{k_-}{\bar{\nu}_{+,1}} \sim b^{-13}$.
Then (using standard basis along $1,2,\bar{1},\bar{2}$) $v^{(\infty)}\equiv v^{(2)} \sim \lambda_{{\cal G}_2}
\left(\begin{array}{c} k_{\text{max}}^{-1} w(\lambda_{{\cal G}_2})_{G_2\to G_1} (Z^{(2)}_{G_1})^{-1} \\
k_{\text{min}}^{-1} w(\lambda_{{\cal G}_2})_{G_2\to G_1} (Z^{(2)}_{G_1})^{-1} \\
 \bar{k}_{\text{max}}^{-1} Z_{G_2}^{-1} \\
 \bar{k}_{\text{min}}^{-1} Z_{G_2}^{-1} \end{array}\right) \sim
 \left(\begin{array}{c} b^{-18} \\ b^{-13} \\ b^{-4} \\ 1 \end{array}\right)$, and $v^{\dagger,(2)}_{1,2} \sim w(\lambda_{{\cal G}_2})_{G_1 \to G_2} \sim \frac{k_{G_1\to G_2}}{\lambda_{{\cal G}_2}}
\sim b^{-10}$ and $v^{\dagger,(2)}_{\bar{1},\bar{2}}\sim 1$.

\bigskip {\em Dynamics.} The model predicts (using the match conditions) the following time-dependent composition.

\medskip (0) For $t\prec 1/k_{\text{max}} \simeq 1$
(compare with Ex. 1, Phase $a$), $X_i(t) = \delta_{i,2}$.

\medskip (1) Let $t:=t^{(2)}_{2\to 1,2}$
be defined as in (\ref{eq:match-condition}) by $e^{\lambda_{{\cal G}_1}t} \sim
v^{\dagger,(2)}_2 e^{\lambda_{{\cal G}_2}t} v^{(2)}_2$, or $t\sim 23\, b^7 \ln(b)\approx 5\times 10^4$. For $1/k_{\text{min}}\prec t \prec t^{(2)}_{2\to 1,2}$,
$\log_b(X_{1,2}(t)) \sim \log_b X_{1,2}^{(1)}(t) \equiv\log_b(e^{\lambda_{{\cal G}_1}t} v^{(1)}_{1,2}) \sim
\frac{\lambda_{{\cal G}_1}}{\ln(b)} t + \left(\begin{array}{c} -5 \\ 0 \end{array}\right)$.

\medskip (2) For $t\succ t^{(2)}_{2\to 1,2}$, the fast increasing
contribution is $X^{(2)}(t) \sim v^{\dagger,(2)}_2 e^{\lambda_{{\cal G}_2}t} v^{(2)}$; thus $\log_b X^{(2)}(t) \sim
\frac{\lambda_{{\cal G}_2}}{\ln(b)} t - 10 + \log_b(v^{(2)}) \sim \frac{\lambda_{{\cal G}_2}}{\ln(b)} t + \left(\begin{array}{c} -28 \\ -23 \\ -14 \\ -10
\end{array}\right)
$.

\Medskip See Suppl. Mat. for agreement with simulations.

\subsection{Example 3: a toy formose model}
\label{sec_example_3}

The formose reaction, observed long ago~\cite{Butlerow1861} -- one of the few known examples of autocatalytic networks of prebiotic chemistry,
easy to obtain in a laboratory at high pH -- is a densely connected
reaction network consisting of oses (sugars) with varying number of carbons, which can merge two by two through
addition reactions called aldol reactions, or conversely, fragment into two pieces by retro-aldol reactions. Dismissing side reactions and possibly important isomer (discussed in Suppl. Mat.) and diastereoisomer effects, leaves
a toy model with only three reaction types coupling abstract molecules $(\mathsf{C}_n)_{n\ge 1}$ with $n$ carbon atoms,

-- $\mathsf{C}_1$-additions {\em (ald1)}, $\mathsf{C}_1 + \mathsf{C}_n \overset{k_{\mathrm{on}}}{\to} \mathsf{C}_{n+1}$ $(n\ge 2)$, allowing
chain polymerization, and the
reverse $\mathsf{C}_1$ retroaldol reactions {$\overline{(ald1)}$}, $ \mathsf{C}_{n+1} \overset{k_{\mathrm{off}}}{\to} \mathsf{C}_1 + \mathsf{C}_n $;

-- $\mathsf{C}_n, n\ge 2$ retroaldol reactions {$\overline{(ald*)}$}, $ \mathsf{C}_{n+m} \overset{\nu_+}{\to} \mathsf{C}_n + \mathsf{C}_m $, $n,m\ge 2$.

In experiments, $\mathsf{C}_1$ (formaldehyde) plays a special role because the addition mechanism does not
allow for $\mathsf{C}_1+\mathsf{C}_1\to \mathsf{C}_2$. Here (as is often the case), $C_1$ is in excess (abundant) and considered as an external species, so that $\mathsf{C}_1$-additions may be written as $C_n \overset{k_{\mathrm{on}}}{\to} C_{n+1}$ up to the
substitution $k_{\mathrm{on}}\to [\mathsf{C}_1]k_{\mathrm{on}}$. In practice one truncates at a given level $n_{max}\ge 4$
by removing all reactions involving $\mathsf{C}_n, n>n_{max}$ as a reactant or a product. Here (see section on formose in
Suppl. Mat. for generalizations) we choose $n_{max}=5$, and consider two regimes, I and II,
depending on the ordering of scales $n_{on},n_{off}$ of $k_{\mathrm{on}},k_{\mathrm{off}}$.

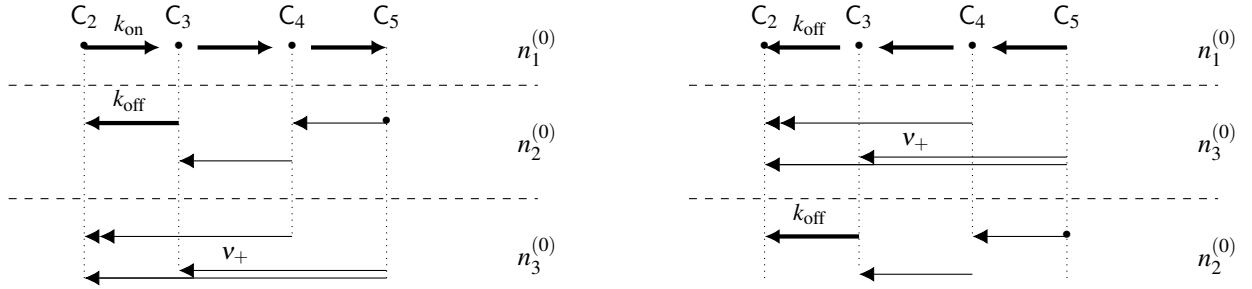
\begin{figure*} \label{fig:ToyFormoseI}
\begin{tikzpicture}
\draw(0,0.4) node {$\mathsf{C}_2$}; \draw(0.6,0.3) node {\small $k_{\mathrm{on}}$};

\draw(1.25,0.4) node {$\mathsf{C}_3$};
\draw(2.75,0.4) node {$\mathsf{C}_4$};
\draw[dotted](0,0)--(0,-3.05);
\draw[dotted](1.25,0)--(1.25,-2.95);
\draw[dotted](2.75,0)--(2.75,-2.5);
\draw[dotted](4,0)--(4,-3.05);

\draw(4,0.4) node {$\mathsf{C}_5$};
\draw(0,0) node {\textbullet};
\draw[->, ultra thick](0,0)--(1,0);
\draw(1.25,0) node {\textbullet};
\draw[->, ultra thick](1.5,0)--(2.5,0);

 \draw(2.75,0) node {\textbullet};
\draw[->, ultra thick](3,0)--(4,0);

\draw(6,0) node {$n_1^{(0)}$};

\draw[dashed](-1,-0.5)--(6,-0.5);

\draw(4,-1) node {\textbullet}; \draw(0.6,-1+0.3) node {\small $k_{\mathrm{off}}$};
\draw[->](4,-1)--(2.75,-1);
\draw[->](2.75,-1.5)--(1.25,-1.5);
\draw[->, ultra thick](1.25,-1)--(0,-1);
\draw(6,-1.25) node {$n_2^{(0)}$};

\draw[dashed](-1,-2)--(6,-2);

\draw[->](2.75,-2.5)--(0.2,-2.5);
\draw[->](0.2,-2.5)--(0,-2.5);

\draw(4,-2.95)--(1.45,-2.95);
\draw[->](1.45,-2.95)--(1.25,-2.95);

\draw(4,-3.05)--(0.2,-3.05);
\draw[->](1.45,-3.05)--(0,-3.05);

\draw(6,-2.75) node {$n_3^{(0)}$};

\draw(2,-2.75) node {\small $\nu_+$};


\begin{scope}[shift={(9,0)}]
\draw(0,0.4) node {$\mathsf{C}_2$}; \draw(0.6,0.3) node {\small $k_{\mathrm{off}}$};
\draw(1.25,0.4) node {$\mathsf{C}_3$};
\draw(2.75,0.4) node {$\mathsf{C}_4$};
\draw[dotted](0,0)--(0,-3.05);
\draw[dotted](1.25,0)--(1.25,-2.95);
\draw[dotted](2.75,0)--(2.75,-2.5);
\draw[dotted](4,0)--(4,-3.05);

\draw(4,0.4) node {$\mathsf{C}_5$};
\draw(0,0) node {\textbullet};
\draw[<-, ultra thick](0,0)--(1,0);
\draw(1.25,0) node {\textbullet};
\draw[<-, ultra thick](1.5,0)--(2.5,0);

\draw(2.75,0) node {\textbullet};
\draw[<-, ultra thick](3,0)--(4,0);

\draw(6,0) node {$n_1^{(0)}$};

\begin{scope}[shift={(0,1.5)}]
\draw[dashed](-1,-2)--(6,-2);

\draw[->](2.75,-2.5)--(0.2,-2.5);
\draw[->](0.2,-2.5)--(0,-2.5);

\draw(4,-2.95)--(1.45,-2.95);
\draw[->](1.45,-2.95)--(1.25,-2.95);

\draw(4,-3.05)--(0.2,-3.05);
\draw[->](1.45,-3.05)--(0,-3.05);

\draw(6,-2.75) node {$n_3^{(0)}$};

\draw(2,-2.75) node {\small $\nu_+$};
\end{scope}

\begin{scope}[shift={(0,-1.5)}]
\draw[dashed](-1,-0.5)--(6,-0.5);

\draw(4,-1) node {\textbullet}; \draw(0.6,-1+0.3) node {\small $k_{\mathrm{off}}$};
\draw[->](4,-1)--(2.75,-1);
\draw[->](2.75,-1.5)--(1.25,-1.5);
\draw[->, ultra thick](1.25,-1)--(0,-1);
\draw(6,-1.25) node {$n_2^{(0)}$};
\end{scope}


\end{scope}
\end{tikzpicture}

\caption{Toy formose I (left), II (right) model}
\end{figure*}

Computations
show that I, resp. II, splits into 3 phases $I_{a,b,c}$, resp. 2 phases $II_{a,b}$. From now on, we simply denote $C_n$ by its index, $n$. By definition, $\eps_2=\eps_3=0$.

For more readability, we represent dominant edges as double arrows $(\Rightarrow)$ instead of
boldface. The source $\sigma$ of a dominant edge $\sigma \Rightarrow \sigma'$ is drawn bold at scale $n_{\sigma} = n_{\sigma\to\sigma'}$.

\medskip {\bf Toy formose (Ia,b,c).} $k_{\mathrm{on}}\succ k_{\mathrm{off}} \succ \nu_+$. Then $\eps_4 \sim \frac{\nu_+}{k_{\mathrm{on}}}$, $\eps_5\sim \frac{\nu_+}{k_{\mathrm{off}}}$. The initial graph ${\mathbb G}(0)$ is
\begin{eqnarray*}
 (k_{\mathrm{on}}) \qquad {\bf 2} \Rightarrow {\bf 3} \Rightarrow {\bf 4} \Rightarrow \textbf{5}
 \\ -----------
 \\ (k_{\mathrm{off}}) \qquad {\bf 5} \Rightarrow 4 \to 3 \to 2
\\
 (\nu_+) \qquad 4\to 2+2, \ 5\to 2+3
\end{eqnarray*}

\medskip
(0) At cut-off scale $n^{(0)}= n_{on}$, $2\to 3 \to 4\to 5$ define a reverse tree rooted at 5. Thus $v^{(0)} \propto \left(\begin{array}{c}
k_{\mathrm{on}} \\ k_{\mathrm{on}} \\ k_{\mathrm{on}}\\ k_{\mathrm{off}} \end{array}\right)^{-1} \, \times\, \left(\begin{array}{c}
0 \\ 0\\ 0 \\ 1 \end{array}\right) = \left(\begin{array}{c} 0 \\ 0 \\ 0 \\ k_{\mathrm{off}}^{-1}
\end{array}\right)$.
(1) At cut-off scale $n^{(1)}=n_{off}$, the cycle $4\underset{k_{\mathrm{off}}}{\overset{k_{\mathrm{on}}}{\rightleftarrows}} 5$ defines a dominant SCC $G_1 = \{4,5\}$ with renormalized weight $Z(0)_{G_1} \sim \max(\frac{k_{4\to 3}}{k_{4\to 5}}, \frac{k_{4\to 2}}{k_{4\to 5}},
\frac{k_{5\to 2}}{k_{5\to 4}}, \frac{k_{5\to 3}}{k_{5\to 4}} ) \sim \max(\frac{k_{\mathrm{off}}}{k_{\mathrm{on}}}, \frac{\nu_+}{k_{\mathrm{off}}})$.

\bigskip {\em Ia phase ($\nu_+ \succ (\frac{k_{\mathrm{off}}}{k_{\mathrm{on}}})^2 k_{\mathrm{on}}$).} Then $Z_{G_1} \sim Z(0)_{{\cal G}_1}\sim
\frac{\nu_+}{k_{\mathrm{off}}}\sim \bar{\eps}_{{\cal G}_1}$. In this transition regime we are unable to decide
whether $G_1$ is autocatalytic or not; in the autocatalytic case, $\lambda^{(1)}\sim \bar{\eps}_{{\cal G}_1} \times k_{\mathrm{off}} \sim \nu_+$.
 The new effective graph is
 \begin{eqnarray*}
(k_{\mathrm{on}}) \qquad {\bf 2} \Rightarrow {\bf 3} \Rightarrow G_1 \\
 (k_{\mathrm{off}}) \qquad 3\to 2 \\
 (k_{G_1}) \qquad {\bf G_1} \Rightarrow 2,3
\end{eqnarray*}
with $k_{G_1\to 2}\sim k_{G_1\to 3} \sim k_{G_1} \sim \frac{k_{5\to 2,3}}{k_5} \times k_{\mathrm{off}} \sim \nu_+$. In both cases, $Z_{G_1}\sim \frac{\nu_+}{k_{\mathrm{off}}}$.
In the autocatalytic case, $ w(\lambda^{(1)})_{G_1\to 2} \sim w(\lambda^{(1)})_{G_1\to 3} \sim 1$,
therefore
$v^{(\infty)}_{Ia} = v^{(1)}_{Ia} \propto \left(\begin{array}{c}
k_{\mathrm{on}} \\ k_{\mathrm{on}} \\ k_{\mathrm{on}}\\ k_{\mathrm{off}} \end{array}\right)^{-1} \, \times\, \left(\begin{array}{c}
1 \\ 1 \\ Z_{G_1}^{-1} \\ Z_{G_1}^{-1} \end{array}\right) \sim \nu_+^{-1} \left(\begin{array}{c} \nu_+/k_{\mathrm{on}}
\\ \nu_+/k_{\mathrm{on}} \\ k_{\mathrm{off}}/ k_{\mathrm{on}} \\ 1 \end{array}\right)$.

\bigskip {\em Ib phase ($(\frac{k_{\mathrm{off}}}{k_{\mathrm{on}}})^3 k_{\mathrm{on}} \prec \nu_+ \prec (\frac{k_{\mathrm{off}}}{k_{\mathrm{on}}})^2 k_{\mathrm{on}}$).} Then $Z_{G_1}\sim Z(0)_{{\cal G}_1}\sim
\frac{k_{\mathrm{off}}}{k_{\mathrm{on}}}$. The new effective graph is
 \begin{eqnarray*}
(k_{\mathrm{on}}) \qquad {\bf 2} \Rightarrow {\bf 3} \Rightarrow G_1 \\
 (k_{\mathrm{off}}) \qquad 3\to 2 \\
 -------- \\
 (k_{G_1}) \qquad {\bf G_1} \Rightarrow 3 \\
 (k_{G_1\to 2}) \qquad G_1 \to 2
\end{eqnarray*}
with $k_{G_1}\sim k_{G_1\to 3}\sim \frac{k_{4\to 3}}{k_4} \times k_{\mathrm{off}} \sim \frac{k_{\mathrm{off}}^2}{k_{\mathrm{on}}} $, $k_{G_1\to 2}\sim \frac{k_{5\to 2}}{k_5} \times k_{\mathrm{off}} \sim \nu_+$ and $\eps_{G_1}\sim \frac{\eps_5}{Z_{G_1}} \sim \frac{\nu_+ k_{\mathrm{on}}}{k_{\mathrm{off}}^2}$. Furthermore, $w_{G_1\to 2} \sim
\frac{k_{G_1\to 2}}{k_{G_1}} \sim
\frac{\nu_+ k_{\mathrm{on}}}{k_{\mathrm{off}}^2}$. Thus $v^{(1)}_{Ib}\propto \left(\begin{array}{c}
k_{\mathrm{on}} \\ k_{\mathrm{on}} \\ k_{\mathrm{on}}\\ k_{\mathrm{off}} \end{array}\right)^{-1} \, \times\, \left(\begin{array}{c}
 w_{G_1\to 2} \\ k_{G_1\to 3}/k_{G_1}\\ Z_{G_1}^{-1} \\ Z_{G_1}^{-1} \end{array}\right) \sim k_{\mathrm{on}}/k_{\mathrm{off}}^2 \left(\begin{array}{c} \nu_+/k_{\mathrm{on}} \\ (k_{\mathrm{off}}/k_{\mathrm{on}})^2 \\ k_{\mathrm{off}}/k_{\mathrm{on}} \\ 1 \end{array}\right) $. (2) At cut-off
 scale $n^{(2)} = n_{G_1}$, the cycle $3\underset{k_{G_1\to 3}}{\overset{k_{\mathrm{on}}}{\rightleftarrows}} G_1$ defines a dominant SCC $G_2 = \{G_1,3\}$ with $Z(0)_{{\cal G}_2} \sim \max(\frac{k_{3\to 2}}{k_3}, \frac{k_{G_1\to 3}}{k_{G_1}}) \sim \max(\frac{k_{\mathrm{off}}}{k_{\mathrm{on}}}, \frac{\nu_+ k_{\mathrm{on}}}{k_{\mathrm{off}}^2})
 \sim \frac{\nu_+ k_{\mathrm{on}}}{k_{\mathrm{off}}^2}$. This is $\sim \bar{\eps}_{{\cal G}_2} \sim \eps_{G_1}$, hence
once again we cannot decide whether it is autocatalytic. In the autocatalytic case, $\lambda^{(2)} \sim \bar{\eps}_{{\cal G}_2} \times k_{G_1} \sim \nu_+$. The new effective graph is
 \begin{eqnarray*}
(k_{\mathrm{on}}) \qquad {\bf 2} \Rightarrow G_2 \\
(k_{G_2}) \qquad {\bf G_2} \Rightarrow 2
\end{eqnarray*}
with $k_{G_2}\sim k_{G_2\to 2} \sim \max(\frac{k_{3\to 2}}{k_3}, \frac{k_{G_1\to 2}}{k_{G_1}}) \, \times \, k_{G_1} \sim \max(\frac{k_{\mathrm{off}}}{k_{\mathrm{on}}},\nu_+ \frac{k_{\mathrm{on}}}{k_{\mathrm{off}}^2})\, \times\, k_{G_1} \sim \nu_+$. In both
 cases, $Z_{G_2} \sim \frac{\nu_+ k_{\mathrm{on}}}{k_{\mathrm{off}}^2}$.
 In the autocatalytic case, $w(\lambda^{(2)})_{G_2\to 2} \sim 1$.
 Thus $v^{(\infty)}_{Ib} \equiv v^{(2)}_{Ib}\propto \left(\begin{array}{c}
k_{\mathrm{on}} \\ k_{\mathrm{on}} \\ k_{\mathrm{on}}\\ k_{\mathrm{off}} \end{array}\right)^{-1} \, \times\, \left(\begin{array}{c}
1 \\ Z_{G_2}^{-1} \\ Z_{G_1}^{-1} Z_{G_2}^{-1} \\ Z_{G_1}^{-1} Z_{G_2}^{-1} \end{array}\right) \sim v^{(\infty)}_{Ia}$.

\medskip {\em Ic phase ($(\nu_+ \prec (\frac{k_{\mathrm{off}}}{k_{\mathrm{on}}})^3 k_{\mathrm{on}}$).} Differences with
Ib phase start with step (2).
Now, $Z_{G_2} \sim Z(0)_{{\cal G}_2} \sim \frac{k_{\mathrm{off}}}{k_{\mathrm{on}}} \succ \bar{\eps}_{{\cal G}_2}$, implying that
 $G_2$ is not autocatalytic. The new effective graph is the same as in phase Ib, but
 now $k_{G_2} \sim k_{G_2\to 2} \sim \frac{k_{3\to 2}}{k_3} \times k_{G_1} \sim \frac{k_{\mathrm{off}}^3}{k_{\mathrm{on}}^2}$ and $\eps_{G_2} \sim \frac{\bar{\eps}_{{\cal G}_2}}{Z(0)_{{\cal G}_2}} \sim
 \frac{\nu_+ k_{\mathrm{on}}^2}{k_{\mathrm{off}}^3}$.
(3) At cut-off scale $n^{(3)} = n_{G_2}$, the cycle $2\underset{k_{G_2\to 2}}{\overset{k_{\mathrm{on}}}{\rightleftarrows}} G_2$ defines an autocatalytic dominant SCC $G_3 = \{G_2,2\}$ with weight
$Z_{G_3}\sim \bar{\eps}_{{\cal G}_3} \sim \eps_{G_2} \sim \frac{\nu_+ k_{\mathrm{on}}^2}{k_{\mathrm{off}}^3}$, exponent
$\lambda^{(3)}\sim \eps_{G_2} \times k_{G_2} \sim \nu_+$ and Lyapunov vector
$v^{(\infty)}_{Ic} \equiv v^{(3)}_{Ic}\propto \left(\begin{array}{c}
k_{\mathrm{on}} \\ k_{\mathrm{on}} \\ k_{\mathrm{on}}\\ k_{\mathrm{off}} \end{array}\right)^{-1} \, \times\, \left(\begin{array}{c}
Z_{G_3}^{-1} \\ Z_{G_2}^{-1}Z_{G_3}^{-1} \\ Z_{G_1}^{-1}Z_{G_2}^{-1}Z_{G_3}^{-1} \\ Z_{G_1}^{-1}Z_{G_2}^{-1}Z_{G_3}^{-1} \end{array}\right)
 \sim \nu_+^{-1} \left(\begin{array}{c} (\frac{k_{\mathrm{off}}}{k_{\mathrm{on}}})^3 \\
 (\frac{k_{\mathrm{off}}}{k_{\mathrm{on}}})^2 \\ \frac{k_{\mathrm{off}}}{k_{\mathrm{on}}} \\
1 \end{array}\right)$.


\bigskip {\bf Toy formose (IIa).} $k_{\mathrm{off}}\succ \nu_+\succ k_{\mathrm{on}}$. Then $\eps_4, \eps_5 \sim \frac{\nu_+}{k_{\mathrm{off}}}$. The initial graph ${\mathbb G}(0)$ is
\begin{eqnarray*}
(k_{\mathrm{off}}) \qquad {\bf 5} \Rightarrow {\bf 4} \Rightarrow {\bf 3} \Rightarrow 2 \\
------------ \\
 (k_{\mathrm{on}}) \qquad {\bf 2} \Rightarrow 3\to 4\to 5 \\
 (\nu_+) \qquad 4\to 2+2, \ 5\to 2+3
\end{eqnarray*}
 (0) At cut-off scale $n_{off}$,
$5\to 4 \to 3\to 2$ define a reverse tree rooted at 2. Thus $v^{(0)}_{IIa} \propto \left(\begin{array}{c}
k_{\mathrm{on}} \\ k_{\mathrm{off}} \\ k_{\mathrm{off}}\\ k_{\mathrm{off}} \end{array}\right)^{-1} \, \times\, \left(\begin{array}{c}
1 \\ 0\\ 0 \\ 0 \end{array}\right) \sim k_{\mathrm{on}}^{-1} \left(\begin{array}{c} 1 \\ 0 \\ 0 \\ 0
\end{array}\right)$. (1) At cut-off scale $n^{(1)}=n_{on}$, the cycle
$3\underset{k_{\mathrm{on}}}{\overset{k_{\mathrm{off}}}{\rightleftarrows}} 2$ defines a non-autocatalytic dominant SCC $G_1 = \{2,3\}$ with renormalized weight $Z_{G_1} \sim \frac{k_{3\to 4}}{k_{3\to 2}}\sim \frac{k_{\mathrm{on}}}{k_{\mathrm{off}}}$. The new effective graph is
\begin{eqnarray*}
(k_{\mathrm{off}}) \qquad {\bf 5} \Rightarrow {\bf 4} \Rightarrow G_1 \\
 (k_{\mathrm{on}}) \qquad 4\to 5 \\
 (k_{G_1}) \qquad {\bf G_1} \Rightarrow 4
\end{eqnarray*}
with $k_{G_1} \sim k_{G_1\to 4} \sim \frac{k_{\mathrm{on}}^2}{k_{\mathrm{off}}}$, and $\eps_{G_1}=0$. Thus
$v^{(1)}_{IIa} \propto \left(\begin{array}{c}
k_{\mathrm{on}} \\ k_{\mathrm{off}} \\ k_{\mathrm{off}}\\ k_{\mathrm{off}} \end{array}\right)^{-1} \, \times\, \left(\begin{array}{c} Z_{G_1}^{-1} \\ Z_{G_1}^{-1} \\ k_{G_1\to 4}/k_{G_1} \\ \frac{k_{G_1\to 4}}{k_{G_1}}\cdot \frac{k_{4\to 5}}{k_4} \end{array}\right) \sim \frac{k_{\mathrm{off}}}{k_{\mathrm{on}}^2} \left(\begin{array}{c} 1 \\ \frac{k_{\mathrm{on}}}{k_{\mathrm{off}}} \\ (\frac{k_{\mathrm{on}}}{k_{\mathrm{off}}})^2 \\ (\frac{k_{\mathrm{on}}}{k_{\mathrm{off}}})^3
 \end{array}\right) $.
 (2) At cut-off scale $n^{(2)}= n_{G_1} $,
the cycle $4\underset{k_{G_1\to 4}}{\overset{k_{\mathrm{off}}}{\rightleftarrows}} G_1$ defines a dominant SCC $G_2 = \{G_1,4\}$, which is autocatalytic since $Z(0)_{{\cal G}_2} \sim \frac{k_{\mathrm{on}}}{k_{\mathrm{off}}}\prec \eps_4$. Thus
$Z_{G_2} \sim \bar{\eps}_{{\cal G}_2}\sim \eps_4\sim\frac{\nu_+}{k_{\mathrm{off}}}$ and $\lambda_{G_2}\sim \eps_4 k_{G_1\to 4} \sim
 (\frac{k_{\mathrm{on}}}{k_{\mathrm{off}}})^2 \nu_+$.
The new effective graph ${\mathbb G}(2)$ is
\begin{eqnarray*}
(k_{\mathrm{off}}) \qquad {\bf 5} \to G_2 \\
(k_{G_2}) \qquad G_2 \to 5
\end{eqnarray*}
 with $k_{G_2} \sim k_{G_2\to 5} \sim \frac{k^3_{on}}{k^2_{off}}$, and $w(\lambda_{G_2})_{G_2\to 5} \sim
 \frac{k_{G_2\to 5}}{\lambda_{G_2}} \sim \frac{k_{\mathrm{on}}}{\nu_+}$. Thus
 $v^{(\infty)}_{IIa} \equiv v^{(2)}_{IIa} \propto \left(\begin{array}{c}
k_{\mathrm{on}} \\ k_{\mathrm{off}} \\ k_{\mathrm{off}}\\ k_{\mathrm{off}} \end{array}\right)^{-1} \times\, \left(\begin{array}{c} Z_{G_1}^{-1} Z_{G_2}^{-1} \\ Z_{G_1}^{-1} Z_{G_2}^{-1} \\ Z_{G_2}^{-1} \\ w(\lambda_{G_2})_{G_2\to 5} \end{array}\right) \sim \frac{k_{\mathrm{off}}}{\nu_+}
v^{(1)}_{IIa}$. Hierarchical graphs are as follows (see \cref{sec_renormalization} for
 color conventions, with cores drawn in red). To the right or at the bottom of each dominant SCC $G_i$, the couple (weight, autocatalytic exponent) $= (Z_{G_i}^{-1},\lambda_{G_i})$ if
 autocatalytic (in red), or simply $Z_{G_i}^{-1}$ if not (in black).

\medskip {\bf Toy formose (IIb).} $k_{\mathrm{off}}\succ k_{\mathrm{on}}\succ \nu_+$. Steps (0), (1) are
identical. (2) $G_2$ is not autocatalytic, $Z_{G_2}:=Z(0)_{{\cal G}_2}$ and $\eps_{G_2}\sim \frac{\bar{\eps}_{{\cal G}_2}}{Z_{G_2}}\sim \frac{\nu_+}{k_{\mathrm{on}}}$. (3) At cut-off scale
$n^{(3)} = n_{G_3}$, the cycle $5 \underset{k_{G_2\to 5}}{\overset{k_{\mathrm{off}}}{\rightleftarrows}} G_2$ defines an auto-catalytic dominant SCC $G_3=\{G_2,5\}$ with $Z_{G_3}\sim
\eps_{G_3}\sim \max(\eps_5, \eps_{G_2}) \sim \frac{\nu_+}{k_{\mathrm{on}}}$, and $\lambda_{G_3} \sim
\frac{\eps_{G_3}}{k_{G_2\to 5}} \sim (\frac{k_{\mathrm{on}}}{k_{\mathrm{off}}})^2 \nu_+$. Thus
$v^{(\infty)}_{IIb} \equiv v^{(3)}_{IIb} \propto \left(\begin{array}{c}
k_{\mathrm{on}} \\ k_{\mathrm{off}} \\ k_{\mathrm{off}}\\ k_{\mathrm{off}} \end{array}\right)^{-1} \, \times\,
\left(\begin{array}{c} Z_{G_1}^{-1} Z_{G_2}^{-1} Z_{G_3}^{-1} \\ Z_{G_1}^{-1} Z_{G_2}^{-1} Z_{G_3}^{-1} \\ (Z_2 Z_3)^{-1} \\ Z_3^{-1} \end{array}\right) \sim \frac{k_{\mathrm{on}}}{\nu_+}
v^{(1)}_{IIa}$. Hierarchical graphs are as follows.

\begin{tikzpicture}[scale=0.45]
\draw(2,3) node {step $2$, $k^{(2)}\sim \frac{k_{\mathrm{on}}^2}{k_{\mathrm{off}}}$};

\draw(-0.3,0) node {$2$}; \draw(2.3,0) node {$3$};
\draw(0,0) node {\textbullet}; \draw(2,0) node {\textbullet};
\draw[->,ultra thick](0,-0.2) arc(210:330:1.15 and 1.15);
\draw[->,ultra thick](2,0.2) arc(30:150:1.15 and 1.15);
\draw(-0.5,-1) rectangle(2.5,1);
\draw(1,1.5) node {$G_{1}$};

\draw[dotted](2.5,1)--(3.5,0.5);
\begin{scope}[shift={(3.5,0.5)}]
\draw(2.3,0) node {$4$};
\draw(0,0) node {\textbullet}; \draw(2,0) node {\textbullet};
\draw[->,ultra thick](0,-0.2) arc(210:330:1.15 and 1.15);
\draw[->,ultra thick](2,0.2) arc(30:150:1.15 and 1.15);
\end{scope}

\draw(-1,-1.5) rectangle(6.5,2); \draw(5,-2) node {$\frac{k_{\mathrm{off}}}{\nu_+}$};

\draw[dotted](-1,-1.5)--(0.5,-3.5);
\begin{scope}[shift={(0.5,-3.5)}]
\draw(2.3,0) node {$5$};
\draw[red](0,0) node {$G_2$};
\draw[->,red](0,-0.2) arc(210:330:1.15 and 1.15);
\draw[red](1,-1.2) node {$k_{\mathrm{on}}/\nu_+$};
\draw[->, ultra thick, blue, dashed](2,0.2) arc(30:150:1.15 and 1.15);
\end{scope}



\begin{scope}[shift={(8.5,0)}]
\draw(3,3.5) node {step $3$, $k^{(3)} \sim \frac{k^3_{on}}{k^2_{off}}$};

\draw(-0.3,0) node {$2$}; \draw(2.3,0) node {$3$};
\draw(0,0) node {\textbullet}; \draw(2,0) node {\textbullet};
\draw[->,ultra thick](0,-0.2) arc(210:330:1.15 and 1.15);
\draw[->,ultra thick](2,0.2) arc(30:150:1.15 and 1.15);
\draw(-0.5,-1) rectangle(2.5,1);
\draw(1,1.5) node {$G_1$};

\draw[dotted](2.5,1)--(3.5,0.5);
\begin{scope}[shift={(3.5,0.5)}]
\draw(2.3,0) node {$4$};
\draw(0,0) node {\textbullet}; \draw(2,0) node {\textbullet};
\draw[->,ultra thick](0,-0.2) arc(210:330:1.15 and 1.15);
\draw[->,ultra thick](2,0.2) arc(30:150:1.15 and 1.15);
\end{scope}

\draw(-1,-1.5) rectangle(6.5,2); \draw(6,-2) node {$G_2$};

\draw[dotted](-1,-1.5)--(0.5,-3.5);
\begin{scope}[shift={(0.5,-3.5)}]
\draw(2.3,0) node {$5$};
\draw(0,0) node {\textbullet}; \draw(2,0) node {\textbullet};
\draw[->, ultra thick](0,-0.2) arc(210:330:1.15 and 1.15);
\draw[->, ultra thick](2,0.2) arc(30:150:1.15 and 1.15);
\end{scope}

\draw[red](-1.5,-5) rectangle(8,2.5); \draw[red](6.5,-6) node {$(\frac{k_{\mathrm{on}}}{\nu_+},
 (\frac{k_{\mathrm{on}}}{k_{\mathrm{off}}})^2 \nu_+)$};

 \draw[dotted](-1.5,-5)--(-0.5,-6); \draw[red](-0.5,-6) node {$G_3$};
\end{scope}
\end{tikzpicture}

Concluding: step $\ge 1$ (in particular, final step) Lyapunov eigenvectors of
 phases $IIa,IIb$ are to leading order proportional to the equilibrium measure, and
$\lambda^*_{IIa,IIb} \sim (\frac{k_{\mathrm{on}}}{k_{\mathrm{off}}})^2 \nu_+$. The two phases, however,
behave differently in case of fluxes or non-uniform degradation. For instance,
consider a selective degradation reaction $5\overset{\beta_5}{\to} \emptyset$. In phase IIa,
the network remains autocatalytic however large is $\beta_5$, since $5$ is not in the support of the
first autocatalytic subgraph met in the algorithm, $G_2$. In phase IIb, however, the edge
$5\to \emptyset$ becomes dominant at step 2 when $\beta_5\succ k_{G_2}$. Thus the autocatalytic
threshold (the maximum value of $\beta_5$ for which the network is autocatalytic) is $\beta_5\sim k_{G_2}$.

A prototype implementation of the algorithm described in \cref{sec_renormalization} was developed in Python to explore the behaviour of the reduction on example networks. A lightweight interface was developed to visualize the reduction process and the resulting graphs. The program takes as input two files: a text file in which each line represents a reaction of the network together with the name of the corresponding parameter, and a tabular file listing all the parameters used in the network, together with their scale and logarithmic base.

These data are first converted into a graph representation where nodes correspond to species and edges correspond to reactions labelled by their parameters' scale. The reduction rules are then applied iteratively until no further reduction is possible, or autocatalytic graphs are found. The program outputs the final reduced graph together with additional structural information, including these graphs.

\begin{figure}[ht]
\centering\includegraphics[width=0.8\linewidth]{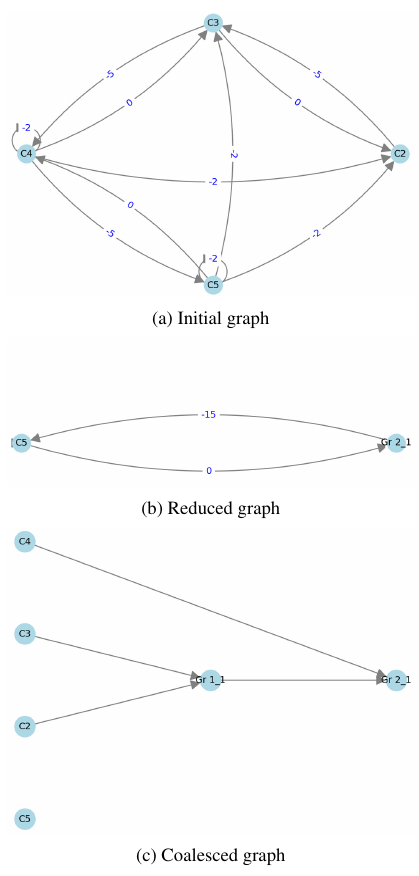}
\caption{Graph reduction and coalescence procedure returned by the program for the Formose IIa model. The initial graph (a) is progressively simplified to obtain the reduced representation (b), followed by a coalesced graph (c) describing the induced dynamics on the aggregated state space.}
\label{fig_streamlit}
\end{figure}

In this example based on the Formose IIa model (see \cref{fig_streamlit}), with parameters chosen as $k_{\mathrm{on}} = 10^{-5}$, $k_{\mathrm{off}} = 1$, and $\nu = 10^{-2}$, the program performs the reduction from the full network representation to the corresponding reduced graph. In addition, a coalescence graph is produced to help identify which species are grouped together during the reduction process. The program then detects an autocatalytic graph $G_{2,1}$ combining the species $C_2, C_3, C_4$. For the chosen parameters, the associated autocatalytic exponent corresponds to a logarithmic scale of $-12$. The algorithm also outputs the symbolic expression of the corresponding autocatalytic exponent, here $\frac{\nu_+ k_{\mathrm{on}}^2}{k_{\mathrm{off}}^2}$, which coincides with the theoretical value of $\lambda_{G_2}$ obtained in the detailed analytical example.

\section{Discussion}
\label{sec_discussion}

We discuss here a few important points that have been left out in the Methods section.

{\em Kinetic scales (see \cref{sec_dominant}). }
 It can be shown that errors due to scale truncation, the \enquote*{sum-max} substitution rule and the perturbative approach underlying the hierarchical algorithm do not accumulate as a result of successive approximations, provided $b$ is large enough, which is the essence of the scale separation hypothesis. How large depends in a complicated way
 on the size (number of species) of the
maximal dominant SCCs; formal mathematical statements
 \cite{Unterberger2025} rely on the limit $b\to\infty$ for a fixed network and are therefore not very useful for real-world applications. However, extended simulations on small- to average-sized networks (the Extended Toy Formose models discussed in Suppl. Mat. contain up to 22 species and 55 reactions) tend to suggest that truncation errors do not invalidate our estimates as soon as $b> 2$ or even less.

{\em Phases (see \cref{sec_phases_thresholds}). } It is possible to represent the rewiring steps by matrices ${\cal M}^{(1)},\ldots,
{\cal M}^{(i_{\mathrm{max}})}$. Choosing any orderings of the rates $k^{(0)},\ldots,k^{(i_{\mathrm{max}})}$, formulas (\ref{eq:ren-rates}) for renormalized rates may be rewritten as
$\log_b(k^{(i)}_p) = \sum_q {\cal M}^{(i)}_{pq} \log_b(k^{(i-1)}_q)$. Iterating, $\log_b(k^{i_{\mathrm{max}}}_p) = {\cal M}^{(i_{\mathrm{max}})}
\cdots {\cal M}^{(1)} \log_b(k^{(0)})$ expresses the last step renormalized rates in terms of the bare ones.
Exploring the various cases described in the algorithm through a decision tree yields a possibly large (the
outcomes of the max rules in the formulas typically vary according to some subset of rate permutations), but finite number of phases. The boundaries of each phase are known and defined by linear inequalities, so that the geometrical domain of a phase is a linear cone. This allows an automated search for phases using linear programming tools. Exhaustive search, however, may be prohibitively time-consuming for large networks.

\medskip {\em Kinetic inference (see \cref{sec_inference}). } The artificial inference example presented here
is a far cry from inference in real experimental conditions.
High-resolution GC-MS$^2$ or LC-MS$^2$ mass spectroscopy allows
a reliable determination of a large list of tabulated species present even in small concentrations in a complex mixture. The signal amplitude allows a determination of the concentrations, up to an unknown multiplicative activation ratio
depending on the binding of the species with the gel, and measurement noise. Finally, in large mixtures there appear untabulated, and by and large unknown, reaction intermediates,
whose raw formulas can however be determined with certainty. These serious caveats may all be categorized as missing data issues, which are in principle within reach of HMM (hidden Markov models) techniques, provided the number of
observations is sufficiently increased.

\section{Perspectives}

Autocatalysis has usually been discussed in relation to stoichiometry alone. The traditional
definition \cite{Blokhuis2020} states that a set of reactions is collectively autocatalytic if there is some
combination whose stoichiometric balance is $>0$ for all species. It completely bypasses
kinetics. The interest of the approach developed in this work is that it does not require the knowledge of kinetic rates, which remains frustratingly scarce and
elusive for reactions in a polar solvent despite the tremendous progress in computational methods. This is also the case of other approaches, that we now review briefly. Parametrizing the dynamics by fluxes instead of concentrations, as in flux-balance
analysis \cite{Palsson2006}, allows various rate-independent bounds on the sensibility of
stationary states on parameters \cite{Kacser1973,Heinrich1974} or on proxy growth rates \cite{Blanco2024}.
The stoichiometric approach has also motivated the search for minimal autocatalytic motifs, which have been classified \cite{Blokhuis2020,Nandan2025}; algorithms have been developed to find them
automatically \cite{Gagrani2024,Kosc2025}. Generally speaking, bioinformatics draws on all these developments to produce
an automated search of synthesis pathways \cite{Andersen2019}. Recent work tends to incorporate thermodynamic constraints to eliminate pathways including reactions with positive Gibbs energy as non viable.

The common drawback of such approaches is precisely that they avoid considering kinetics. Having
applications in mind to prebiotic chemistry, this is however at odds with the fact that most
relevant phenomena are out-of-equilibrium, due to fluxes, chemical gradients, dry-wet cycles, etc., all conjectural but physically natural conditions to leave equilibrium in quest for
possible chemical evolution towards complexity.

Putting kinetics back at the heart of the matter as we do here suggests very different directions
for research. We summarize our findings, which are for now limited to dilute regimes, but should
remain generally valid (though considerably more involved) without this much simplifying assumption. First, the reaction graph is not a static object. Much in the way that the laws of physics depend on the observation scale, we have proved that the effective reaction graph depends on
the kinetic scale. Though Perron-Frobenius matrices are way simpler objects than e.g. particle physics Lagrangians, applying our algorithm to a large battery of examples has taught us how unpredictably
our renormalization rules iteratively transform the \enquote*{bare} network $G^{(0)}\equiv G$. Let us consider a large, very connected network and discuss extreme cases.
If only one scale is present, then it is safe to presume that the spectrum of the generator will
look like that of a random matrix. Probability conservation in absence of irreversible one-to-many and degradation reactions forces the Lyapunov exponent to be zero, but in general, the average Lyapunov exponent will essentially depend on the number of the latter. Our approach has nothing to say about this case, and becomes useful only if rates are split over several scales. The Toy Formose/Extended Toy Formose models involve only three scales, but these are enough
to define a large number of phases with different asymptotic compositions, including $n-2$ phases $I$ for networks containing
carbon chains up to $n$ carbons if $n\ge 4$. For $\nu_+$ ($k_-$) large enough (phase Ia), the maximal dominant SCC $G_1$ appearing at step 1 is already autocatalytic, and it is protected from any degradation acting on external species, i.e. it remains autocatalytic for all values of $\beta_{\sigma}$,
$\sigma\not \in G_1$. This may be seen as a primitive sort of compartmentalization of $G_1$
inside the network. As $\nu_+$ ($k_-$) decreases, exploring successive steps leads to larger
autocatalytic hierarchical subgraphs which are more complex, i.e. contain a larger number
of organization levels materialized by the nested structure. This self-organization certainly
 leads to a low composition entropy level, since it reduces the number of effective vertices. Such subgraphs have a lower
growth rate, but are more robust, in the sense that there is a larger number of \enquote*{bare} wiring
possibilities leading to the same hierarchical graph after the successive rewirings. By contrast, minimal single-scale autocatalytic graphs, i.e. cycles, acquire robustness only by adding bypasses or shunts when available. Phases II
are simpler; they apparently do not depend so much in the cut-off in terms of number of carbon atoms, involve only 2 or 3 steps, and
the asymptotic composition is a perturbation of the power-law equilibrium measure typical of step-growth polymerization \cite{Gnanou2008}.

Obviously, linear models can only have a limited complexity if the latter is intended as measuring the unpredictability of
time trajectories, since ultimately the \enquote*{fittest} wins, i.e. the subgraph
with largest growth rate becomes the only core if the network is connected, precluding any
multistability. There is however a path to go beyond this limitation: given a general
reaction network with non-linear mass-action dynamics, locate the stationary composition scales, which are \enquote*{fat} stationary points; linearize the equations of motion around these; determine the (complex-valued, in general) Lyapunov data of the linearized generators; and predict trajectories
away from them. This program -- a far cry from the hopeless topological classification of dynamical systems --, which deliberately shies away from chaotic regimes, is under way, but how generic and applicable to complex, real-life chemical systems it will be remains to be determined.

Finally, this program must be coupled with a reverse network inference method to make the bridge with actual chemistry. We conjecture that varying fluxes makes it possible in theory to determine completely the hierarchical structure and characteristic elements. Starting from there, we have designed a finely tuned statistical inference procedure that must be tested and coupled with an experiment design scheme. Let us just mention that it will allow, in the case of large experimental networks, not to explore systematically all phases, whose number grows a priori exponentially with the number of reactions. Also, the Bayesian framework will produce a statistical superposition of neighboring hierarchical networks instead of a single one, thus allowing a good fit in spite of the precision cut-offs inherent in the definition of the scales.


\begin{acknowledgments}
J.U. thanks Philippe Nghe for his kind help in the preparation of this manuscript through numerous discussions and suggestions for improvement.
\end{acknowledgments}

\section*{Data Availability Statement}

An implementation of the scale-splitting renormalization algorithm in Python, along with examples and a graphical user interface, are available at \url{https://github.com/Unterberger/ChemNetInference}.

Data sharing is not applicable to this article as no new data were created or analyzed in this study.

\appendix

\section{Main notation}
\label{sec_main_notation}

We discuss here in more detail the notations, Markov chain and graph-theoretic concepts used in \cref{sec_hierarchical_formulas}.

\medskip
{\em Open reaction networks} (\cref{sec_framework}). See e.g. \onlinecite{Rao2016}. An open chemical reaction network is specified
by (i) a set of external species, $\Sigma^{ext}$; (ii) a set of (internal) species, $\Sigma$; a
stoichiometry matrix ${\mathbb S} = ({\mathbb S}_{\sigma,\rho})_{\sigma\in \Sigma\cup\Sigma^{ext},\rho \in R}$ with
integer coefficients. Decompose ${\mathbb S}$ as a difference ${\mathbb S}^+ - {\mathbb S}^-$ by letting
${\mathbb S}^+_{i,j} = \max(0, {\mathbb S}_{i,j})$, ${\mathbb S}^-_{i,j} = \max(0, -{\mathbb S}_{i,j})$. Then each column $j$ of ${\mathbb S}$ is interpreted as a reaction $\rho_j:
\sum_{i\in \Sigma} {\mathbb S}^-_{i,j} \sigma_i \to \sum_{i\in \Sigma} {\mathbb S}^+_{i,j} \sigma_i$
involving species in the set $(\sigma_i)_i \simeq \Sigma\cup \Sigma^{ext}$,
 while $(\rho_j)_j \simeq R$. (Internal) species $\sigma\in \Sigma$ such that
 ${\mathbb S}^-_{\sigma,\rho}>0$ are called
(internal) reactants of $\rho$. We generally assume the reaction rates to be mass-action rates, that is,
we consider the associated ODE system for the species concentrations,
$ dX_{\sigma}/dt = \sum_{\rho \in R} {\mathbb S}_{\sigma,\rho} {\cal J}^{\rho}(X) $, where
${\cal J}^{\rho}(X) = k^{\rho} \prod_{\sigma\in\Sigma\cup\Sigma^{ext}} X_{\sigma}^{{\mathbb S}^-_{\sigma,\rho}}$ are
the fluxes of the system. The constants $(k_{\rho})$ are the kinetic rates.

\medskip {\em Deficiency rate.} The {\em order} $p(\rho)$ of a uni-molecular reaction is by definition
the number of its products. Let
\begin{equation} \kappa_{\sigma} := \sum_{\rho\ |\ p(\rho)=2,\ \rho: X_{\sigma}\to \cdots}
k^{\rho} \label{eq:deficiency-rate}
\end{equation}
be the sum of the kinetic rates of all $1\to 2$ reactions $\rho$ with reactant $\sigma$.
If $\kappa_{\sigma}>0$, we add to $G$ the {\em self-edge} $\sigma\overset{\kappa_{\sigma}}{\to} \sigma$ (then edges $\sigma\to\sigma'\not=\sigma$ are called {\em non-trivial}). By
construction,
 $\sum_{\sigma'} A_{\sigma',\sigma} = \kappa_{\sigma}$. If there are no $1\to 2$ reactions,
 then $\kappa_{\sigma}=0$, and $\sum_{\sigma'} A_{\sigma',\sigma}=0$ expresses Markov chain probability
 conservation.

\medskip {\em Associated Markov chain.} We let $\tilde{A}=(\tilde{A}_{\sigma',\sigma})_{\sigma',\sigma\in \Sigma}$
be the matrix with off-diagonal coefficients $\tilde{A}_{\sigma',\sigma} =A_{\sigma',\sigma}$ $(\sigma\not=\sigma')$ and negative diagonal coefficients

\begin{equation} \tilde{A}_{\sigma,\sigma}=-\sum_{\sigma'\not=\sigma} A_{\sigma',\sigma} = A_{\sigma,\sigma}-\kappa_{\sigma} = -k_{\sigma}.
\end{equation}

\medskip {\em Overall degradation rate.} Let $\alpha\ge 0$ be a constant. Then
$A(\alpha):= A - \alpha \Id$ has same off-diagonal coefficients as $A$, and diagonal coefficients
$|A(\alpha)_{\sigma,\sigma}|= |A_{\sigma,\sigma}| +\alpha$

\bibliography{bibliography}

\begin{thebibliography}{38}%
\makeatletter
\providecommand \@ifxundefined [1]{%
 \@ifx{#1\undefined}
}%
\providecommand \@ifnum [1]{%
 \ifnum #1\expandafter \@firstoftwo
 \else \expandafter \@secondoftwo
 \fi
}%
\providecommand \@ifx [1]{%
 \ifx #1\expandafter \@firstoftwo
 \else \expandafter \@secondoftwo
 \fi
}%
\providecommand \natexlab [1]{#1}%
\providecommand \enquote  [1]{``#1''}%
\providecommand \bibnamefont  [1]{#1}%
\providecommand \bibfnamefont [1]{#1}%
\providecommand \citenamefont [1]{#1}%
\providecommand \href@noop [0]{\@secondoftwo}%
\providecommand \href [0]{\begingroup \@sanitize@url \@href}%
\providecommand \@href[1]{\@@startlink{#1}\@@href}%
\providecommand \@@href[1]{\endgroup#1\@@endlink}%
\providecommand \@sanitize@url [0]{\catcode `\\12\catcode `\$12\catcode
  `\&12\catcode `\#12\catcode `\^12\catcode `\_12\catcode `\%12\relax}%
\providecommand \@@startlink[1]{}%
\providecommand \@@endlink[0]{}%
\providecommand \url  [0]{\begingroup\@sanitize@url \@url }%
\providecommand \@url [1]{\endgroup\@href {#1}{\urlprefix }}%
\providecommand \urlprefix  [0]{URL }%
\providecommand \Eprint [0]{\href }%
\providecommand \doibase [0]{https://doi.org/}%
\providecommand \selectlanguage [0]{\@gobble}%
\providecommand \bibinfo  [0]{\@secondoftwo}%
\providecommand \bibfield  [0]{\@secondoftwo}%
\providecommand \translation [1]{[#1]}%
\providecommand \BibitemOpen [0]{}%
\providecommand \bibitemStop [0]{}%
\providecommand \bibitemNoStop [0]{.\EOS\space}%
\providecommand \EOS [0]{\spacefactor3000\relax}%
\providecommand \BibitemShut  [1]{\csname bibitem#1\endcsname}%
\let\auto@bib@innerbib\@empty
\bibitem [{\citenamefont {Semenov}\ \emph {et~al.}(2016)\citenamefont
  {Semenov}, \citenamefont {Kraft}, \citenamefont {Ainla}, \citenamefont
  {Zhao}, \citenamefont {Baghbanzadeh}, \citenamefont {Campbell}, \citenamefont
  {Kang}, \citenamefont {Fox},\ and\ \citenamefont {Whitesides}}]{Semenov2016}%
  \BibitemOpen
  \bibfield  {author} {\bibinfo {author} {\bibfnamefont {S.~N.}\ \bibnamefont
  {Semenov}}, \bibinfo {author} {\bibfnamefont {L.~J.}\ \bibnamefont {Kraft}},
  \bibinfo {author} {\bibfnamefont {A.}~\bibnamefont {Ainla}}, \bibinfo
  {author} {\bibfnamefont {M.}~\bibnamefont {Zhao}}, \bibinfo {author}
  {\bibfnamefont {M.}~\bibnamefont {Baghbanzadeh}}, \bibinfo {author}
  {\bibfnamefont {V.~E.}\ \bibnamefont {Campbell}}, \bibinfo {author}
  {\bibfnamefont {K.}~\bibnamefont {Kang}}, \bibinfo {author} {\bibfnamefont
  {J.~M.}\ \bibnamefont {Fox}},\ and\ \bibinfo {author} {\bibfnamefont {G.~M.}\
  \bibnamefont {Whitesides}},\ }\href {https://doi.org/10.1038/nature19776}
  {\bibfield  {journal} {\bibinfo  {journal} {Nature}\ }\textbf {\bibinfo
  {volume} {537}},\ \bibinfo {pages} {656} (\bibinfo {year}
  {2016})}\BibitemShut {NoStop}%
\bibitem [{\citenamefont {Muchowska}\ \emph {et~al.}(2019)\citenamefont
  {Muchowska}, \citenamefont {Varma},\ and\ \citenamefont
  {Moran}}]{Muchowska2019}%
  \BibitemOpen
  \bibfield  {author} {\bibinfo {author} {\bibfnamefont {K.~B.}\ \bibnamefont
  {Muchowska}}, \bibinfo {author} {\bibfnamefont {S.~J.}\ \bibnamefont
  {Varma}},\ and\ \bibinfo {author} {\bibfnamefont {J.}~\bibnamefont {Moran}},\
  }\href {https://doi.org/10.1038/s41586-019-1151-1} {\bibfield  {journal}
  {\bibinfo  {journal} {Nature}\ }\textbf {\bibinfo {volume} {569}},\ \bibinfo
  {pages} {104} (\bibinfo {year} {2019})}\BibitemShut {NoStop}%
\bibitem [{\citenamefont {Robinson}\ \emph {et~al.}(2022)\citenamefont
  {Robinson}, \citenamefont {Daines}, \citenamefont {Van~Duppen}, \citenamefont
  {De~Jong},\ and\ \citenamefont {Huck}}]{Robinson2022}%
  \BibitemOpen
  \bibfield  {author} {\bibinfo {author} {\bibfnamefont {W.~E.}\ \bibnamefont
  {Robinson}}, \bibinfo {author} {\bibfnamefont {E.}~\bibnamefont {Daines}},
  \bibinfo {author} {\bibfnamefont {P.}~\bibnamefont {Van~Duppen}}, \bibinfo
  {author} {\bibfnamefont {T.}~\bibnamefont {De~Jong}},\ and\ \bibinfo {author}
  {\bibfnamefont {W.~T.~S.}\ \bibnamefont {Huck}},\ }\href
  {https://doi.org/10.1038/s41557-022-00956-7} {\bibfield  {journal} {\bibinfo
  {journal} {Nature Chemistry}\ }\textbf {\bibinfo {volume} {14}},\ \bibinfo
  {pages} {623} (\bibinfo {year} {2022})}\BibitemShut {NoStop}%
\bibitem [{\citenamefont {Grassi}\ \emph {et~al.}(2022)\citenamefont {Grassi},
  \citenamefont {Nauman}, \citenamefont {Ramsey}, \citenamefont {Bovino},
  \citenamefont {Picogna},\ and\ \citenamefont {Ercolano}}]{Grassi2022}%
  \BibitemOpen
  \bibfield  {author} {\bibinfo {author} {\bibfnamefont {T.}~\bibnamefont
  {Grassi}}, \bibinfo {author} {\bibfnamefont {F.}~\bibnamefont {Nauman}},
  \bibinfo {author} {\bibfnamefont {J.~P.}\ \bibnamefont {Ramsey}}, \bibinfo
  {author} {\bibfnamefont {S.}~\bibnamefont {Bovino}}, \bibinfo {author}
  {\bibfnamefont {G.}~\bibnamefont {Picogna}},\ and\ \bibinfo {author}
  {\bibfnamefont {B.}~\bibnamefont {Ercolano}},\ }\href
  {https://doi.org/10.1051/0004-6361/202039956} {\bibfield  {journal} {\bibinfo
   {journal} {Astronomy \& Astrophysics}\ }\textbf {\bibinfo {volume} {668}},\
  \bibinfo {pages} {A139} (\bibinfo {year} {2022})}\BibitemShut {NoStop}%
\bibitem [{\citenamefont {Lee}\ and\ \citenamefont {Othmer}(2010)}]{Lee2010}%
  \BibitemOpen
  \bibfield  {author} {\bibinfo {author} {\bibfnamefont {C.~H.}\ \bibnamefont
  {Lee}}\ and\ \bibinfo {author} {\bibfnamefont {H.~G.}\ \bibnamefont
  {Othmer}},\ }\href {https://doi.org/10.1007/s00285-009-0269-4} {\bibfield
  {journal} {\bibinfo  {journal} {Journal of Mathematical Biology}\ }\textbf
  {\bibinfo {volume} {60}},\ \bibinfo {pages} {387} (\bibinfo {year}
  {2010})}\BibitemShut {NoStop}%
\bibitem [{\citenamefont {Kan}\ \emph {et~al.}(2016)\citenamefont {Kan},
  \citenamefont {Lee},\ and\ \citenamefont {Othmer}}]{Kan2016}%
  \BibitemOpen
  \bibfield  {author} {\bibinfo {author} {\bibfnamefont {X.}~\bibnamefont
  {Kan}}, \bibinfo {author} {\bibfnamefont {C.~H.}\ \bibnamefont {Lee}},\ and\
  \bibinfo {author} {\bibfnamefont {H.~G.}\ \bibnamefont {Othmer}},\ }\href
  {https://doi.org/10.1007/s00285-016-0980-x} {\bibfield  {journal} {\bibinfo
  {journal} {Journal of Mathematical Biology}\ }\textbf {\bibinfo {volume}
  {73}},\ \bibinfo {pages} {1081} (\bibinfo {year} {2016})}\BibitemShut
  {NoStop}%
\bibitem [{\citenamefont {E}\ \emph {et~al.}(2007)\citenamefont {E},
  \citenamefont {Liu},\ and\ \citenamefont {{Vanden-Eijnden}}}]{E2007}%
  \BibitemOpen
  \bibfield  {author} {\bibinfo {author} {\bibfnamefont {W.}~\bibnamefont {E}},
  \bibinfo {author} {\bibfnamefont {D.}~\bibnamefont {Liu}},\ and\ \bibinfo
  {author} {\bibfnamefont {E.}~\bibnamefont {{Vanden-Eijnden}}},\ }\href
  {https://doi.org/10.1016/j.jcp.2006.06.019} {\bibfield  {journal} {\bibinfo
  {journal} {Journal of Computational Physics}\ }\textbf {\bibinfo {volume}
  {221}},\ \bibinfo {pages} {158} (\bibinfo {year} {2007})}\BibitemShut
  {NoStop}%
\bibitem [{\citenamefont {Sinitsyn}\ \emph {et~al.}(2009)\citenamefont
  {Sinitsyn}, \citenamefont {Hengartner},\ and\ \citenamefont
  {Nemenman}}]{Sinitsyn2009}%
  \BibitemOpen
  \bibfield  {author} {\bibinfo {author} {\bibfnamefont {N.~A.}\ \bibnamefont
  {Sinitsyn}}, \bibinfo {author} {\bibfnamefont {N.}~\bibnamefont
  {Hengartner}},\ and\ \bibinfo {author} {\bibfnamefont {I.}~\bibnamefont
  {Nemenman}},\ }\href {https://doi.org/10.1073/pnas.0809340106} {\bibfield
  {journal} {\bibinfo  {journal} {Proceedings of the National Academy of
  Sciences}\ }\textbf {\bibinfo {volume} {106}},\ \bibinfo {pages} {10546}
  (\bibinfo {year} {2009})}\BibitemShut {NoStop}%
\bibitem [{\citenamefont {Sinanoglu}(1975)}]{Sinanoglu1975}%
  \BibitemOpen
  \bibfield  {author} {\bibinfo {author} {\bibfnamefont {O.}~\bibnamefont
  {Sinanoglu}},\ }\href {https://doi.org/10.1021/ja00842a001} {\bibfield
  {journal} {\bibinfo  {journal} {Journal of the American Chemical Society}\
  }\textbf {\bibinfo {volume} {97}},\ \bibinfo {pages} {2309} (\bibinfo {year}
  {1975})}\BibitemShut {NoStop}%
\bibitem [{\citenamefont {Hirono}\ \emph {et~al.}(2021)\citenamefont {Hirono},
  \citenamefont {Okada}, \citenamefont {Miyazaki},\ and\ \citenamefont
  {Hidaka}}]{Hirono2021}%
  \BibitemOpen
  \bibfield  {author} {\bibinfo {author} {\bibfnamefont {Y.}~\bibnamefont
  {Hirono}}, \bibinfo {author} {\bibfnamefont {T.}~\bibnamefont {Okada}},
  \bibinfo {author} {\bibfnamefont {H.}~\bibnamefont {Miyazaki}},\ and\
  \bibinfo {author} {\bibfnamefont {Y.}~\bibnamefont {Hidaka}},\ }\href
  {https://doi.org/10.1103/PhysRevResearch.3.043123} {\bibfield  {journal}
  {\bibinfo  {journal} {Physical Review Research}\ }\textbf {\bibinfo {volume}
  {3}},\ \bibinfo {pages} {043123} (\bibinfo {year} {2021})}\BibitemShut
  {NoStop}%
\bibitem [{\citenamefont {Katsoulakis}\ and\ \citenamefont
  {Vilanova}(2020)}]{Katsoulakis2020}%
  \BibitemOpen
  \bibfield  {author} {\bibinfo {author} {\bibfnamefont {M.~A.}\ \bibnamefont
  {Katsoulakis}}\ and\ \bibinfo {author} {\bibfnamefont {P.}~\bibnamefont
  {Vilanova}},\ }\href {https://doi.org/10.1016/j.jcp.2019.108997} {\bibfield
  {journal} {\bibinfo  {journal} {Journal of Computational Physics}\ }\textbf
  {\bibinfo {volume} {401}},\ \bibinfo {pages} {108997} (\bibinfo {year}
  {2020})}\BibitemShut {NoStop}%
\bibitem [{\citenamefont {Gabrielli}\ \emph {et~al.}(2025)\citenamefont
  {Gabrielli}, \citenamefont {Garlaschelli}, \citenamefont {Patil},\ and\
  \citenamefont {Serrano}}]{Gabrielli2025}%
  \BibitemOpen
  \bibfield  {author} {\bibinfo {author} {\bibfnamefont {A.}~\bibnamefont
  {Gabrielli}}, \bibinfo {author} {\bibfnamefont {D.}~\bibnamefont
  {Garlaschelli}}, \bibinfo {author} {\bibfnamefont {S.~P.}\ \bibnamefont
  {Patil}},\ and\ \bibinfo {author} {\bibfnamefont {M.~{\'A}.}\ \bibnamefont
  {Serrano}},\ }\href {https://doi.org/10.1038/s42254-025-00817-5} {\bibfield
  {journal} {\bibinfo  {journal} {Nature Reviews Physics}\ }\textbf {\bibinfo
  {volume} {7}},\ \bibinfo {pages} {203} (\bibinfo {year} {2025})}\BibitemShut
  {NoStop}%
\bibitem [{\citenamefont {Villegas}\ \emph {et~al.}(2023)\citenamefont
  {Villegas}, \citenamefont {Gili}, \citenamefont {Caldarelli},\ and\
  \citenamefont {Gabrielli}}]{Villegas2023}%
  \BibitemOpen
  \bibfield  {author} {\bibinfo {author} {\bibfnamefont {P.}~\bibnamefont
  {Villegas}}, \bibinfo {author} {\bibfnamefont {T.}~\bibnamefont {Gili}},
  \bibinfo {author} {\bibfnamefont {G.}~\bibnamefont {Caldarelli}},\ and\
  \bibinfo {author} {\bibfnamefont {A.}~\bibnamefont {Gabrielli}},\ }\href
  {https://doi.org/10.1038/s41567-022-01866-8} {\bibfield  {journal} {\bibinfo
  {journal} {Nature Physics}\ }\textbf {\bibinfo {volume} {19}},\ \bibinfo
  {pages} {445} (\bibinfo {year} {2023})}\BibitemShut {NoStop}%
\bibitem [{\citenamefont {Villegas}\ \emph {et~al.}(2025)\citenamefont
  {Villegas}, \citenamefont {Gabrielli}, \citenamefont {Poggialini},\ and\
  \citenamefont {Gili}}]{Villegas2025}%
  \BibitemOpen
  \bibfield  {author} {\bibinfo {author} {\bibfnamefont {P.}~\bibnamefont
  {Villegas}}, \bibinfo {author} {\bibfnamefont {A.}~\bibnamefont {Gabrielli}},
  \bibinfo {author} {\bibfnamefont {A.}~\bibnamefont {Poggialini}},\ and\
  \bibinfo {author} {\bibfnamefont {T.}~\bibnamefont {Gili}},\ }\href
  {https://doi.org/10.1103/PhysRevResearch.7.013065} {\bibfield  {journal}
  {\bibinfo  {journal} {Physical Review Research}\ }\textbf {\bibinfo {volume}
  {7}},\ \bibinfo {pages} {013065} (\bibinfo {year} {2025})}\BibitemShut
  {NoStop}%
\bibitem [{\citenamefont {Henze}\ \emph {et~al.}(2019)\citenamefont {Henze},
  \citenamefont {Mu}, \citenamefont {Puljiz}, \citenamefont {Kamaleson},
  \citenamefont {Huwald}, \citenamefont {Haslegrave}, \citenamefont
  {Di~Fenizio}, \citenamefont {Parker}, \citenamefont {Good}, \citenamefont
  {Rowe}, \citenamefont {Ibrahim},\ and\ \citenamefont {Dittrich}}]{Henze2019}%
  \BibitemOpen
  \bibfield  {author} {\bibinfo {author} {\bibfnamefont {R.}~\bibnamefont
  {Henze}}, \bibinfo {author} {\bibfnamefont {C.}~\bibnamefont {Mu}}, \bibinfo
  {author} {\bibfnamefont {M.}~\bibnamefont {Puljiz}}, \bibinfo {author}
  {\bibfnamefont {N.}~\bibnamefont {Kamaleson}}, \bibinfo {author}
  {\bibfnamefont {J.}~\bibnamefont {Huwald}}, \bibinfo {author} {\bibfnamefont
  {J.}~\bibnamefont {Haslegrave}}, \bibinfo {author} {\bibfnamefont {P.~S.}\
  \bibnamefont {Di~Fenizio}}, \bibinfo {author} {\bibfnamefont
  {D.}~\bibnamefont {Parker}}, \bibinfo {author} {\bibfnamefont
  {C.}~\bibnamefont {Good}}, \bibinfo {author} {\bibfnamefont {J.~E.}\
  \bibnamefont {Rowe}}, \bibinfo {author} {\bibfnamefont {B.}~\bibnamefont
  {Ibrahim}},\ and\ \bibinfo {author} {\bibfnamefont {P.}~\bibnamefont
  {Dittrich}},\ }\href {https://doi.org/10.1038/s41598-019-40648-w} {\bibfield
  {journal} {\bibinfo  {journal} {Scientific Reports}\ }\textbf {\bibinfo
  {volume} {9}},\ \bibinfo {pages} {3902} (\bibinfo {year} {2019})}\BibitemShut
  {NoStop}%
\bibitem [{\citenamefont {Peng}\ \emph {et~al.}(2022)\citenamefont {Peng},
  \citenamefont {Linderoth},\ and\ \citenamefont {Baum}}]{Peng2022}%
  \BibitemOpen
  \bibfield  {author} {\bibinfo {author} {\bibfnamefont {Z.}~\bibnamefont
  {Peng}}, \bibinfo {author} {\bibfnamefont {J.}~\bibnamefont {Linderoth}},\
  and\ \bibinfo {author} {\bibfnamefont {D.~A.}\ \bibnamefont {Baum}},\ }\href
  {https://doi.org/10.1371/journal.pcbi.1010498} {\bibfield  {journal}
  {\bibinfo  {journal} {PLOS Computational Biology}\ }\textbf {\bibinfo
  {volume} {18}},\ \bibinfo {pages} {e1010498} (\bibinfo {year}
  {2022})}\BibitemShut {NoStop}%
\bibitem [{\citenamefont {Anderson}(1991)}]{Anderson1991}%
  \BibitemOpen
  \bibfield  {author} {\bibinfo {author} {\bibfnamefont {W.~J.}\ \bibnamefont
  {Anderson}},\ }\href {https://doi.org/10.1007/978-1-4612-3038-0} {\emph
  {\bibinfo {title} {Continuous-time {{Markov}} chains: {{An}}
  applications-oriented approach}}},\ Springer {{Series}} in {{Statistics}}\
  (\bibinfo  {publisher} {Springer New York},\ \bibinfo {address} {New York,
  NY},\ \bibinfo {year} {1991})\BibitemShut {NoStop}%
\bibitem [{\citenamefont {Unterberger}\ and\ \citenamefont
  {Nghe}(2022)}]{Unterberger2022}%
  \BibitemOpen
  \bibfield  {author} {\bibinfo {author} {\bibfnamefont {J.}~\bibnamefont
  {Unterberger}}\ and\ \bibinfo {author} {\bibfnamefont {P.}~\bibnamefont
  {Nghe}},\ }\href {https://doi.org/10.1007/s00285-022-01798-0} {\bibfield
  {journal} {\bibinfo  {journal} {Journal of Mathematical Biology}\ }\textbf
  {\bibinfo {volume} {85}},\ \bibinfo {pages} {26} (\bibinfo {year}
  {2022})}\BibitemShut {NoStop}%
\bibitem [{\citenamefont {Eigen}\ \emph {et~al.}(1988)\citenamefont {Eigen},
  \citenamefont {McCaskill},\ and\ \citenamefont {Schuster}}]{Eigen1988}%
  \BibitemOpen
  \bibfield  {author} {\bibinfo {author} {\bibfnamefont {M.}~\bibnamefont
  {Eigen}}, \bibinfo {author} {\bibfnamefont {J.}~\bibnamefont {McCaskill}},\
  and\ \bibinfo {author} {\bibfnamefont {P.}~\bibnamefont {Schuster}},\ }\href
  {https://doi.org/10.1021/j100335a010} {\bibfield  {journal} {\bibinfo
  {journal} {The Journal of Physical Chemistry}\ }\textbf {\bibinfo {volume}
  {92}},\ \bibinfo {pages} {6881} (\bibinfo {year} {1988})}\BibitemShut
  {NoStop}%
\bibitem [{\citenamefont {Pross}\ and\ \citenamefont
  {Pascal}(2023)}]{Pross2023}%
  \BibitemOpen
  \bibfield  {author} {\bibinfo {author} {\bibfnamefont {A.}~\bibnamefont
  {Pross}}\ and\ \bibinfo {author} {\bibfnamefont {R.}~\bibnamefont {Pascal}},\
  }\href {https://doi.org/10.3390/life13112171} {\bibfield  {journal} {\bibinfo
   {journal} {Life}\ }\textbf {\bibinfo {volume} {13}},\ \bibinfo {pages}
  {2171} (\bibinfo {year} {2023})}\BibitemShut {NoStop}%
\bibitem [{\citenamefont {Wilson}(1983)}]{Wilson1983}%
  \BibitemOpen
  \bibfield  {author} {\bibinfo {author} {\bibfnamefont {K.~G.}\ \bibnamefont
  {Wilson}},\ }\href {https://doi.org/10.1103/RevModPhys.55.583} {\bibfield
  {journal} {\bibinfo  {journal} {Reviews of Modern Physics}\ }\textbf
  {\bibinfo {volume} {55}},\ \bibinfo {pages} {583} (\bibinfo {year}
  {1983})}\BibitemShut {NoStop}%
\bibitem [{\citenamefont {Mastropietro}(2008)}]{Mastropietro2008}%
  \BibitemOpen
  \bibfield  {author} {\bibinfo {author} {\bibfnamefont {V.}~\bibnamefont
  {Mastropietro}},\ }\href {https://doi.org/10.1142/6748} {\emph {\bibinfo
  {title} {Non-perturbative renormalization}}}\ (\bibinfo  {publisher} {WORLD
  SCIENTIFIC},\ \bibinfo {year} {2008})\BibitemShut {NoStop}%
\bibitem [{\citenamefont {Unterberger}(2012)}]{Unterberger2012}%
  \BibitemOpen
  \bibfield  {author} {\bibinfo {author} {\bibfnamefont {J.}~\bibnamefont
  {Unterberger}},\ }\bibfield  {journal} {\bibinfo  {journal} {Confluentes
  Mathematici}\ }\textbf {\bibinfo {volume} {4}},\ \href
  {https://doi.org/10.1142/S179374421240004X} {10.1142/S179374421240004X}
  (\bibinfo {year} {2012})\BibitemShut {NoStop}%
\bibitem [{\citenamefont {Li}\ \emph {et~al.}(2008)\citenamefont {Li},
  \citenamefont {Shen},\ and\ \citenamefont {Li}}]{Li2008}%
  \BibitemOpen
  \bibfield  {author} {\bibinfo {author} {\bibfnamefont {B.}~\bibnamefont
  {Li}}, \bibinfo {author} {\bibfnamefont {Y.}~\bibnamefont {Shen}},\ and\
  \bibinfo {author} {\bibfnamefont {B.}~\bibnamefont {Li}},\ }\href
  {https://doi.org/10.1021/jp077597q} {\bibfield  {journal} {\bibinfo
  {journal} {The Journal of Physical Chemistry A}\ }\textbf {\bibinfo {volume}
  {112}},\ \bibinfo {pages} {2311} (\bibinfo {year} {2008})}\BibitemShut
  {NoStop}%
\bibitem [{\citenamefont {Rao}\ and\ \citenamefont {Esposito}(2016)}]{Rao2016}%
  \BibitemOpen
  \bibfield  {author} {\bibinfo {author} {\bibfnamefont {R.}~\bibnamefont
  {Rao}}\ and\ \bibinfo {author} {\bibfnamefont {M.}~\bibnamefont {Esposito}},\
  }\href {https://doi.org/10.1103/PhysRevX.6.041064} {\bibfield  {journal}
  {\bibinfo  {journal} {Physical Review X}\ }\textbf {\bibinfo {volume} {6}},\
  \bibinfo {pages} {041064} (\bibinfo {year} {2016})}\BibitemShut {NoStop}%
\bibitem [{\citenamefont {Blokhuis}\ \emph {et~al.}(2020)\citenamefont
  {Blokhuis}, \citenamefont {Lacoste},\ and\ \citenamefont
  {Nghe}}]{Blokhuis2020}%
  \BibitemOpen
  \bibfield  {author} {\bibinfo {author} {\bibfnamefont {A.}~\bibnamefont
  {Blokhuis}}, \bibinfo {author} {\bibfnamefont {D.}~\bibnamefont {Lacoste}},\
  and\ \bibinfo {author} {\bibfnamefont {P.}~\bibnamefont {Nghe}},\ }\href
  {https://doi.org/10.1073/pnas.2013527117} {\bibfield  {journal} {\bibinfo
  {journal} {Proceedings of the National Academy of Sciences}\ }\textbf
  {\bibinfo {volume} {117}},\ \bibinfo {pages} {25230} (\bibinfo {year}
  {2020})}\BibitemShut {NoStop}%
\bibitem [{\citenamefont {Norris}(1997)}]{Norris1997}%
  \BibitemOpen
  \bibfield  {author} {\bibinfo {author} {\bibfnamefont {J.~R.}\ \bibnamefont
  {Norris}},\ }\href {https://doi.org/10.1017/CBO9780511810633} {\emph
  {\bibinfo {title} {Markov chains}}},\ \bibinfo {edition} {1st}\ ed.\
  (\bibinfo  {publisher} {Cambridge University Press},\ \bibinfo {year}
  {1997})\BibitemShut {NoStop}%
\bibitem [{\citenamefont {Unterberger}(2025)}]{Unterberger2025}%
  \BibitemOpen
  \bibfield  {author} {\bibinfo {author} {\bibfnamefont {J.}~\bibnamefont
  {Unterberger}},\ }\href {https://doi.org/10.48550/ARXIV.2511.11073} {\bibinfo
  {title} {General multi-scale estimates for {{Lyapunov}} data of
  {{Perron-Frobenius}} matrices. {{The}} case of diluted autocatalytic chemical
  reaction networks}} (\bibinfo {year} {2025})\BibitemShut {NoStop}%
\bibitem [{\citenamefont {Butlerow}(1861)}]{Butlerow1861}%
  \BibitemOpen
  \bibfield  {author} {\bibinfo {author} {\bibfnamefont {A.}~\bibnamefont
  {Butlerow}},\ }\href {https://doi.org/10.1002/jlac.18611200308} {\bibfield
  {journal} {\bibinfo  {journal} {Justus Liebigs Annalen der Chemie}\ }\textbf
  {\bibinfo {volume} {120}},\ \bibinfo {pages} {295} (\bibinfo {year}
  {1861})}\BibitemShut {NoStop}%
\bibitem [{\citenamefont {Palsson}(2006)}]{Palsson2006}%
  \BibitemOpen
  \bibfield  {author} {\bibinfo {author} {\bibfnamefont {B.~{\O}.}\
  \bibnamefont {Palsson}},\ }\href {https://doi.org/10.1017/CBO9780511790515}
  {\emph {\bibinfo {title} {Systems biology: {{Properties}} of reconstructed
  networks}}},\ \bibinfo {edition} {1st}\ ed.\ (\bibinfo  {publisher}
  {Cambridge University Press},\ \bibinfo {year} {2006})\BibitemShut {NoStop}%
\bibitem [{\citenamefont {Kacser}\ and\ \citenamefont
  {Burns}(1973)}]{Kacser1973}%
  \BibitemOpen
  \bibfield  {author} {\bibinfo {author} {\bibfnamefont {H.}~\bibnamefont
  {Kacser}}\ and\ \bibinfo {author} {\bibfnamefont {J.~A.}\ \bibnamefont
  {Burns}},\ }\href@noop {} {\bibfield  {journal} {\bibinfo  {journal}
  {Symposia of the Society for Experimental Biology}\ }\textbf {\bibinfo
  {volume} {27}},\ \bibinfo {pages} {65} (\bibinfo {year} {1973})}\BibitemShut
  {NoStop}%
\bibitem [{\citenamefont {Heinrich}\ and\ \citenamefont
  {Rapoport}(1974)}]{Heinrich1974}%
  \BibitemOpen
  \bibfield  {author} {\bibinfo {author} {\bibfnamefont {R.}~\bibnamefont
  {Heinrich}}\ and\ \bibinfo {author} {\bibfnamefont {T.~A.}\ \bibnamefont
  {Rapoport}},\ }\href {https://doi.org/10.1111/j.1432-1033.1974.tb03318.x}
  {\bibfield  {journal} {\bibinfo  {journal} {European Journal of
  Biochemistry}\ }\textbf {\bibinfo {volume} {42}},\ \bibinfo {pages} {89}
  (\bibinfo {year} {1974})}\BibitemShut {NoStop}%
\bibitem [{\citenamefont {Blanco}\ \emph {et~al.}(2024)\citenamefont {Blanco},
  \citenamefont {Gonz{\'a}lez},\ and\ \citenamefont {Gagrani}}]{Blanco2024}%
  \BibitemOpen
  \bibfield  {author} {\bibinfo {author} {\bibfnamefont {V.}~\bibnamefont
  {Blanco}}, \bibinfo {author} {\bibfnamefont {G.}~\bibnamefont
  {Gonz{\'a}lez}},\ and\ \bibinfo {author} {\bibfnamefont {P.}~\bibnamefont
  {Gagrani}},\ }\href {https://doi.org/10.48550/ARXIV.2412.15776} {\bibinfo
  {title} {Identifying self-amplifying hypergraph structures through
  mathematical optimization}} (\bibinfo {year} {2024})\BibitemShut {NoStop}%
\bibitem [{\citenamefont {Nandan}\ \emph {et~al.}(2025)\citenamefont {Nandan},
  \citenamefont {Nghe},\ and\ \citenamefont {Unterberger}}]{Nandan2025}%
  \BibitemOpen
  \bibfield  {author} {\bibinfo {author} {\bibfnamefont {P.}~\bibnamefont
  {Nandan}}, \bibinfo {author} {\bibfnamefont {P.}~\bibnamefont {Nghe}},\ and\
  \bibinfo {author} {\bibfnamefont {J.}~\bibnamefont {Unterberger}},\ }\href
  {https://doi.org/10.48550/ARXIV.2507.15546} {\bibinfo {title} {Autocatalytic
  cores in the diluted regime: classification and properties}} (\bibinfo {year}
  {2025})\BibitemShut {NoStop}%
\bibitem [{\citenamefont {Gagrani}\ \emph {et~al.}(2024)\citenamefont
  {Gagrani}, \citenamefont {Blanco}, \citenamefont {Smith},\ and\ \citenamefont
  {Baum}}]{Gagrani2024}%
  \BibitemOpen
  \bibfield  {author} {\bibinfo {author} {\bibfnamefont {P.}~\bibnamefont
  {Gagrani}}, \bibinfo {author} {\bibfnamefont {V.}~\bibnamefont {Blanco}},
  \bibinfo {author} {\bibfnamefont {E.}~\bibnamefont {Smith}},\ and\ \bibinfo
  {author} {\bibfnamefont {D.}~\bibnamefont {Baum}},\ }\href
  {https://doi.org/10.1007/s10910-024-01576-x} {\bibfield  {journal} {\bibinfo
  {journal} {Journal of Mathematical Chemistry}\ }\textbf {\bibinfo {volume}
  {62}},\ \bibinfo {pages} {1012} (\bibinfo {year} {2024})}\BibitemShut
  {NoStop}%
\bibitem [{\citenamefont {Kosc}\ \emph {et~al.}(2025)\citenamefont {Kosc},
  \citenamefont {Kuperberg}, \citenamefont {Rajon},\ and\ \citenamefont
  {Charlat}}]{Kosc2025}%
  \BibitemOpen
  \bibfield  {author} {\bibinfo {author} {\bibfnamefont {T.}~\bibnamefont
  {Kosc}}, \bibinfo {author} {\bibfnamefont {D.}~\bibnamefont {Kuperberg}},
  \bibinfo {author} {\bibfnamefont {E.}~\bibnamefont {Rajon}},\ and\ \bibinfo
  {author} {\bibfnamefont {S.}~\bibnamefont {Charlat}},\ }\href
  {https://doi.org/10.1073/pnas.2421274122} {\bibfield  {journal} {\bibinfo
  {journal} {Proceedings of the National Academy of Sciences}\ }\textbf
  {\bibinfo {volume} {122}},\ \bibinfo {pages} {e2421274122} (\bibinfo {year}
  {2025})}\BibitemShut {NoStop}%
\bibitem [{\citenamefont {Andersen}\ \emph {et~al.}(2019)\citenamefont
  {Andersen}, \citenamefont {Flamm}, \citenamefont {Merkle},\ and\
  \citenamefont {Stadler}}]{Andersen2019}%
  \BibitemOpen
  \bibfield  {author} {\bibinfo {author} {\bibfnamefont {J.~L.}\ \bibnamefont
  {Andersen}}, \bibinfo {author} {\bibfnamefont {C.}~\bibnamefont {Flamm}},
  \bibinfo {author} {\bibfnamefont {D.}~\bibnamefont {Merkle}},\ and\ \bibinfo
  {author} {\bibfnamefont {P.~F.}\ \bibnamefont {Stadler}},\ }\href
  {https://doi.org/10.1109/TCBB.2017.2781724} {\bibfield  {journal} {\bibinfo
  {journal} {IEEE/ACM Transactions on Computational Biology and
  Bioinformatics}\ }\textbf {\bibinfo {volume} {16}},\ \bibinfo {pages} {510}
  (\bibinfo {year} {2019})}\BibitemShut {NoStop}%
\bibitem [{\citenamefont {Gnanou}\ and\ \citenamefont
  {Fontanille}(2008)}]{Gnanou2008}%
  \BibitemOpen
  \bibfield  {author} {\bibinfo {author} {\bibfnamefont {Y.}~\bibnamefont
  {Gnanou}}\ and\ \bibinfo {author} {\bibfnamefont {M.}~\bibnamefont
  {Fontanille}},\ }\href {https://doi.org/10.1002/9780470238127} {\emph
  {\bibinfo {title} {Organic and physical chemistry of polymers}}},\ \bibinfo
  {edition} {1st}\ ed.\ (\bibinfo  {publisher} {Wiley},\ \bibinfo {year}
  {2008})\BibitemShut {NoStop}%
\end{thebibliography}%

\end{document}